\documentclass[twoside,11pt]{article}
\usepackage{jmlr2e}
\usepackage{lastpage}
\usepackage{amsmath,color}
\usepackage{graphicx,psfrag,epsf}
\usepackage{enumerate}
\usepackage{natbib}
\usepackage{url} % not crucial - just used below for the URL 
\usepackage{amssymb}
\usepackage{multirow}
\usepackage{array}
\usepackage{bm}
\usepackage{booktabs}
\usepackage{mathtools}
\usepackage{setspace}
\usepackage{algorithm2e,algpseudocode}

\newcommand\independent{\protect\mathpalette{\protect\independenT}{\perp}}
\def\independenT#1#2{\mathrel{\rlap{$#1#2$}\mkern2mu{#1#2}}}
\usepackage[font=small,labelfont=bf]{caption}
\usepackage{lastpage}
\newcommand{\Mean}{{\mbox{E}}}

\newcommand{\Cov}{{\mbox{cov}}}

\newcommand{\prob}{{\mbox{Pr}}}
\newtheorem{thm}{Theorem}

\DeclareMathOperator*{\argmax}{arg\,max}

\begin{document}:

\title{An Online Sequential Test for Qualitative Treatment Effects}

\author{\name Chengchun Shi \email{c.shi7@lse.ac.uk}\\ 
%\affil{
\addr Department of Statistics, London School of Economics and Political Science\\ %, London\\ WC2A 2AE, U.K\\
%	\email{c.shi7@lse.ac.uk} \email{T.Xu12@lse.ac.uk} \email{w.p.bergsma@lse.ac.uk}}
\AND
\name Shikai Luo \email{sluo198912@163.com}\\
\addr Tecent PCG\\
\AND
\name Hongtu Zhu \email{htzhu@email.unc.edu}\\
\addr Department of Biostatistics, University of North-Carolina\\
\AND
\name Rui Song \email{rsong@ncsu.edu}\\
\addr Department of Statistics, North-Carolina State University} %\email{lexinli@berkeley.edu}}

\editor{}%Xiaotong Shen}%Hui Zou}%KY}

\maketitle	
\graphicspath{{fig/}}
%\def\spacingset#1{\renewcommand{\baselinestretch}%
%{#1}\small\normalsize} \spacingset{1}
%%
%%
%%%%%%%%%%%%%%%%%%%%%%%%%%%%%%%%%%%%%%%%%%%%%%%%%%%%%%%%%%%%%%%%%%%%%%%%%%%%%%%%
%\if0\blind
%{
%			\title{\bf Determining the number of latent factors in multi-relational learning}
%%			\author{Chengchun Shi, Wenbin Lu and Rui Song\thanks{Chengchun Shi is graduate student (E-mail: cshi4@ncsu.edu), Wenbin Lu is Professor (E-mail: lu@stat.ncsu.edu), and Rui Song is Associate Professor (rsong@ncsu.edu), Department of Statistics, North Carolina State University, Raleigh, NC 27695.}	
%			\date{\empty}
%%		}
%			\maketitle
%	} \fi
%	
%	
%
%
%\baselineskip=21pt

\begin{abstract}%
	Tech companies (e.g., Google or Facebook) often use randomized online experiments and/or A/B testing primarily 
	based on the average treatment effects to compare their new product with an old one. However,   it is also critically important to detect qualitative treatment effects such that 
	the new one  may significantly outperform   the existing one only  under some specific circumstances.   
	The aim of this paper is to develop a powerful testing procedure to efficiently  detect such qualitative treatment effects.  We  propose a scalable online updating algorithm to implement our test procedure. 
	It has three novelties including   adaptive randomization, sequential monitoring,  and online updating with guaranteed type-I error control. We also thoroughly examine the theoretical properties of our testing procedure including the limiting distribution of test statistics and the justification of an efficient bootstrap method. Extensive empirical studies are conducted to examine the finite sample performance of our test procedure. %An implementation is available at  {\url{https://drive.google.com/drive/folders/1EA6P3-Wset1lXf7DoIPnAV_x7eAc3miv?usp=sharing}}
\end{abstract}


\section{Introduction}
Tech companies use randomized online experiments, or A/B testing to compare their new product with a well-established one. Most works  in the literature focus on the average treatment effects (ATE) between the new and existing products \citep[see][and the references therein]{kharitonov2015,johari2015,johari2017,yang2017,ju2019}. In addition to ATE, sometimes we are interested in locating the subgroup (if exists) that the new product performs significantly better than the existing one, as early as possible. Consider a ride-hailing company (e.g., Uber). Suppose some passengers are in the recession state (at a high risk of stopping using the company’s app) and the company comes up with certain strategy to intervene the recession process. We would like to {\color{black}test} if there are some subgroups that are sensitive to the strategy and pin-point these subgroups if exists. 
%However, even if the new product is no better than the existing one on average, there could be some circumstances under which the new one performs significantly better. Applying the new product under these circumstances could further   increase  company's profit. %(or improve the customer satisfaction).
It motivates us to consider
%%conditional on a set of auxiliary variables  \cite{Athey2016,taddy2016,Wager2018}. 
%%In this paper,   we consider 
the null hypothesis that the treatment effect is nonpositive for all passenger. 

Such a null hypothesis is closely related to the notion of qualitative treatment effects in medical studies \citep[QTE,][]{gail1985,gunter2007variable, gunter2011,Roth2018,shi202sparse}, and conditional moment inequalities in economics \citep[see for example,][]{AandS2013,AandS2014,Chernozhukov2013,Armstrong2016,Lee2015,hsu2017}. However, these tests %statistics in the literature  \cite{Athey2016,taddy2016,Wager2018} 
are computed offline and might not be suitable to implement in online settings. Moreover, it is assumed in those  papers that observations are independent. In online experiment, one may wish to adaptively allocate the treatment based on the observed data stream in order to maximize the cumulative reward or to detect the alternative more efficiently. The independence assumption is thus violated. In addition, 
%Moreover, in tech companies, 
an online experiment is desired to be terminated as early as possible in order to save time and budget. %To the best of our knowledge,
Sequential testing for qualitative treatment effects has been less explored. %yet. 

\begin{figure}[!t]
	\centering
	\includegraphics[width=12cm]{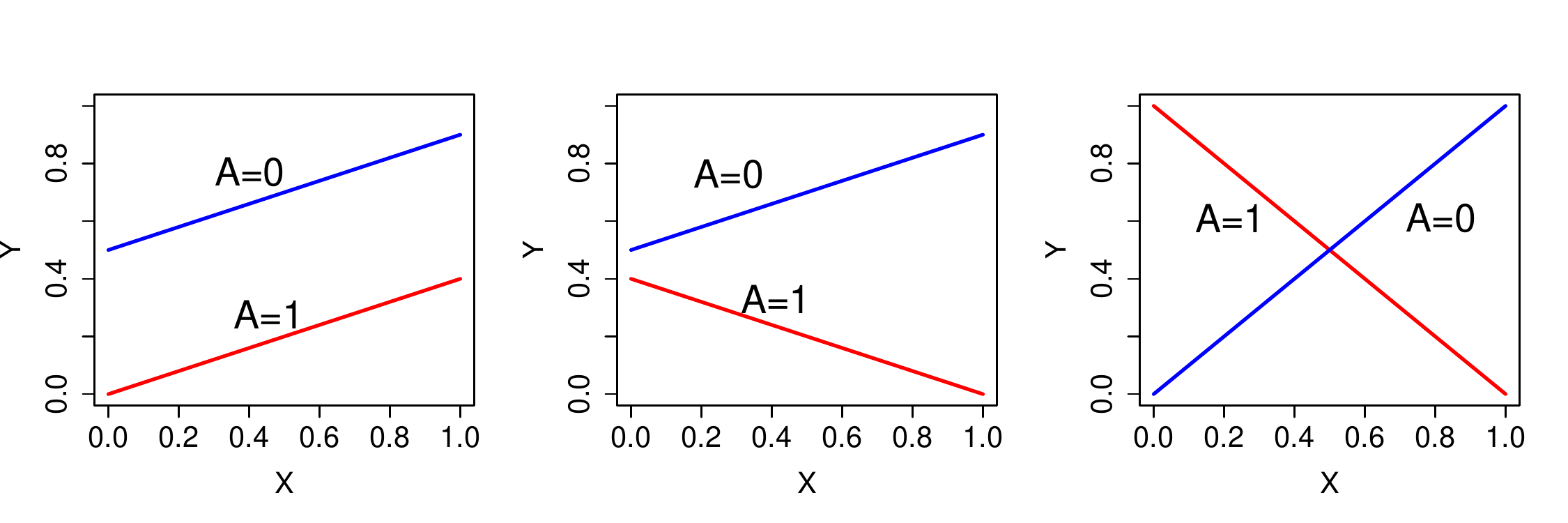}
	
	\caption{Plots demonstrating QTE. $X$ denotes the observed covariates, $A$ denotes the received treatment and $Y$ denotes the associated reward. In the ride-hailing example, $X$ is a feature vector describing the characteristics of a passenger, $A$ is a binary strategy indicator and $Y$ is the passenger's number of rides in the following two weeks. In the left panel, the treatment effect does not depend on $X$. Neither HTE nor QTE exists in this case. In the middle panel, HTE exists. However, the treatment effect is always negative. As such, QTE does not exist. In the right penal, both QTE and HTE exist. 
	}\label{fig0}
\end{figure}

{\color{black}In the literature, there is a line of research on estimation and inference of the heterogeneous treatment effects \citep[HTE,][]{Athey2016,taddy2016,Wager2018,yu2020new}. In particular, \cite{yu2020new} proposed an online test for HTE. We remark that HTE and QTE are related yet fundamentally different hypotheses. There are cases where HTE exists whereas QTE does not. See Figure \ref{fig0} for an illustration. Consequently, applying their test will fail in our setting.}

%The focus of this paper is to propose a valid testing procedure for QTE that allows online updating, adaptive allocation and sequential monitoring. 
The contributions of this paper are summarized as follows. First, we propose a new testing procedure for treatment comparison based on the notion of QTE. 
When the null hypothesis is not rejected, the new product is no better than the control for any realization of covariates, and  thus it is not useful at all. Otherwise, the company could implement different products according to the auxiliary covariates observed, to maximize the average reward obtained. {\color{black}We remark that there are plenty cases where the treatment effects are always
	nonpositive \citep[see Section 5 of ][]{Lee2015,shi202sparse}. A by-product of our test is that it yields a decision rule to implement personalization when the null is rejected (see Section \ref{sec:testlimit} for details).}
Although we primarily focus on QTE in this paper, our procedure can be easily extended to testing ATE as well (see Section \ref{sec:ATE} for details). 
%when no covariates are collected, our test compares ATE between the new and existing products. 

Second, we propose a scalable online updating algorithm to implement our test. To allow for sequential monitoring, our procedure leverages idea from the $\alpha$ spending function approach \citep{Lan1983} originally designed for sequential analysis in a clinical trial \citep[see][for an overview]{jennison1999}. Classical sequential tests focus on ATE. The test statistic at each interim stage is asymptotically normal and the stopping boundary can be recursively updated via numerical integration. However, the limiting distribution of the proposed test statistic does not have a tractable analytical form, making the numerical integration method difficult to apply. To resolve this issue, we propose a scalable bootstrap-assisted procedure to determine the stopping boundary. 
%To determine its critical value at each interim stage, we propose a scalable bootstrap algorithm. 

Third, %theoretical properties of the proposed test are thoroughly investigated. 
we adopt a theoretical framework that allows   the maximum number of interim analyses $K$ to diverge as the number of observations increases, since tech companies might analyze the results every few minutes (or hours) to determine whether to stop the experiment or continue collecting more data. 
It is ultimately different from   classical sequential analysis where  $K$ is fixed.  
Moreover, the derivation of the asymptotic property of the proposed test is further complicated due to the adaptive randomization procedure,  which makes  observations dependent of each other. Despite these technical challenges, we %thoroughly studied the theoretical properties of the proposed test. We not only prove that its type-I error rate can be well controlled, but also
%Our major theorems (Theorem \ref{thm1} and \ref{thm4}) 
establish a nonasymptotic upper bound on the type-I error rate by  explicitly characterizing the conditions needed on randomization procedure, $K$ and the number of samples observed at the initial decision point to ensure the validity of our test.
\section{Background and problem formulation}\label{secback}
We propose a potential outcome framework to formulate our problem. Suppose that we have two products including the control and the treatment. The observed data at time point $t$ consists of a sequence of triples $\{(X_i,A_i,Y_i)\}_{i=1}^{N(t)}$, where $N(\cdot)$ is a counting process that is independent of the data stream $\{(X_i,A_i,Y_i)\}_{i=1}^{+\infty}$, $A_i$ is a binary random variable indicating the product executed for the $i$-th experiment, $X_i\in \mathbb{R}^p$ denotes the associated covariates, and $Y_i$ stands for the associated reward (the larger the better by convention). We allow $A_i$ to depend on $X_i$ and past observations $\{(X_j,A_j,Y_j)\}_{j<i}$ so that the randomization procedure can be adaptively changed. 
In addition, define $Y_i^*(0)$ and $Y_i^*(1)$ to be the potential outcome that would have been observed if the corresponding product is executed for the $i$-th experiment. Suppose that $\{(X_i,Y_i^*(0),Y_i^*(1))\}_{i=1}^{+\infty}$ are independently and identically distributed copies of $(X,Y^*(0), Y^*(1))$.  %where the covariance matrix of the random vector $X$ is non-degenerate. 
Let $\mathbb{X}$ be the support of $X$ and $Q_0(x,a)=\Mean \{Y^*(a)|X=x\}$ for $a=0,1$, we focus on testing the following hypotheses:
\begin{eqnarray*}
	H_0: Q_0(x,1) \le Q_0(x,0),\forall x\in \mathbb{X} \,\,\,\,\hbox{versus}\,\,\,\,H_1: Q_0(x,1)> Q_0(x,0),\exists x\in \mathbb{X}.
\end{eqnarray*}
Notice that when there are no covariates, i.e., $\mathbb{X}=\emptyset$,  the hypotheses are reduced to $H_0: \tau_0\le 0$ versus $H_1:\tau_0 > 0$,  where $\tau_0$ corresponds to   ATE, i.e, $\tau_0=\Mean \{Y^*(1)-Y^*(0)\}$. %Thus, the proposed test includes 
%When $H_0$ holds, the treatment is no better than the control for any realization of $X$, and is thus not of practical interests. When $H_1$ holds, the company could implement different products according to the decision function $\argmax_{a\in \{0,1\}} \Mean \{Y^*(a)|X=x\}$,
%to maximize the average reward obtained. 
%according to the auxiliary information contained in $X$. When no covariates are collected, i.e, $\mathbb{X}=\emptyset$, it reduces to test the unconditional hypothesis: $H_0:\Mean Y^*(0)\le \Mean Y^*(1)$ vs $H_1:\Mean Y^*(0)>\Mean Y^*(1)$.
%To simplify the computation, we assume%posit linear models for the conditional mean of the potential outcomes. Specifically, assume
In general, we require $\mathbb{X}$ to be a compact set. 
We consider a large linear approximation space $\mathcal{Q}$ for the conditional mean function $Q_0$. Specifically, let $\mathcal{Q}=\{Q(x,a;\beta_0,\beta_1)=\varphi^\top(x) \beta_a:\beta_0,\beta_1\in \mathbb{R}^{q} \}$ be the approximation space, where $\varphi(x)$ is a $q$-dimensional vector composed of basis functions on $\mathbb{X}$. The dimension $q$ is allowed to {\color{black}diverge with the number of observations} in order to alleviate the effects of model misspecification. The use of linear approximation space simplifies the computation of our testing procedure. %When \eqref{md} holds, it is equivalent to test
{\color{black}When $Q_0(x,0)$ and $Q_0(x,1)$ are well approximated by $\varphi^\top(x)\beta_0^*$ and $\varphi^\top(x)\beta_1^*$ for some $\beta_0^*$ and $\beta_1^*$,} %(see Assumption A3 in Section \ref{sectest}), 
it suffices to test
\begin{eqnarray}\label{hnull}
	H_0: \varphi^\top(x) (\beta_1^*-\beta_0^*) \le 0,\forall x\in \mathbb{X}\,\,\,\,\hbox{versus}\,\,\,\,H_1: \varphi^\top(x) (\beta_1^*-\beta_0^*)>0,\exists x\in \mathbb{X}.
\end{eqnarray}
{\color{black}We require the approximation error $\hbox{err}=\inf_{\beta_0,\beta_1\in \mathbb{R}^p}\sup_{x\in \mathbb{X}, a\in \{0,1\}} |Q_0(x,a)-Q(x,a;\beta_0,\beta_1)|$ to decay to zero at a rate of $o\{N^{-1/2}(T)\}$ to ensure the validity of the proposed test. See Appendix \ref{secAE} for details.}
%For clarity,  here we assume $Q_0(x,a)=Q(x,a;\beta_0^*,\beta_1^*)$ for some $\beta_0^*$ and $\beta_1^*$. In Appendix \ref{secAE}, we allow the approximation error $\inf_{\beta_0,\beta_1\in \mathbb{R}^p}\sup_{x\in \mathbb{X}, a\in \{0,1\}} |Q_0(x,a)-Q(x,a;\beta_0,\beta_1)|$ to be nonzero. %and decrease with the sample size. 

Let $\mathcal{F}_j$ denote the sub-dataset $\{(X_i,A_i,Y_i)\}_{1\le i\le j}$ for $j\ge 1$ and $\mathcal{F}_0=\emptyset$. Throughout this paper, we assume that  the following two assumptions hold. 

\smallskip 
\noindent (A1) $Y_i=A_i Y_i^*(1)+(1-A_i) Y_i^*(0)$ for $\forall i\ge 1$. \\
\noindent (A2)  $A_i$ is independent of $Y_i^*(0),Y_i^*(1)$, $\{(X_k,Y_k^*(0),Y_k^*(1))\}_{k>i}$ given $X_i$ and $\mathcal{F}_{i-1}$, for any $i$. %holds for $\forall i\ge 1$, where the notation $Z_1\independent Z_2|Z_3$ means $Z_1$ and $Z_2$ are independent conditional on $Z_3$. 
%\noindent (A3) $\forall i\ge 1,a\in \{0,1\},x\in \mathbb{X}$, $\pi_{i-1}(a,x)=\prob(A_i=a|X_i=x,\mathcal{F}_{i-1})>0$, almost surely.

\smallskip 

Assumption (A1) is referred to be the stable unit treatment value assumption and Assumption (A2) is the sequential randomization assumption \citep{zhang2013robust} and is automatically satisfied in a randomized study where the treatments are independently generated of the observed data. %and (A3) is the positivity assumption. 
(A2) essentially assumes there is no unmeasured confounders. These assumptions guarantee that both regression coefficients (defined through potential outcomes) are estimable from the observed dataset as shown in the following lemma.
%For brevity, 
%We summarize this result in Lemma \ref{lemmakeyequation} of the supplementary article. %Under (A1)-(A3), we have
\begin{lemma}\label{lemmakeyequation}
	Let $\mathbb{I}(\cdot)$ denotes the indicator function. Under (A1)-(A2), we have
	\begin{eqnarray*}
		%\Mean (Y_i|A_i=a,X_i=x)=\Mean \{Y^*(a)|X=x\}=\varphi^\top(x) \beta_a,\,\,\,\,\forall a\in \{0,1\}, i\ge 1.
		\Mean [\mathbb{I}(A_i=a) \{Y_i-% \varphi^\top(X_i) \beta_a^*\}
		{\color{black}Q_0(X_i,a)}\}]=0,\,\,\,\,\forall a\in \{0,1\}, i\ge 1.
	\end{eqnarray*} 
\end{lemma}

\section{Online sequential testing for QTE}\label{sectest}
\subsection{Test statistics and their limiting distribution}\label{sec:testlimit}
%At each time point $t$, 
We first present our test statistic for testing $H_0$. 
In view of Lemma \ref{lemmakeyequation}, we estimate $\beta_a^*$ by using the ordinary least squares estimator
\begin{eqnarray*}
	\widehat{\beta}_a(t)=\widehat{\Sigma}_a^{-1}(t) \left\{ \frac{1}{N(t)}\sum_{i=1}^{N(t)} \mathbb{I}(A_i=a) \varphi(X_i) Y_i \right\}\,\,\,\,
\end{eqnarray*} 
at each time point $t$ for  $a\in \{0,1\}$, where  $\widehat{\Sigma}_a(t)=N^{-1}(t)\sum_{i=1}^{N(t)} \mathbb{I}(A_i=a) \varphi(X_i) \varphi^\top(X_i)$. A generalized inverse might be used even if $\widehat{\Sigma}_a(t)$ is  not invertible. Consider the following normalized test statistic
\begin{eqnarray*}
	S(t)=\sup_{x\in \mathbb{X}} \frac{\varphi^\top(x) \{\widehat{\beta}_1(t)-\widehat{\beta}_0(t)\}}{\widehat{s.e.}[\varphi^\top(x) \{\widehat{\beta}_1(t)-\widehat{\beta}_0(t)\}]},
\end{eqnarray*} 
for some standard error estimator $\widehat{s.e.}[\cdot]$ whose explicit form will be presented below. The benefits of working with normalized statistics are two folds. First, there is an efficiency gain compared to the unnormalized statistics without standardization, i.e., $\sup_{x\in \mathbb{X}} \varphi^\top(x) \{\widehat{\beta}_1(t)-\widehat{\beta}_0(t)\}$. In cases where $\varphi^\top(x) \{\widehat{\beta}_1(t)-\widehat{\beta}_0(t)\}$ is not consistently estimated for some value of $x$, the supremum might be attained by these values. However, due to the large estimation error, they might differ significant from the oracle maximizer, $\arg\max_x \{Q_0(x,1)-Q_0(x,0)\}$, thus lowering the power. We conduct some simulation studies (results are not included in the paper) and find that the normalized test statistic has much better power properties compared to the unnormalized one. Second, it requires weaker assumptions than the unnormalized test statistic. We will discuss this in detail in Appendix \ref{secAE}. 

Meanwhile, we remark that the studentized supremum type statistics have been used in the economics literature. For instance, \citet{Chen2015} proposed to use studentization for controlling the bias term in nonparametric series regression. \citet{belloni2015some} proposed to construct uniform confidence bands in nonparametric regression based on studentized supremum type statistics. \citet{chen2018optimal} developed a uniform inference on nonlinear functionals of nonparametric instrumental variables regression using studentized supremum type statistics. The benefits of using these statistics have been discussed in these papers as well. 

%as we will elaborate in  commented, the proposed test requires the approximation error to decay at a rate of $o\{(qN(T))^{-1/2}\}$. In contrast, the self-normalized test  requires the approximation error to decay at $o\{(N(T))^{-1/2}\}$. This is because the denominator is of the order $\sqrt{N(t_k)} \|\varphi(x)\|_2$ under mild conditions. It helps eliminate the bias of the test resulting from the model approximation error. } 
%The supremum in \eqref{statistic} can be efficiently computed. Specifically, suppose $\mathbb{X}$ is a product space, i.e, $\mathbb{X}=\prod_{j=1}^p\mathbb{X}_j$,  where $\mathbb{X}_j$ is a bounded subset of $\mathbb{R}$ with $L_j=\inf_{z\in \mathbb{X}_j}>-\infty$ and $U_j=\sup_{z\in \mathbb{X}_j}<+\infty$, for $j=1,\dots,p$. The $S(t)$ takes the form of $\sup_{x\in \mathbb{X}} \varphi^\top(x) \nu$ for some $(p+1)$-dimensional vector $\nu=(\nu^{(1)},\dots,\nu^{(p+1)})^\top$. The supreme can be calculated as 
%\begin{eqnarray*}
%	\sup_{x\in \mathbb{X}} \varphi^\top(x) \nu=\nu^{(1)}+\sum_{j=2}^{p+1} \nu^{(j+1)} \{U_j \mathbb{I}(\nu^{(j+1)}\ge 0)+L_j \mathbb{I}(\nu^{(j+1)}<0)\}. 
%\end{eqnarray*} 

Under $H_0$, we expect $S(t)$ to be small. A large $S(t)$ can be interpreted as the evidence against $H_0$. As such, we reject $H_0$ for large $S(t)$. {\color{black}We remark that when $H_0$ is rejected, we can apply the decision rule $d(x)=\argmax_{a\in \{0,1\}}\varphi^\top(x) \widehat{\beta}_a(t)$ for personalized recommendation.}
%we have

To determine the rejection region, we next discuss the limiting distribution of $S(t)$. Under $H_0$, 
{\color{black}\begin{eqnarray}
		\begin{split}
			S(t)\le \sup_{x\in \mathbb{X}} \frac{\varphi^\top(x) \{\widehat{\beta}_1(t)-\beta_1^*-\widehat{\beta}_0(t)+\beta_0^*\}+\sup_{x\in \mathbb{X}} \varphi^\top(x) (\beta_1^*-\beta_0^*)}{\widehat{s.e.}[\varphi^\top(x) \{\widehat{\beta}_1(t)-\widehat{\beta}_0(t)\}]}\\ \label{Stlimiting}
			\le \sup_{x\in \mathbb{X}} \frac{\varphi^\top(x)  \{\widehat{\beta}_1(t)-\beta_1^*-\widehat{\beta}_0(t)+\beta_0^*\}}{\widehat{s.e.}[\varphi^\top(x) \{\widehat{\beta}_1(t)-\widehat{\beta}_0(t)\}]}.
		\end{split}
\end{eqnarray}}
Both equalities hold when $\beta_0^*=\beta_1^*$. Suppose there exists some function $\pi^*(\cdot,\cdot)$ defined on $\{0,1\}\times\mathbb{X}$ that satisfies $\Mean^X |\sum_{i=1}^n n^{-1}\pi_{i-1}(a,X)-\pi^*(a,X)|\stackrel{P}{\to} 0,\forall a\in\{0,1\}$ as $n\to \infty$,  where $\pi_n(\cdot,\cdot)=\prob(A_n=a|X_n=x,\mathcal{F}_{n-1})$, and the expectation $\Mean^X$ is taken with respect to $X$. This condition implies that the treatment assignment mechanism cannot be arbitrary (see the discussion below Theorem \ref{thm1} for details). Then we will show %that
\begin{eqnarray}\label{approximation0}
	B(t)\equiv\sqrt{N(t)}\{\widehat{\beta}_1(t)-\beta_1^*-\widehat{\beta}_0(t)+\beta_0^*\}\stackrel{d}{\to} N(0, \sum_{a\in \{0,1\}} \Sigma_a^{-1} \Phi_a  \Sigma_a^{-1} ),\,\,\,\hbox{as}\,\,N(t)\to \infty,
\end{eqnarray}
where $\Sigma_a=\Mean \pi^*(a,X)\varphi(X)\varphi^\top(X)$, $\Phi_a=\Mean \pi^*(a,X)\sigma^2(a,X)\varphi(X) \varphi^\top(X)$,   and $\sigma^2(a,x)=\Mean [\{Y^*(a)-Q_0(X,a)\}^2|X=x]$, for any $ x\in \mathbb{X}$. {\color{black}Consequently, we set the denominator of $S(t)$ to
	\begin{eqnarray*}
		\widehat{s.e.}[\varphi^\top(x) \{\widehat{\beta}_1(t)-\widehat{\beta}_0(t)\}]=\left\{\frac{1}{N(t)}\sum_{a\in \{0,1\}}\varphi^\top(x) \widehat{\Sigma}_a^{-1}(t) \widehat{\Phi}_a(t) \widehat{\Sigma}_a^{-1}(t)\right\}^{1/2},
	\end{eqnarray*}
	where $\widehat{\Phi}_a(t)$ denotes the sandwich estimator for $\Phi_a$ computed using time points up to time $t$. Please refer to Algorithm \ref{alg1} for a detailed definition.
}

In addition, according to \eqref{approximation0}, the right-hand-side (RHS) of \eqref{Stlimiting} is to converge in distribution to the maximum of some Gaussian random variables. This observation forms the basis of our test.
%As a result, we expect $S(t)$ to be small under $H_0$. 

%The proposed testing procedure can be sequentially monitored. 
We next discuss the sequential implementation of our test. Assume that the interim analyses are conducted at time points $t_1,t_2,\dots,t_K\in [0,\dots,T]$ such that $0<t_1<t_2<\cdots<t_K=T$. %These time points and the maximum number of interim analyses $K$ can be random variables as well. Moreover, 
%
%Since $N$ is a counting process, the number of samples collected at the end of each interim analysis $N(t_k)$ is a random variable. 
We allow $K$ to grow with the number of observations. In the most extreme case, one may set $t_k=\inf_t \{N(t)\ge N(t_{k-1})+1\},\forall k\ge 2$. %and thus $N(t_K)=N(t_1)+K-1$. 
That is, we make a decision regarding the null hypothesis upon the arrival of each observation. %However, we do require that $\{t_k\}_{k}$ and $K$ are independent of $\{(X_i,A_i,Y_i)\}_{i\ge 1}^{+\infty}$. 
In addition, we assume that $t_1$ is large so that there are enough number of samples $N(t_1)$ to guarantee the validity of the normal approximation for $B(t_1)$. {\color{black}We remark that in typical tech companies such as Amazon, Facebook, etc., massive data are collected even within a short time interval. Large sample approximation is validated in these applications.}

%Based on $\{S(t)\}_{t>0}$, 
To guarantee our test controls the type-I error, 
we reject $H_0$ and terminate the experiment at $t_k$ if $\sqrt{N(t_k)}S(t_k)\ge z_k$ for some $k=1,\dots,K$ with some suitably chosen $z_1,\dots,z_K>0$ that satisfy
\begin{eqnarray*}
	\prob\left(\max_{k\in\{1,\dots,K\}} \{\sqrt{N(t_k)}S(t_k)-z_k\}> 0 \right)\le \alpha+o(1)
\end{eqnarray*}
for a given significance level $\alpha>0$ under $H_0$. In view of \eqref{Stlimiting}, it suffices to find $\{z_k\}_{k}$ that satisfy
{\color{black}\begin{eqnarray}\label{aimtoapproximate}
		\prob\left\{\max_{k\in\{1,\dots,K\}} \left(\sup_{x\in \mathbb{X}} \frac{\varphi^\top(x) B(t_k)}{\sqrt{N(t)}\widehat{s.e.}[\varphi^\top(x) \{\widehat{\beta}_1(t)-\widehat{\beta}_0(t)\}]} -z_k\right)>0 \right\}\le \alpha+o(1),
\end{eqnarray}}
where the stochastic process $B(\cdot)$ is defined in \eqref{approximation0}. 

To determine $\{z_k\}_k$, we need to derive the asymptotic distribution of the left-hand-side (LHS) of \eqref{aimtoapproximate}. To this end, define a mean-zero Gaussian process $G(t)$ with covariance function
\begin{eqnarray*}
	\Cov(G(t), G(t'))=N^{1/2}(t) N^{-1/2}(t') \sum_{a\in \{0,1\}} \Sigma_a^{-1} \Phi_a  \Sigma_a^{-1},\,\,\,\,\forall 0< t\le t'. 
\end{eqnarray*}
In the following, we show that the LHS of \eqref{aimtoapproximate} can be uniformly approximated based on $G(\cdot)$, for any $\{z_k\}_{k=1,\dots,K}$. To establish our theoretical results, we need some regularity conditions on $\varphi(\cdot)$. To save space, we summarize these assumptions in (A3)  and put them in Appendix \ref{secAE}. {\color{black}A random variable $Z$ is said to have a sub-Gaussian tail if there exist some constants $C,\nu>0$ such that
	\begin{eqnarray*}
		P(|Z|>z)\le C\exp(-\nu z^2),
	\end{eqnarray*}
	for any $z$. 
} 
%To simplify the analysis, we assume $\mathbb{X}=[0,1]^d$ where $d$ is the dimension of the covariates. In addition, we need to impose some conditions on $\varphi(\cdot)$. We summarize these assumptions in (A3)  and put them in Appendix \ref{secAE}. 

%For two nonnegative sequences $\{s_{1,n}\}_n$ and $\{s_{2,n}\}_n$, we use the notation $s_{1,n}\asymp s_{2,n}$ to represent that $\bar{c}^{-1} s_{2,n}\le s_{1,n}\le \bar{c} s_{2,n}$ for some universal constant $\bar{c}\ge 1$. Let $\lambda_{\min}[\hbox{Mat}]$ and $\lambda_{\max}[\hbox{Mat}]$ denote the minimum and maximum eigenvalue of a matrix $\hbox{Mat}$, respectively. 

\begin{thm}\label{thm1}
	%Assume $\lambda_{\min}[\Mean \varphi(X)\varphi^\top(X)]\asymp 1$,  $\lambda_{\max}[\Mean \varphi(X)\varphi^\top(X)]\asymp 1$, $\sup_{x} \|\varphi(x)\|_2=O(p^{1/2})$ and $N(t_1)/\log N(t_1)\gg p$ almost surely. 
	Assume (A1)-(A3) hold. 
	For $a=0,1$, assume $\inf_{x\in\mathbb{X}} \pi^*(a,x)>0$, $\Mean [\{Y^*(a)\}^2|X]$ is a bounded random variable, {\color{black}and $|Y^*(a)|$ has a sub-Gaussian tail}. Assume there exists some $0<\alpha_0\le 1$ such that for any sequence $\{j_n\}_n$ that satisfies $j_n^{\alpha_0}/\log^{\alpha_0} j_n\gg q^2$, the following event occurs with probability at least $1-O(j_n^{-\alpha_0})$,
	\begin{eqnarray}\label{condthm1}
		%\sum_{a\in\{0,1\}} \Mean^X 
		\sup_{\substack{a\in \{0,1\}} }\Mean \left|\sum_{i=1}^k \{\pi_{i-1}(a,x)-\pi^*(a,x)\}\right| \le O(1) q k^{1-\alpha_0}\log^{\alpha_0} k,\,\,\,\,\forall k\ge j_n,
	\end{eqnarray}  
	where $O(1)$ denotes some positive constant. Assume $q=O(N^{\alpha^*}(t_1))$ for some $0\le \alpha^*<\min(1/3,\alpha_0/2)$ and $N(t_1)\gg \log N(T)$ almost surely. Then conditional on the counting process $N(\cdot)$, there exists some constant $c>0$ such that
	\begin{eqnarray*}
		\sup_{z_1,\dots,z_K} \left|\prob\left\{\max_{k\in\{1,\dots,K\}} \left(\sup_{x\in \mathbb{X}}\frac{  \varphi^\top(x) B(t_k)}{\sqrt{N(t)}\widehat{s.e.}[\varphi^\top(x) \{\widehat{\beta}_1(t)-\widehat{\beta}_0(t)\}]} -z_k\right)> 0 \right\}\right.\\-\left.\prob\left\{\max_{k\in\{1,\dots,K\}} \left(\sup_{x\in \mathbb{X}} \frac{\varphi^\top(x) G(t_k )}{\sqrt{\sum_{a\in \{0,1\}}\varphi^\top(x) \Sigma_a^{-1} \Phi_a \Sigma_a^{-1} \varphi(x)}} -z_k\right)> 0 \right\} \right|\\
		\le c \left[N^{-1/8}(t_1) \log^{15/8} \{K N(t_1) \}+\{N^{-\alpha_0/3}(t_1)+q^{3/2} N^{-\alpha_0}(t_1)\} \log^{(5+\alpha_0)/3} \{K N(t_1)\}\right.\\+\left. {\color{black}\textrm{err}\log^{1/2} \{KN(t_1)\} } \right].
	\end{eqnarray*} 
\end{thm}

Theorem \ref{thm1} implies that the approximation error 
depends on the number of observations obtained up to the first decision point $N(t_1)$, the maximum number of interim analyses $K$, the total number of basis functions $q$, err, and $\alpha_0$,  which characterizes the convergence rate of the treatment assignment mechanism $\sum_{i=1}^n n^{-1} \pi_{i-1}$. Clearly, the error will decay to zero when the followings hold with probability tending to $1$,
%\begin{eqnarray}\label{adaptiveallocationerror}
%q=O(N^{\alpha_*}(t_1)),\,\,\,\,\hbox{for~some}~0\le \alpha^*<\min(1/6,\alpha_0/3),
%%\log(K) \ll N^{-1/15}(t_1)+ N^{-\alpha_0/5}(t_1)+\delta_{N(t_1)}^{1/3}.
%\end{eqnarray}
\begin{eqnarray}\label{adaptiveallocationerror2}
	\log(K) \ll \min\{N^{1/15-2\alpha^*/5}(t_1), N^{(\alpha_0-3\alpha^*)/(5+\alpha_0)}(t_1)\}.
\end{eqnarray}
%with probability tending to $1$. 
%When \eqref{adaptiveallocationerror2} holds, $K$ is allowed to grow exponentially fast with respect to $N(t_1)$. 

In Section \ref{secadaptiverandomize}, we show that $\alpha_0=1/2$, when an $\epsilon$-greedy strategy is used for randomization to balance the trade-off between exploration and exploitation. %In this case, \eqref{adaptiveallocationerror} requires $q$ to grow at a slower rate than $N^{1/6}(t_1)$. This condition is automatically satisfied when $q$ is bounded. 
Condition \eqref{adaptiveallocationerror2} is satisfied when $K$ grows polynomially fast with respect to $N(t_1)$. In addition to $\epsilon$-greedy, other adaptive allocation procedures (e.g., upper confidence bound or  Thompson sampling) could be applied as well.
%In addition to the $\epsilon$-greedy strategy, other adaptive allocation procedures can also be considered \citep[see for example,][]{Zhang2007, Hu2015, Metel2017}.

As discussed in the introduction, the derivation of Theorem \ref{thm1} is nontrivial. One way to obtain the magnitude of the approximation error is to apply the strong approximation theorem for multidimensional martingales \cite[see][]{Morrow1982,Zhang2004}. However, the rate of approximation typically depends on the dimension and decays fast as the dimension increases. To derive Theorem \ref{thm1}, we view $\{\varphi^\top(x) B(t_K)\}_{x\in \mathbb{X}, k\in \{1,\dots,\kappa\}}$ as a high-dimensional martingale and adopt the Gaussian approximation techniques that have been recently developed by \cite{belloni2018}. In view of \eqref{Stlimiting}, an application of Theorem \ref{thm1} yields the following result. 

\begin{thm}\label{thm2}
	Assume that the conditions of  Theorem \ref{thm1} hold,  \eqref{adaptiveallocationerror2} holds with probability tending to $1$. Then for any $z_1,\dots,z_k$ that satisfy
	\begin{eqnarray}\label{threshold}
		\prob\left\{\max_{k\in\{1,\dots,K\}} \left(\sup_{x\in \mathbb{X}} \frac{\varphi^\top(x) G(t_k)}{{\sqrt{\sum_{a\in \{0,1\}}\varphi^\top(x) \Sigma_a^{-1} \Phi_a \Sigma_a^{-1} \varphi(x)}}} -z_k\right)> 0 \right\}=\alpha+o(1),
	\end{eqnarray}
	as $N(t_1)$ diverges to infinity, we have under $H_0$,
	\begin{eqnarray*}
		\prob\left(\max_{k\in\{1,\dots,K\}} \{S(t_k)-z_k\}> 0 \right)\le \alpha+o(1).
	\end{eqnarray*}
	The above equality holds when $\beta_0^*=\beta_1^*$. 
\end{thm}

Theorem \ref{thm2} suggests that the type-I error rate of the proposed test can be well controlled. It remains to find critical values $\{z_k\}_{1\le k\le K}$ that satisfy \eqref{threshold}. In the next section, we propose a bootstrap-assisted procedure to determine these critical values.
%This can be achieved by specifying an $\alpha$ spending function $\alpha(t)$ that is non-increasing on $[0,T]$ and satisfies $\alpha(0)=0,\alpha(T)=\alpha$, and iteratively calculate $z_k,k=1,\dots,K$ as the solution of
%\begin{eqnarray}\label{estimatingequation}
%	\prob\left( \max_{j\in \{1,\dots,k-1\} } \left( \sup_{x\in \mathbb{X}} \varphi^\top(x) G(t_j)-z_j \right)\le 0, \sup_{x\in \mathbb{X}} \varphi^\top(x) G(t_K)>z_k \right)=\alpha(t_K),
%\end{eqnarray}
%%$\{z_k\}_{k}$ 
%until $\sqrt{N(t_K)}S(t_K)>z_k$ for some $k$. Popular choices of the spending function $\alpha(\cdot)$ includes
%\begin{eqnarray*}
%	&&\alpha_1(t)=2-2\Phi\left(\frac{\Phi^{-1}(1-\alpha/2) \sqrt{T} }{\sqrt{t}}\right),\,\,\,\,\,\,\,\,\,\,\,\,\,\,\,\,\,\,\,\,\,\,\,\alpha_2(t)=\alpha\log\left(1+(e-1)\frac{t}{T}\right),\\
%	&&\alpha_3(t)=\alpha \left( \frac{t}{T} \right)^{\theta},\,\,\,\,\hbox{for}~~\theta>0,\,\,\,\,\,\,\,\,\,\,\,\,\,\,\,\,\,\,\,\alpha_4(t)=\alpha\frac{1-\exp(-\gamma t/T)}{1-\exp(-\gamma)},\,\,\,\,\hbox{for}~~\gamma\neq 0,
%\end{eqnarray*}
%where $\Phi(\cdot)$ denotes the cumulative distribution function of a standard normal variable and $\Phi^{-1}(\cdot)$ its quantile function. 
%\vspace{-0.5cm}
\subsection{Bootstrap stopping boundary}
%The asymptotic joint distribution of $\sup_{x\in \mathbb{X}} \varphi^\top(x) G(t_1), \dots, \sup_{x\in \mathbb{X}} \varphi^\top(x) G(t_K)$ is very complicated and does not have a tractable analytical form. We propose to compute $\{z_k\}_{k=1}^K$ based on the multiplier bootstrap. 
%We consider the wild bootstrap \citep{Wu1986}. 
%To search for $\{z_k\}_k$ that satisfies \eqref{threshold}, 
We first outline a method based on the wild bootstrap \citep{Wu1986} to approximate the limiting distribution of $\{S(t_k)\}_k$. Then we discuss its limitation and present our proposal, a scalable bootstrap algorithm to determine the stopping boundary. 

%A key observation is that  $\{\widehat{\beta}_1(t_k)-\beta_1^*-\widehat{\beta}_0(t_k)+\beta_0^*$ is asymptotically multivariate normal. %, as shown in 
%Similar to \eqref{approximation0}, we can show that $\{\widehat{\beta}_1(t_k)-\beta_1^*-\widehat{\beta}_0(t_k)+\beta_0^*\}_k$ is asymptotically multivariate normal. 
The idea is to generate bootstrap samples $\{\widehat{\beta}_a^{\tiny{\hbox{MB}}}(t_k)\}_{a,k}$ that have asymptotically the same joint distribution as $\{\widehat{\beta}_a(t_k)-\beta_a^*\}_{a,k}$. Then the joint distribution of $\{S(t_k)\}_k$ can be well-approximated by the conditional distribution of $\{ \widehat{S}^{\tiny{\hbox{MB}}}(t_k)\}_{k}$ given the data,  where $\widehat{S}^{\tiny{\hbox{MB}}}(t)=\sup_{x\in \mathbb{X}} \varphi^\top(x) \{\widehat{\beta}_1^{\tiny{\hbox{MB}}}(t)- \widehat{\beta}_0^{\tiny{\hbox{MB}}}(t)\}$ for any $t$.
Specifically, let $\{\xi_i\}_{i=1}^{+\infty}$ be a sequence of i.i.d. standard normal random variables independent of $\{(X_i,A_i,Y_i)\}_{i=1}^{+\infty}$. For $a\in \{0,1\}$, define %$\widehat{S}^{\tiny{\hbox{MB}}}(t)=\sup_{x\in \mathbb{X}} \varphi^\top(x) \{\widehat{\beta}_1^{\tiny{\hbox{MB}}}(t)- \widehat{\beta}_0^{\tiny{\hbox{MB}}}(t)\}$, 
%\begin{eqnarray*}
%	\widehat{S}^{\tiny{\hbox{MB}}}(t)=\sup_{x\in \mathbb{X}} \varphi^\top(x) \{\widehat{\beta}_1^{\tiny{\hbox{MB}}}(t)- \widehat{\beta}_0^{\tiny{\hbox{MB}}}(t)\}, 
%\end{eqnarray*}
%where
\begin{eqnarray*}
	\widehat{\beta}_a^{\tiny{\hbox{MB}}}(t)=\widehat{\Sigma}_{a}^{-1}(t) \left[\frac{1}{N(t)} \sum_{i=1}^{N(t)} \mathbb{I}(A_i=a)\varphi(X_i) \{Y_i-\varphi^\top(X_i) \widehat{\beta}(t)\} \xi_i \right],\,\,\,\,\forall a\in \{0,1\}. 
\end{eqnarray*}
Both the asymptotic means of $\sqrt{N(t)}\widehat{\beta}_a^{\tiny{\hbox{MB}}}(t)$ and $\sqrt{N(t)}(\widehat{\beta}_a(t)-\beta_a^*)$ are zero. 
In addition, their covariance functions are asymptotically the same. By design, $\{\widehat{\beta}_a^{\tiny{\hbox{MB}}}(t_k)\}_{a,k}$ is multivariate normal. Similar to \eqref{approximation0}, we can show $\{\widehat{\beta}_a(t_k)-\beta_a^*\}_{a,k}$ is asymptotically multivariate normal. Consequently, the limiting distributions of $\{\widehat{\beta}_a^{\tiny{\hbox{MB}}}(t_k)\}_{a,k}$ and $\{\widehat{\beta}_a(t_k)-\beta_a^*\}_{a,k}$ are asymptotically equivalent. As such, the bootstrap approximation is valid. 

{\color{black}However, calculating $\widehat{\beta}_a^{\tiny{\hbox{MB}}}(t_k)$ requires $O( N(t_k)q+q^3)$ operations. The time complexity of the resulting bootstrap algorithm is $O(BN(t_k)q+q^3)$ up to the $k$-th interim stage}, where $B$ is the total number of bootstrap samples. This can be time consuming when $\{N(t_k)-N(t_{k-1})\}_{k=1}^K$ are large.  
To facilitate the computation, we observe that in the calculation of $\widehat{\beta}_a^{\tiny{\hbox{MB}}}$, the random noise is generated upon the arrival of each observation. This is unnecessary as we aim to 
%It is unnecessary to 
approximate the distribution of $\widehat{\beta}_a(\cdot)$ only at finitely many time points.  

We next present our proposal. Let $\{e_{i,a}\}_{i=1,\dots,K,a=0,1}$ be a sequence of i.i.d $N(0,I_{q})$ random vectors independent of the observed data,  where $I_{q}$ denotes the $q\times q$ identity matrix. At the $k$-th interim stage, %after obtaining $\widehat{\beta}_a(t_K)$ and $\widehat{\Sigma}_a(t_K)$, 
we compute
{\color{black}\begin{eqnarray*}
		\widehat{S}^{\tiny{\hbox{MB}}*}(t_k)= \sup_{x\in \mathbb{X}} \frac{\varphi^\top(x) \{\widehat{\beta}_1^{\tiny{\hbox{MB}}*}(t_k)- \widehat{\beta}_0^{\tiny{\hbox{MB}}*}(t_k)\}}{\widehat{s.e.}[\varphi^\top(x) \{\widehat{\beta}_1(t)-\widehat{\beta}_0(t)\} ]},
\end{eqnarray*}}
where $\widehat{\beta}_a^{\tiny{\hbox{MB}}*}(t_k)$ equals
\begin{eqnarray*}
	\frac{1}{N(t_k)}\sum_{j=1}^{k} \left(\sum_{i=N(t_{j-1})+1}^{N(t_j)} \widehat{\Sigma}_a^{-1}(t_j) \mathbb{I}(A_i=a) \varphi(X_i) \varphi^\top(X_i) \{Y_i-\varphi(X_i)^\top \widehat{\beta}_a(t_j)\}^2 \widehat{\Sigma}_a^{-1}(t_j) \right)^{1/2} e_{j,a}.
\end{eqnarray*}
For any $k_1$ and $k_2$, the conditional covariance of $\sqrt{N(t_{k_1})}\{\widehat{\beta}_1^{\tiny{\hbox{MB}}*}(t_{k_1})-\widehat{\beta}_0^{\tiny{\hbox{MB}}*}(t_{k_1})\}$ and $\sqrt{N(t_{k_2})}\{\widehat{\beta}_1^{\tiny{\hbox{MB}}*}(t_{k_2})-\widehat{\beta}_0^{\tiny{\hbox{MB}}*}(t_{k_2})\}$ equals
\begin{eqnarray*}
	\frac{1}{\sqrt{N(t_{k_1})N(t_{k_2})}}\sum_{a=0}^1 \sum_{j=1}^{k_1} \sum_{i=N(t_{j-1})+1}^{N(t_j)}\widehat{\Sigma}_a^{-1}(t_j) \mathbb{I}(A_i=a) \varphi(X_i) \varphi^\top(X_i) \{Y_i-\varphi^\top(X_i) \widehat{\beta}_a(t_j)\}^2 \widehat{\Sigma}_a^{-1}(t_j).
\end{eqnarray*}
Under the given conditions in Theorem \ref{thm1}, it is to converge to
\begin{eqnarray}\nonumber
	\frac{\sqrt{N(t_{k_1})}}{\sqrt{N(t_{k_2})}} \sum_{a=0}^1 \Sigma_a^{-1}\Phi_a\Sigma_a^{-1}=\Cov(G(t_{k_1}), G(t_{k_2})).
\end{eqnarray}
%where $\Cov^*(Z_1,Z_2)$ denotes the conditional covariance of $Z_1$ and $Z_2$ given the data $\{(X_i,A_i,Y_i)\}_{i=1}^{+\infty}$. 
This means $\{\sqrt{N(t_k)}(\widehat{\beta}_1^{\tiny{\hbox{MB*}}}(t_k)-\widehat{\beta}_0^{\tiny{\hbox{MB*}}}(t_k))\}_{k}$ and $\{G(t_{k})\}_{k}$ have the same asymptotic distribution. Consequently, 
%This means that  
$\{\sqrt{N(t_k)}\widehat{S}^{\tiny{\hbox{MB}}*}(t_k) \}_{k=1}^{K}$ can be used to approximate the joint distribution of $\{\sup_{x\in \mathbb{X}} \varphi^\top(x) G(t_k)/{\sqrt{\sum_{a\in \{0,1\}}\varphi^\top(x) \Sigma_a^{-1} \Phi_a \Sigma_a^{-1} \varphi(x)}}\}_{k=1}^K$. %Theorem 3 provides formal justification for our proposed bootstrap method and is put in the appendix.  

%The moment condition on the potential outcomes in Theorem \ref{thm3} is slightly stronger compared to Theorem \ref{thm1}. This condition is imposed in order to establish a nonasymptotic error bound for
%\begin{eqnarray*}
%	\max_{k_1,k_2}\left\|\sqrt{N(t_{k_1})N(t_{k_2})}\Cov^*\left(\widehat{\beta}_1^{\tiny{\hbox{MB}}*}(t_{k_1})-\widehat{\beta}_0^{\tiny{\hbox{MB}}*}(t_{k_1}), \widehat{\beta}_1^{\tiny{\hbox{MB}}*}(t_{k_2})-\widehat{\beta}_0^{\tiny{\hbox{MB}}*}(t_{k_2})\right)-\Cov(G(t_{k_1}), G(t_{k_2}))\right\|_2,
%\end{eqnarray*}
%
%\vspace{-0.5cm}
%where $\Cov^*(Z_1,Z_2)$ denotes the conditional covariance of $Z_1$ and $Z_2$ given the data $\{(X_i,A_i,Y_i)\}_{i=1}^{+\infty}$. 
To choose $\{z_k\}_k$ that satisfies \eqref{threshold}, we adopt the $\alpha$-spending approach that allocates the total allowable type I error at each interim stage according to an error-spending function. This guarantees our test controls the type-I error. We begin by specifying an $\alpha$ spending function $\alpha(t)$ that is non-increasing and satisfies $\alpha(0)=0$, $ \alpha(T)=\alpha$. Popular choices of $\alpha(\cdot)$ include 
\begin{eqnarray}\label{alphaspend}
	\begin{split}
		&&\alpha_1(t)=\alpha\log\left(1+(e-1)\frac{t}{T}\right),\,\,\,\,\,\,\,\,\,\,\,\,\,\,\,\,\,\,\,\,\,\,\,\alpha_2(t)=2-2\Phi\left(\frac{\Phi^{-1}(1-\alpha/2) \sqrt{T} }{\sqrt{t}}\right),\\
		&&\alpha_3(t)=\alpha \left( \frac{t}{T} \right)^{\theta},\,\,\,\,\hbox{for}~~\theta>0,\,\,\,\,\,\,\,\,\,\,\,\,\,\,\,\,\,\,\,\alpha_4(t)=\alpha\frac{1-\exp(-\gamma t/T)}{1-\exp(-\gamma)},\,\,\,\,\hbox{for}~~\gamma\neq 0,
	\end{split}
\end{eqnarray}
where $\Phi(\cdot)$ denotes the cumulative distribution function of a standard normal variable and $\Phi^{-1}(\cdot)$ is  its quantile function. 
%In Figure \ref{fig:alpha_spending_functions}, we depict these error-spending functions.  
%\begin{figure}[!t]
%	\centering
%	\includegraphics[width=0.5\textwidth,height=0.25\textheight]{fig/alpha_spending_functions.png}
%	\caption{Alpha spending functions when $\theta=0.5, \gamma=1.0$.}
%	\label{fig:alpha_spending_functions}
%\end{figure}

Based on $\alpha(\cdot)$, we iteratively calculate $\widehat{z}_k,k=1,\dots,K$ as the solution of
\begin{eqnarray}\label{estimatingequation}
	\prob^*\left\{ \max_{j\in \{1,\dots,k-1\} } \left( \sqrt{N(t_j)}\widehat{S}^{\tiny{\hbox{MB}}*}(t_j)-\widehat{z}_j \right)\le 0, \sqrt{N(t_k)}\widehat{S}^{\tiny{\hbox{MB}}*}(t_k)>\widehat{z}_k \right\}=\alpha(t_k)-\alpha(t_{k-1}),
\end{eqnarray}
%$\{z_k\}_{k}$ 
and reject $H_0$ when $\sqrt{N(t_k)}S(t_k)>\widehat{z}_k$ holds for some $k$. 

The validity of the bootstrap test is summarized in Theorems \ref{thm3} and \ref{thm4} below. %in the appendix. 

\begin{thm}\label{thm3}
	Assume the conditions in Theorem \ref{thm1} hold. %Assume that $\max_{a\in\{0,1\}}\Mean \{Y^*(a)\}^5<+\infty$. 
	%Assume $q=O(N^{\alpha^*}(t_1))$ for some $0<\alpha^*<1/3$, almost surely. 
	Then conditional on the counting process $N(\cdot)$, we have
	\begin{eqnarray*}
		\sup_{z_1,\dots,z_K} \left|\prob^*\left\{\max_{k\in\{1,\dots,K\}} \left(\sqrt{N(t_k)}\widehat{S}^{\tiny{\hbox{MB}}*}(t_k) -z_k\right)> 0 \right\}\right.\\
		-\left.\prob\left\{\max_{k\in\{1,\dots,K\}} \left(\sup_{x\in \mathbb{X}} \frac{\varphi^\top(x) G(t_k)}{\sqrt{\sum_a \varphi^\top(x) \Sigma_a^{-1} \Phi_a \Sigma_a^{-1} \varphi(x) } } -z_k\right)> 0 \right\} \right|\\
		\le c \left[ q^{6/5}N^{-1/6}(t_1) \log^{11/6} \{K N(t_1)\}+q^{5/3}N^{-\alpha_0/3}(t_1) \log^{(5+\alpha_0)/3} \{K N(t_1) \}\right]
	\end{eqnarray*}
	for some constant $c>0$  with probability at least $1-O(N^{-\alpha_0}(t_1))$,  where $\prob^*(\cdot)$ denotes the probability measure conditional on the data stream $\{X_i,A_i,Y_i\}_{i=1}^{+\infty}$. 
\end{thm}

\begin{thm}\label{thm4}
	Assume the conditions in Theorem \ref{thm3} hold. Then conditional on $N(\cdot)$, the critical values $\{\widehat{z}_k\}_k$ satisfy
	\begin{eqnarray}\label{errorbound}
		\begin{split}
			&\left|\prob\left\{\max_{k\in\{1,\dots,K\}} \left(\sup_{x\in \mathbb{X}} \frac{\varphi^\top(x) G(t_k)}{\sqrt{\sum_a \varphi^\top(x) \Sigma_a^{-1} \Phi_a \Sigma_a^{-1} \varphi(x) } } -\widehat{z}_k\right)> 0 \right\}-\alpha\right|\\ 
			\le& c \left[ q^{6/5}N^{-1/6}(t_1) \log^{11/6} \{K N(t_1)\}+q^{5/3}N^{-\alpha_0/3}(t_1) \log^{(5+\alpha_0)/3} \{K N(t_1) \}\right],
		\end{split}
	\end{eqnarray}
	for some constant $c>0$. 
\end{thm}
When the RHS of \eqref{errorbound} is $o_p(1)$, it follows from Theorems \ref{thm2} and \ref{thm4} that our test is valid. 
The conditional distribution in \eqref{estimatingequation} can be approximated by the empirical distribution of Bootstrap samples. 

{\color{black}Next, we study the power property and the stopping time of the proposed test. Let sig$=\sup_x \{Q(x,1)-Q(x,0)\}$ denote the qualitative treatment effect signal. We will show that the power of the proposed test approaches to one as long as $\sqrt{N(t_K)}$sig$\gg \sqrt{q\log \{ N(t_1)\}}$. In addition, the stopping time depends crucially on sig. Specifically, we will show that the proposed test will reject the null at the $k$th interim stage as long as $\sqrt{N(t_k)}$sig$\gg \sqrt{q\log \{N(t_1)\}}$.
	\begin{thm}\label{thm6}
		Assume the conditions in Theorem \ref{thm2} %and \ref{thm3} 
		hold, there exists some sufficiently large constant $C>0$ such that $\min\{\alpha(t_1), \min_k \alpha(t_k)-\alpha(t_{k-1})\}\gg N^{-C}(t_1)$, and that $\sqrt{N(t_K)}$sig$\gg \sqrt{q\log \{ N(t_1)\}}$. Then the power of the proposed test approaches one as $N(t_1)$ diverges to infinity. In addition, for any $k$ such that $\sqrt{N(t_k)}$sig$\gg \sqrt{q\log \{ N(t_1)\}}$, the stopping time will be smaller than or equal to $t_k$, with probability tending to $1$. 
	\end{thm}	
}
%}

Finally, we remark that our test can be online updated as batches of observations arrive at the end of each interim stage. A pseudocode summarizing our procedure is given in Algorithm 1. In Algorithm 1, we use $O_{p+1}$ to denote a $(p+1)\times (p+1)$ zero matrix and $0_{p+1}$ to denote a $(p+1)$-dimensional zero vector. %For any vector $\phi$, $\phi^{(\ell)}$ denotes the $\ell$-th element of $\phi$. 
{\color{black}The spatial complexity of the proposed algorithm is $O(Bq+q^2)$,  where $B$ is the number of bootstrap samples. The time complexity is $O(Bkq^2+N(t_k)q^2)$ up to the $k$-th interim stage. Suppose $N(t_j)-N(t_{j-1})=n$ for any $1\le j\le K$, we have $Bkq^2+N(t_k)q^2=(B+n)kq^2\ll Bnkq=BN(t_k)q$ when $Bn\gg (B+n)q$, or equivalently, $\min(B,n)\gg q$. Hence, our procedure is much faster compared to the standard wild bootstrap as long as the number of bootstrap samples and the number of observations per batch are much large than the number of basis functions.}
%Set $N(t_0)=0$. We summarize our procedure in the following algorithm.

\begin{algorithm}[!t]
	\begin{algorithmic}
		%\Procedure{Generate the optimal I2DR}{}
		%\State \textbf{Global:} data $\{(X_i,A_i,Y_i):i=1,\dots,n\}$; sample size $n$; covariates dimension $p$; number of initial intervals $m$; penalty terms $\gamma_n$ and $\lambda_n$.
		%		\footnotesize
		\State \textbf{Input:} Number of bootstrap samples $B$, an $\alpha$ spending function $\alpha(\cdot)$. 
		%integers $l, r \in  N$; cost $c \in  \mathbb{R}$; a vector of integers $\tau\in N^m$; Bellman function $B\in  \mathbb{R}^m$;  vectors $\beta, \phi \in \mathbb{R}^{p+1}$; matrix $\Phi\in \mathbb{R}^{(p+1)\times (p+1)}$. 
		%\State \textbf{Output:} %cut points $\widehat{\mathcal{C}}\subset  \{1,\cdots,m\}$; cut point locations on the entire dose range $J(\widehat{\mathcal{C}}) \in  \mathbb{R}^{|\widehat{\mathcal{C}}|} $; the optimal dose rule for each partition $\hat{\beta} \in \mathbb{R}^{p+1}\times\mathbb{R}^{|\widehat{\mathcal{C}}|}$.
		%$\widehat{\mathcal{P}}$ and $\{\widehat{\beta}_{\mathcal{I}}:\mathcal{I}\in \widehat{\mathcal{P}} \}$.
		\State \textbf{Initialize: } $n=0$, $\widehat{\Sigma}_{0}=\widehat{\Sigma}_1=O_{p+1}$, $\widehat{\gamma}_0=\widehat{\gamma}_1=0_{p+1}$, $\widehat{\beta}_{0,b}=\widehat{\beta}_{1,b}=0_{p+1}$, and a set $\mathcal{I}=\{1,\dots,B\}$. 
		%	\State \textbf{Begin:}   %\text{      \% \small{find the optimal partition}}
		
		\State \textbf{For} $k= 1$ to $K$ \textbf{do}
		%		\For{$k= 1$ to $K$}   
		\State \quad \textbf{Initialize: }$m=0$ and $\widehat{\Phi}_0=\widehat{\Phi}_1=O_{p+1}$.
		\State \quad \textbf{Step 1: Online update of $\widehat{\beta}_a$}
		\State \quad \textbf{For} $i=N(t_{k-1})+1$ to $N(t_k)$ \textbf{do} 
		\State \qquad $n=n+1$ and $m=m+1$;
		\State \qquad $\widehat{\Sigma}_a=(1-n^{-1})\widehat{\Sigma}_a+n^{-1} \varphi(X_i) \varphi^\top(X_i) \mathbb{I}(A_i=a), a=0, 1$; 
		\State \qquad $\widehat{\gamma}_a=(1-n^{-1})\widehat{\gamma}_a+n^{-1} \varphi(X_i) Y_i \mathbb{I}(A_i=a), a=0, 1$;
		\State \quad \textbf{Compute }$\widehat{\beta}_a=\widehat{\Sigma}_a^{-1} \widehat{\gamma}_a$ for $a\in \{0,1\}$; %and $S=\sup_{x\in \mathbb{X}} \varphi^\top(x) (\widehat{\beta}_1-\widehat{\beta}_0)$;
		\State \quad \textbf{Step 2: Bootstrap}
		\State \quad \textbf{For} $i=N(t_{k-1})+1$ to $N(t_k)$ \textbf{do} 
		\State \qquad $\widehat{\Phi}_a=\widehat{\Phi}_a+\widehat{\Sigma}_a^{-1}\varphi(X_i)\varphi^\top(X_i) \{Y_i-\varphi^\top(X_i) \widehat{\beta}_a\}^2 \widehat{\Sigma}_a^{-1} \mathbb{I}(A_i=a), a=0,1$;
		\State \quad \textbf{Compute} {\color{black}$S=\sup_{x\in \mathbb{X}} [\varphi^\top(x) (\widehat{\beta}_1-\widehat{\beta}_0)/\sqrt{\sum_a \varphi^\top(x) \widehat{\Sigma}_a^{-1} \widehat{\Phi}_a \widehat{\Sigma}_a^{-1}}]$};
		\State \quad \textbf{For} $b=1,\dots,B$ \textbf{do}
		\State \qquad Generate two independent $N(0,I_{p+1})$ Gaussian vectors $e_0$, $e_1$;
		\State \qquad $\widehat{\beta}_{a,b}=(1-m n^{-1}) \widehat{\beta}_{a,b}+n^{-1} \widehat{\Phi}_a^{1/2} e_a, a=0, 1$;
		\State \qquad \textbf{Compute} {\color{black}$\widehat{S}_b=\sup_{x\in \mathbb{X}}[\varphi^\top(x) (\widehat{\beta}_{1,b}-\widehat{\beta}_{0,b})/\sqrt{\sum_a \varphi^\top(x) \widehat{\Sigma}_a^{-1} \widehat{\Phi}_a \widehat{\Sigma}_a^{-1}}]$};
		\State \quad \textbf{Step 3: Reject or not}
		\State \quad Set $z$ to be the upper $\{\alpha(t)-|\mathcal{I}|^c/B|\}/(1-|\mathcal{I}^c|/B)$-th percentile of $\{\widehat{S}_b\}_{b\in \mathcal{I}}$; 
		\State \quad Update $\mathcal{I}$ as $\mathcal{I}\leftarrow \{b\in \mathcal{I}: \widehat{S}_b\le z \}$. 
		\State \quad \textbf{If} {${S}>z$}: Reject $H_0$ and terminate the experiment.
		%		\EndIf
		%		\EndFor
		%		\EndFor
	\end{algorithmic}
	\caption{the Pseudocode that summarizing the online bootstrap testing procedure.}\label{alg1}
\end{algorithm}	

\subsection{Adaptive randomization}\label{secadaptiverandomize}
In practice, the company might want to allocate more traffic to a better treatment based on the observed data stream. The $\epsilon$-greedy strategy is commonly used to balance the trade-off between exploration and exploitation. For a given $0<\varepsilon_0<1$, consider the following randomization procedure: for some integer $N_0>0$ and any $j\ge N_0$, $a\in \{0,1\}$, $x\in \mathbb{X}$, we set %$\pi_{j-1}(a,x)=(1-\varepsilon_0) \mathbb{I}\{a=\mathbb{I}(\varphi^\top(x) \widehat{\beta}_{a,j}>0)\}+\varepsilon_0 \mathbb{I}\{a\neq \mathbb{I}(\varphi^\top(x) \widehat{\beta}_{a,j}>0)\}$ 
\begin{eqnarray*}
	\pi_{j-1}(a,x)=(1-\varepsilon_0) a\mathbb{I}\{\varphi^\top(x) (\widehat{\beta}_{1,j-1}-\widehat{\beta}_{0,j-1})>0 \}+\varepsilon_0 (1-a) \mathbb{I}\{\varphi^\top(x) (\widehat{\beta}_{1,j-1}-\widehat{\beta}_{0,j-1})\le 0 \},
\end{eqnarray*}
where %$\widehat{\beta}_{a,i}=\widehat{\beta}_a(t_i)$ for some $t_i$ such that $N(t_i)=i$. 
\begin{eqnarray*}
	\widehat{\beta}_{a,j}=\widehat{\Sigma}_{a,j}^{-1}  \frac{1}{j}\sum_{i=1}^j \{\mathbb{I}(A_i=a)\varphi(X_i) Y_i\}\,\,\,\,\hbox{and}\,\,\,\,\widehat{\Sigma}_{a,j}=\frac{1}{j}\sum_{i=1}^{j} \mathbb{I}(A_i=a) \varphi(X_i) \varphi^\top(X_i).
\end{eqnarray*}
It is immediate to see that $\widehat{\Sigma}_a(t)=\widehat{\Sigma}_{a,n(t)}$ and $\widehat{\beta}_{a}(t)=\widehat{\beta}_{a,n(t)}$. 
Define %$\pi^*(a,x)=(1-\varepsilon_0) \mathbb{I}\{a=\mathbb{I}(\varphi^\top(x) \beta_a>0)\}+\varepsilon_0 \mathbb{I}\{a\neq \mathbb{I}(\varphi^\top(x) \beta_a>0)\}$, 
\begin{eqnarray*}
	\pi^*(a,x)=(1-\varepsilon_0) a\mathbb{I}\{\varphi^\top(x) (\beta_1-\beta_0)>0\}+\varepsilon_0 (1-a)\mathbb{I}\{\varphi^\top(x) (\beta_1-\beta_0)\le 0\}
\end{eqnarray*}
for any $a\in \{0,1\}$ and $x\in \mathbb{X}$. 
\begin{lemma}\label{thm5}
	Assume (A1)-(A3) hold. Assume $\inf_{x\in\mathbb{X}} \pi^*(a,x)>0$ and $|Y^*(a)|$ is bounded almost surely, for $a\in \{0,1\}$. Assume {\color{black}$\prob(|Q_0(X,1)-Q_0(X,0)|\le \epsilon)\le L_0 \epsilon$}, for some constant $L_0>0$ and any $\epsilon>0$. Then for any $\{j_n\}_n$ that satisfies $\sqrt{j_n}/\sqrt{\log j_n}\gg q^2$, the following event occurs with probability at least $1-O(j_n^{-1})$,
	\begin{eqnarray}\label{eqn:lemma2}
		\sum_{a\in\{0,1\}} \Mean^{\mathcal{F}_{i-1}} \left|\sum_{i=1}^k \{\pi_{i-1}(a,X)-\pi^*(a,X)\}\right|\preceq q \sqrt{k \log k},\,\,\,\,\forall k\ge j_n.
	\end{eqnarray} 
\end{lemma}

{\color{black}We make a few remarks. First, Lemma \ref{thm5} implies that Condition \eqref{condthm1} in Theorem \ref{thm1} automatically holds with $\alpha_0=1/2$, when the epsilon-greedy strategy is used. Nonetheless, Condition (5) is weaker than \eqref{eqn:lemma2}, as it only requires the average estimated policy aggregated over different interim stages to converge at certain rate. Second, by setting $\epsilon=0$, the assumption $\prob(|Q_0(X,1)-Q_0(X,0)|\le \epsilon)\le L_0 \epsilon$ requires $\prob(Q_0(X,1)=Q_0(X,0))=0$, almost surely. It essentially requires that the difference between the two Q-functions is nonzero, almost surely. This implies that the optimal decision rule is uniquely defined and is thus identifiable. This condition is necessary to guarantee that the estimated policy converges to the oracle optimal policy. Without this identifiability condition, the estimated optimal decision rule could fluctuate randomly and will not stabilize \citep{Alex2016}. Third, our condition also requires $\prob(0<|Q_0(X,1)-Q_0(X,0))| \le \epsilon)\le L_0 \epsilon$ for any $\epsilon>0$. The latter condition is well-known as the margin condition that is commonly imposed in the literature to bound the difference between the expect return under the estimated and the optimal decision rule \citep{qian2011,Alex2016,shi2020breaking}. The identifiability assumption is not needed to establish the rate of convergence of the expected return under the estimated decision rule. }

\section{Numerical studies}
\label{secnumer}%\vspace{-0.2cm}
\subsection{Simulation studies}
\subsubsection{Testing QTE}
In this section, we conduct Monte Carlo simulations to examine the finite sample properties of the proposed test.
We generated the potential outcomes as 
%\begin{align*}
$Y _i^*(a)= 1 + (X_{i1} - X_{i2})/2 + a \tau(X_i) + \varepsilon_i$, 
%\end{align*}
where $\varepsilon_i$'s are i.i.d $N(0, 0.5^2)$. 
The covariates $X_i=(X_{i1}, X_{i2}, X_{i3})^\top$ were generated as follows. 
We first generated $X_i^*=(X_{i1}^*, X_{i2}^*, X_{i3}^*)^\top$ from a multivariate normal distribution with zero mean and covariance matrix equal to $\{ 0.5^{|i-j|} \}_{i,j}$.
%$(1,  0.5, 0.25\|		0.5,  1, 0.5 \| 0.25,  0.5,  1)$. 
Then we set $X_{ij}=X_{ij}^* \mathbb{I}(X_{ij}^*|\le 2)+2\hbox{sgn}(X_{ij}^*) \mathbb{I}(X_{ij}^*|> 2)$. %to -2 if $X_{ij}^*< -2$ 2 for $X_{ij}^*>2$, and $X_{ij}^*$ for  $.$ 
We consider two randomization designs. 
%both complete randomization and adaptive randomization. 
In the first design, the treatment assignment is nondynamic and completely random. Specifically we set $\pi_{i}(a,x)=0.5$, for any $a,x$ and $i$. In the second design, we use an $\epsilon$-greedy strategy to generate the treatment with $\varepsilon = 0.3$. %for adaptive randomization. 
In addition, we set $N(T_1) = 2000$ and $N(T_j)-N(T_{j-1})=2n$ for $2\le j \le K$ and some $n>0$. We consider two combinations of $(n, K)$, corresponding to $(n, K) = (200, 5)$ and  $(20, 50)$.

We set the significance level $\alpha=0.05$ and choose $B = 10000$.
We set $\tau(X_i) = \phi_\delta\{(X_{i1}+X_{i2})/\sqrt{2}\} X_{i3}^2$
for some function $\phi_\delta$ parameterized by some $\delta \ge 0$. 
We consider two scenarios for $\phi_\delta$. Specifically, we set $\phi_\delta(x) = \delta x^2/3 $ in Scenario 1 and $\phi_\delta = \delta \cos(\pi x)$ in Scenario 2. For each setting, we further consider four cases by setting $\delta = 0, 0.05, 0.10, $ and  $0.15$. When $\delta=0$, $H_0$ holds. Otherwise, $H_1$ holds.  For all settings, we construct the basis function $\varphi(\cdot)$ using additive cubic splines. For each univariate spline, we set the number of internal knots to be $4$. These knots are equally spaced between $\left[-2, 2\right]$.

\begin{figure}[!t]
	\centering
	\includegraphics[width=\textwidth,height=0.27\textheight]{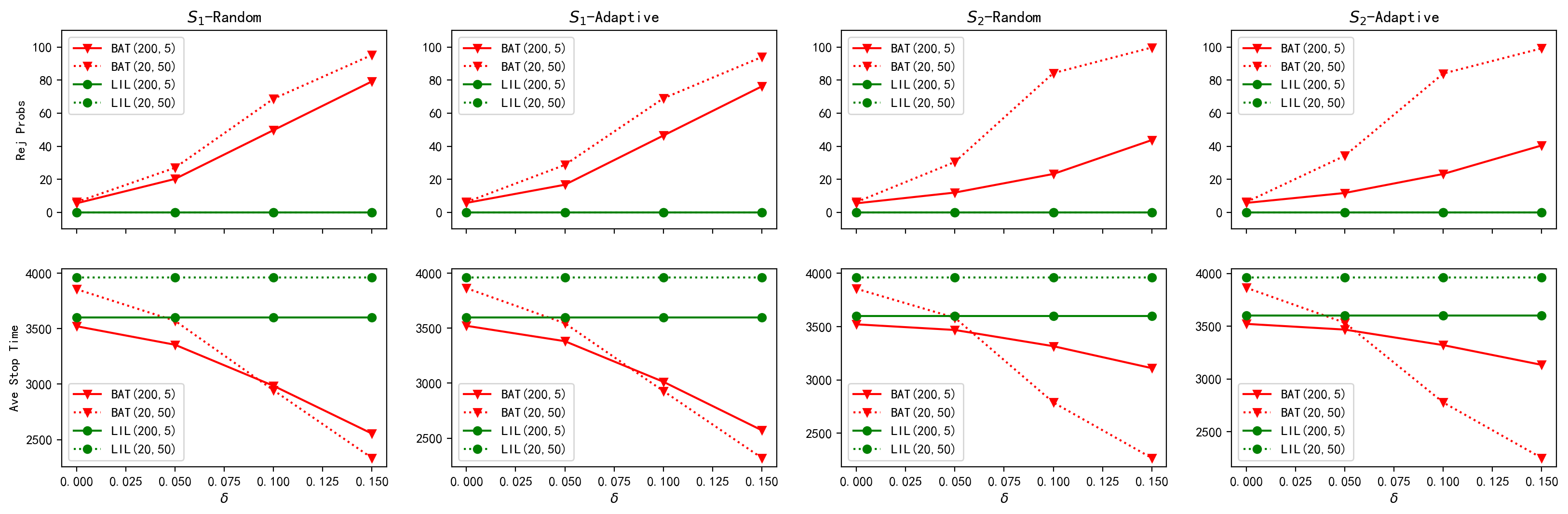}
	%	\vspace{-0.6cm}
	\footnotesize
	\caption{Rejection probabilities and average stopping times of the proposed test
		when $\alpha_1(\cdot)$ is chosen as the spending function. From left to right: Scenario 1 with random design, Scenario 1 with $\epsilon$-greedy design, Scenario 2 with random design and Scenario 2 with $\epsilon$-greedy design. }\label{fig:hte}
	%	\vspace{-0.3cm}
\end{figure}

We denote our test by BAT, short for bootstrap-assisted test. We run our experiments on a single computer instance with 40 Intel(R) Xeon(R) 2.20GHz CPUs. It takes 1-2 seconds on average to compute each test. 
%In Table \ref{tab:hte}, we report the rejection probabilities and average stopping times (defined as the average number of samples consumed when the experiment is terminated) of the proposed test aggregated over 400 simulations when $\alpha_1(\cdot)$ is chosen as the spending function. 
In Figure \ref{fig:hte}, we plot the rejection probabilities of our tests and the average stopping times (defined as the average number of samples consumed when the experiment is terminated), aggregated over 400 simulations when $\alpha_1(\cdot)$ is chosen as the spending function. {\color{black}The detailed values of these rejection probabilities and average stopping times can be found in Table \ref{tab:hte}.} 
It can be seen that the type-I error rates are close to the nominal level in all cases. The power of our test increases as $\delta$ increases, demonstrating its consistency. In addition, when $\delta>0$, our experiments are stopped early in all cases. 

To further evaluate our method, we compare it with a test based on the law of iterated logarithm (denoted by LIL). LIL determines the decision boundary based on an always valid finite error bound (see Appendix \ref{secbaseline} for details about the competing method). It can be seen from Figure \ref{fig:hte} that our method has good power properties, whereas LIL fails to detect the alternative and does not have any power at all. 
%much larger power than the law of iterated logarithm approach.

\subsubsection{Testing ATE}\label{sec:ATE}
We extend our proposal to testing ATE in this section. Specifically, we focus on testing the following hypothesis,
\begin{eqnarray*}
	H_0: \Mean Y_i^*(1)\le \Mean Y_i^*(0) \,\,\,\,\hbox{versus}\,\,\,\,H_1: \Mean Y_i^*(1)> \Mean Y_i^*(0).
\end{eqnarray*}
Under (A1) and (A2), it suffices to test
\begin{eqnarray*}
	H_0: \Mean Q(X_i,1)\le \Mean Q(X_i,0) \,\,\,\,\hbox{versus}\,\,\,\,H_1: \Mean Q(X_i,1)> \Mean Q(X_i,0).
\end{eqnarray*}
We similarly use basis approximations to model the Q-function. The proposed method is very similar to that in Section \ref{sectest}. {\color{black}The main difference lies in that instead of employing a supremum type statistics, we aggregate the estimated treatment effect and set $S(t)=N^{-1}\sum_{i=1}^N \varphi^\top(X_i)\{\widehat{\beta}_1(t)-\widehat{\beta}_0(t)\}$. Its asymptotic distribution can be approximated by the corresponding bootstrap statistic $N^{-1}\sum_{i=1}^N \varphi^\top(X_i)\{\widehat{\beta}_1^{\textrm{\tiny{MB}}*}(t)-\widehat{\beta}_0^{\textrm{\tiny{MB}}*}(t)\}$. The proposed algorithm can then be applied to determine the rejection boundary.}
To save space, we summarize our proposal in the following algorithm. %This validity of this algorithm requires $\Var(\varphi(X_i))\to 0$. This assumption is automatically satisfied when tensor-product B-spline is used. 
We next conduct simulation studies to evaluate this algorithm.

\begin{algorithm}[!h]
	\begin{algorithmic}
		%\Procedure{Generate the optimal I2DR}{}
		%\State \textbf{Global:} data $\{(X_i,A_i,Y_i):i=1,\dots,n\}$; sample size $n$; covariates dimension $p$; number of initial intervals $m$; penalty terms $\gamma_n$ and $\lambda_n$.
		%		\footnotesize
		\State \textbf{Input:} Number of bootstrap samples $B$, an $\alpha$ spending function $\alpha(\cdot)$. 
		%integers $l, r \in  N$; cost $c \in  \mathbb{R}$; a vector of integers $\tau\in N^m$; Bellman function $B\in  \mathbb{R}^m$;  vectors $\beta, \phi \in \mathbb{R}^{p+1}$; matrix $\Phi\in \mathbb{R}^{(p+1)\times (p+1)}$. 
		%\State \textbf{Output:} %cut points $\widehat{\mathcal{C}}\subset  \{1,\cdots,m\}$; cut point locations on the entire dose range $J(\widehat{\mathcal{C}}) \in  \mathbb{R}^{|\widehat{\mathcal{C}}|} $; the optimal dose rule for each partition $\hat{\beta} \in \mathbb{R}^{p+1}\times\mathbb{R}^{|\widehat{\mathcal{C}}|}$.
		%$\widehat{\mathcal{P}}$ and $\{\widehat{\beta}_{\mathcal{I}}:\mathcal{I}\in \widehat{\mathcal{P}} \}$.
		\State \textbf{Initialize: } $n=0$, $\widehat{\Sigma}_{0}=\widehat{\Sigma}_1=O_{p+1}$, $\widehat{\gamma}_0=\widehat{\gamma}_1=0_{p+1}$, $\widehat{\beta}_{0,b}=\widehat{\beta}_{1,b}=0_{p+1}$, $\bar{\varphi}=0$ and a set $\mathcal{I}=\{1,\dots,B\}$. 
		%	\State \textbf{Begin:}   %\text{      \% \small{find the optimal partition}}
		\State \textbf{For} $k= 1$ to $K$ \textbf{do}
		%		\For{$k= 1$ to $K$}   
		\State \quad \textbf{Initialize: }$m=0$, $\widehat{\phi}=0$ and $\widehat{\Phi}_0=\widehat{\Phi}_1=O_{p+1}$.
		\\
		\State \quad \textbf{For} $i=N(t_{k-1})+1$ to $N(t_k)$ \textbf{do} 
		\State \qquad $n=n+1$, $m=m+1$ and $\bar{\varphi}=n^{-1}(n-1) \bar{\varphi}+n^{-1} \varphi(X_i)$;
		\State \qquad$\widehat{\Sigma}_a=(1-n^{-1})\widehat{\Sigma}_a+n^{-1} \varphi(X_i) \varphi^\top(X_i) \mathbb{I}(A_i=a), a=0, 1$; 
		\State \qquad$\widehat{\gamma}_a=(1-n^{-1})\widehat{\gamma}_a+n^{-1} \varphi(X_i) Y_i \mathbb{I}(A_i=a), a=0, 1$;
		\State \quad \textbf{Compute }$\widehat{\beta}_a=\widehat{\Sigma}_a^{-1} \widehat{\gamma}_a$ for $a\in \{0,1\}$ and $S= \bar{\varphi}^\top (\widehat{\beta}_1-\widehat{\beta}_0)$;
		\\
		\State \quad \textbf{For} $i=N(t_{k-1})+1$ to $N(t_k)$ \textbf{do} 
		\State \qquad $\widehat{\phi}=\widehat{\phi}+[\{\varphi(X_i)-\bar{\varphi}\}^\top(\widehat{\beta}_1-\widehat{\beta}_0)]^2$. 
		\State \qquad$\widehat{\Phi}_a=\widehat{\Phi}_a+\widehat{\Sigma}_a^{-1}\varphi(X_i)\varphi^\top(X_i) \{Y_i-\varphi^\top(X_i) \widehat{\beta}_a\}^2 \widehat{\Sigma}_a^{-1} \mathbb{I}(A_i=a), a=0, 1$;
		\State \quad \textbf{For} $b=1,\dots,B$ \textbf{do}
		\State \qquad Generate two independent $N(0,I_{p+1})$ Gaussian vectors $e_0$, $e_1$, $N(0,1)$ random variable $e_2$;
		\State \qquad $\widehat{\beta}_{a,b}=(1-m n^{-1}) \widehat{\beta}_{a,b}+n^{-1} \widehat{\Phi}_a^{1/2} e_a+n^{-1} \widehat{\phi}^{1/2} e_2, a=0, 1$;
		\State \qquad \textbf{Compute} $\widehat{S}_b=\bar{\varphi}^\top (\widehat{\beta}_{1,b}-\widehat{\beta}_{0,b})$;
		\\
		\State \quad Set $z$ to be the upper $\{\alpha(t)-|\mathcal{I}|^c/B|\}/(1-|\mathcal{I}^c|/B)$-th percentile of $\{\widehat{S}_b\}_{b\in \mathcal{I}}$; 
		\State \quad Update $\mathcal{I}$ as $\mathcal{I}\leftarrow \{b\in \mathcal{I}: \widehat{S}_b\le z \}$. 
		\State \quad \textbf{If} {$S>z$}:
		\State \qquad Reject $H_0$ and terminate the experiment; 
		%		\EndIf
		%		\EndFor
		%		\EndFor
	\end{algorithmic}
\end{algorithm}	 
{\color{black}We compare our procedure with the always valid test \citep[AVT,][]{johari2017} that extends the two-sample t-test for sequential monitoring. AVT requires to impose a parametric likelihood model assumption (in addition to the conditional mean model) to construct a likelihood-ratio-based ``always valid p-value". To implement the test, we follow the proposal detailed in Section 4.3 of \cite{johari2017}, assume the responses are normal with constant means and known variances, and compute the mixture sequential probability ratio test statistic accordingly. Please refer to Appendix \ref{secbaseline} for details. We remark that the validity of the resulting test requires no confounding variables exist, as the test statistic is derived without adjusting for confounders.} 

We generate the potential outcomes with the same model, except that $\varepsilon_i$'s are i.i.d $N(0, 1)$.  However, we set $N(T_1) = 1000$ and $N(T_j)-N(T_{j-1})=2n$ for $2\le j \le K$ and some $n>0$.  We consider two combinations of $(n, K)$, corresponding to $(n, K) = (100, 5)$ and  $(10, 50)$. For all settings, we use a linear function to approximate $Q$. 
%we consider the $\varphi(x) = x$, since the correctness of statistical inference on average treatment effect does not depend on how we choose $\varphi(\cdot)$ under our framework.

\begin{figure}[t]
	\centering
	\includegraphics[width=\textwidth,height=0.27\textheight]{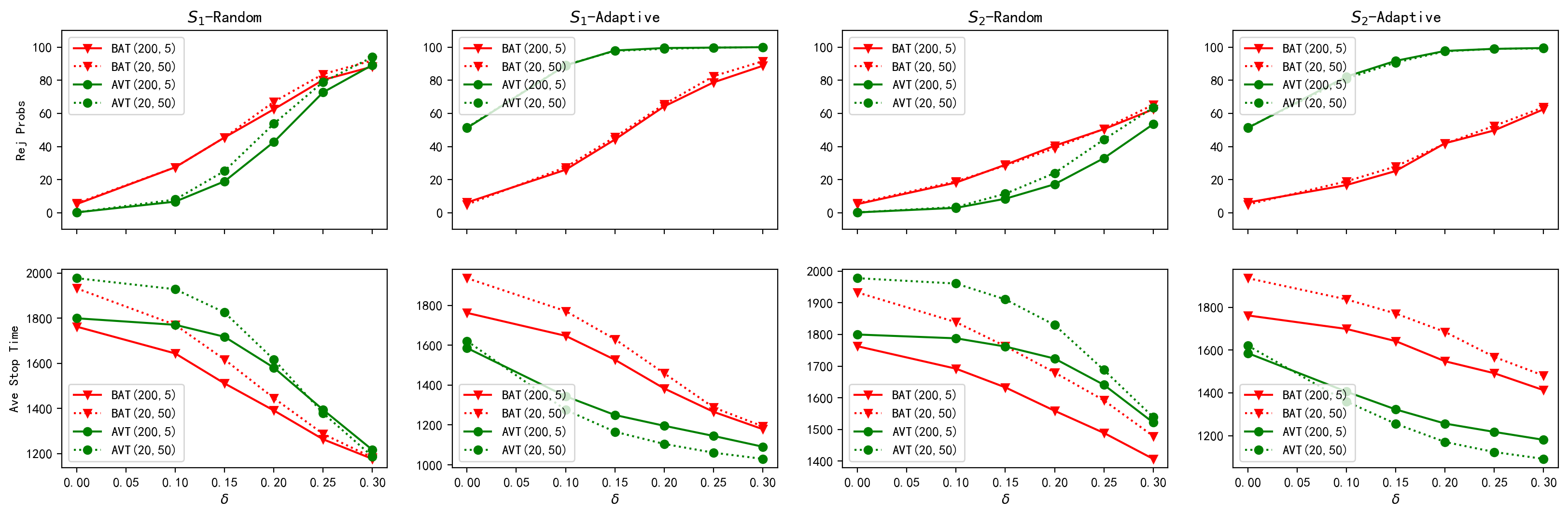}
	\vspace{-0.6cm}
	\footnotesize
	\caption{\small Rejection probabilities and average stopping times of the proposed test
		when $\alpha_1(\cdot)$ is chosen as the spending function. From left to right: Scenario 1 with random design, Scenario 1 with $\epsilon$-greedy design, Scenario 2 with random design and Scenario 2 with $\epsilon$-greedy design.}\label{fig:ate}
\end{figure}

In Table \ref{tab:ate} and Figure \ref{fig:ate}, we show the rejection probabilities and average stopping times of the proposed test aggregated over 400 simulations,
when $\alpha_1(\cdot)$ is chosen as the spending function. %We also compare our test with the always valid test. 
It can be seen that our method behaves better than
the always valid test when the effect size is small, and comparable when the effect size is large. The always valid test fails in the adaptive randomization settings, as the type-I error rates are around 50\% under the null hypothesis. {\color{black}This is because under such a design, the time-varying variables will confound the treatment and the outcome. As commented before, the always valid test is established under settings where no confounders exist. It is expected to fail under the adaptive design.}

\label{sectable}
\begin{table}[h]
	\centering
	\footnotesize
	\scalebox{0.9}{
		\begin{tabular}{|c|c|c|r|r|r|r|r|r|r|r|}
			& & \multirow{2}{*}{method} & \multicolumn{4}{|c|}{BAT} & \multicolumn{4}{|c|}{LIL}  \\
			\cline{4-11}
			& &              & \multicolumn{2}{|c|}{Random} & \multicolumn{2}{|c|}{Adaptive} 
			& \multicolumn{2}{|c|}{Random} & \multicolumn{2}{|c|}{Adaptive} \\
			\hline
			& $(n, K)$ & $\delta$ & rej probs & E[stop] & rej probs & E[stop] & rej probs & E[stop] & rej probs & E[stop] \\
			\hline
			\multirow{8}{*}{S1} & \multirow{4}{*}{(200, 5)} 
			& 0.00 & 5.5(1.1)&3522(16)&5.8(1.2)&3522(16)&0.0(0.0)&3600(0)&0.0(0.0)&3600(0) \\
			&& 0.05 & 20.2(2.0)&3355(27)&16.8(1.9)&3382(26)&0.0(0.0)&3600(0)&0.0(0.0)&3600(0) \\
			&& 0.10 & 0.49.8(2.5)&2985(36)&46.5(2.5)&3013(36)&0.0(0.0)&3600(0)&0.0(0.0)&3600(0) \\
			&& 0.15 & 79.2(2.0)&2554(35)&76.2(2.1)&2572(35)&0.0(0.0)&3600(0)&0.0(0.0)&3600(0) \\
			\cline{2-11}
			& \multirow{4}{*}{(20, 50)} 
			& 0.00 & 6.2(1.2)&3856(20)&6.2(1.2)&3864(19)&0.0(0.0)&3960(0)&0.0(0.0)&3960(0) \\
			&& 0.05 & 27.0(2.2)&3571(35)&28.7(2.3)&3545(36)&0.0(0.0)&3960(0)&0.0(0.0)&3960(0) \\
			&& 0.10 & 68.8(2.3)&2945(42)&69.0(2.3)&2929(42)&0.0(0.0)&3960(0)&0.0(0.0)&3960(0) \\
			&& 0.15 & 95.2(1.1)&2334(29)&94.0(1.2)&2320(28)&0.0(0.0)&3960(0)&0.0(0.0)&3960(0) \\
			\hline 
			\multirow{8}{*}{S2} & \multirow{4}{*}{(200, 5)} 
			& 0.00 & 5.5(1.1)&3522(16)&5.8(1.2)&3522(16)&0.0(0.0)&3600(0)&0.0(0.0)&3600(0) \\
			&& 0.05 & 12.0(1.6)&3469(20)&11.8(1.6)&3468(20)&0.0(0.0)&3600(0)&0.0(0.0)&3600(0) \\
			&& 0.10 & 23.2(2.1)&3317(28)&23.2(2.1)&3321(28)&0.0(0.0)&3600(0)&0.0(0.0)&3600(0)\\
			&& 0.15 & 43.8(2.5)&3111(33)&40.5(2.5)&3134(33)&0.0(0.0)&3600(0)&0.0(0.0)&3600(0)\\
			\cline{2-11}
			& \multirow{4}{*}{(20, 50)}
			& 0.00 & 6.2(1.2)&3856(20)&6.2(1.2)&3864(19)&0.0(0.0)&3960(0)&0.0(0.0)&3960(0)\\
			&& 0.05 & 30.5(2.3)&3584(32)&34.2(2.4)&3533(34)&0.0(0.0)&3960(0)&0.0(0.0)&3960(0)\\
			&& 0.10 & 84.2(1.8)&2789(35)&84.0(1.8)&2778(36)&0.0(0.0)&3960(0)&0.0(0.0)&3960(0)\\
			&& 0.15 & 99.8(0.2)&2268(18)&99.2(0.4)&2252(17)&0.0(0.0)&3960(0)&0.0(0.0)&3960(0) \\
			\hline
	\end{tabular}}
	\caption{QTE: rejection probabilities (multiplied by 100) and average stopping times under Scenarios 1 and 2 when $\alpha_1(\cdot)$ is chosen as the spending function. Standard errors are reported in the parentheses.}\label{tab:hte}
\end{table}

\begin{table}[!t]
	\centering
	\footnotesize
	\scalebox{0.85}{
		\begin{tabular}{|c|c|c|r|r|r|r|r|r|r|r|}
			& & \multirow{2}{*}{method} & \multicolumn{4}{|c|}{BAT} & \multicolumn{4}{|c|}{AVT}  \\
			\cline{4-11}
			& &              & \multicolumn{2}{|c|}{Random} & \multicolumn{2}{|c|}{Adaptive} 
			& \multicolumn{2}{|c|}{Random} & \multicolumn{2}{|c|}{Adaptive} \\
			\hline
			& $(n, K)$ & $\delta$ & rej probs & E[stop] & rej probs & E[stop] & rej probs & E[stop] & rej probs & E[stop] \\
			\hline
			\multirow{12}{*}{S1} & \multirow{6}{*}{(200, 5)}
			&0.00&5.2(1.1)&1763(8)&6.2(1.2)&1762(8)&0.2(0.2)&1800(0)&51.2(2.5)&1586(11)\\
			&&0.10&27.5(2.2)&1644(15)&26.0(2.2)&1647(14)&6.8(1.3)&1771(6)&89.0(1.6)&1344(9)\\
			&&0.15&45.5(2.5)&1511(17)&44.2(2.5)&1527(17)&19.0(2.0)&1718(10)&98.0(0.7)&1250(6)\\
			&&0.20&62.5(2.4)&1391(18)&64.2(2.4)&1383(18)&42.8(2.5)&1581(15)&99.5(0.4)&1196(6)\\
			&&0.25&80.2(2.0)&1263(17)&78.8(2.0)&1266(17)&72.8(2.2)&1394(17)&99.8(0.2)&1145(5)\\
			&&0.30&88.2(1.6)&1176(14)&88.8(1.6)&1179(14)&89.2(1.5)&1216(14)&100.0(0.0)&1091(5)\\
			\cline{2-11}
			& \multirow{6}{*}{(20, 50)}
			&0.00&5.8(1.2)&1933(9)&5.0(1.1)&1936(9)&0.2(0.2)&1978(1)&51.5(2.5)&1621(17)\\
			&&0.10&27.5(2.2)&1771(18)&27.5(2.2)&1771(18)&8.0(1.4)&1929(9)&89.2(1.5)&1276(13)\\
			&&0.15&45.5(2.5)&1617(22)&45.8(2.5)&1630(21)&25.2(2.2)&1826(15)&97.8(0.7)&1166(7)\\
			&&0.20&67.0(2.4)&1446(22)&65.5(2.4)&1459(22)&53.8(2.5)&1617(20)&99.0(0.5)&1105(6)\\
			&&0.25&83.8(1.8)&1287(19)&82.5(1.9)&1288(19)&79.0(2.0)&1379(19)&99.8(0.2)&1061(4)\\
			&&0.30&92.0(1.4)&1182(16)&91.5(1.4)&1193(16)&94.0(1.2)&1187(15)&100.0(0.0)&1030(2)\\
			\hline
			\multirow{12}{*}{S2} & \multirow{6}{*}{(200, 5)} &0.00&5.2(1.1)&1763(8)&6.2(1.2)&1762(8)&0.2(0.2)&1800(0)&51.2(2.5)&1586(11) \\
			&&0.10&18.2(1.9)&1692(12)&16.8(1.9)&1699(12)&3.0(0.9)&1788(3)&82.2(1.9)&1406(10) \\
			&&0.15&29.0(2.3)&1633(15)&25.2(2.2)&1642(15)&8.5(1.4)&1762(7)&91.8(1.4)&1323(9) \\
			&&0.20&40.5(2.5)&1559(17)&42.0(2.5)&1548(17)&17.2(1.9)&1724(10)&97.8(0.7)&1257(7)\\
			&&0.25&50.5(2.5)&1489(18)&49.8(2.5)&1492(18)&33.0(2.4)&1641(14)&99.0(0.5)&1218(6)\\
			&&0.30&62.5(2.4)&1407(18)&62.5(2.4)&1413(18)&53.5(2.5)&1522(16)&99.5(0.4)&1181(6)\\
			\cline{2-11}
			& \multirow{6}{*}{(20, 50)} 
			&0.00&5.8(1.2)&1933(9)&5.0(1.1)&1936(9)&0.2(0.2)&1978(1)&51.5(2.5)&1621(17)\\
			&&0.10&19.0(2.0)&1839(16)&19.0(2.0)&1837(16)&3.5(0.9)&1961(5)&81.0(2.0)&1360(15)\\
			&&0.15&28.5(2.3)&1763(19)&28.0(2.2)&1771(18)&11.5(1.6)&1911(10)&90.8(1.4)&1256(12)\\
			&&0.20&39.0(2.4)&1680(21)&41.8(2.5)&1685(20)&24.0(2.1)&1830(15)&97.5(0.8)&1171(8)\\
			&&0.25&50.7(2.5)&1592(22)&52.5(2.5)&1568(22)&44.2(2.5)&1688(19)&99.0(0.5)&1124(6)\\
			&&0.30&65.2(2.4)&1479(22)&63.7(2.4)&1481(22)&63.5(2.4)&1539(20)&99.2(0.4)&1092(5)\\
			\hline
	\end{tabular}}
	\caption{ATE: rejection probabilities (multiplied by 100) and average stopping times under Scenarios 1 and 2 when $\alpha_1(\cdot)$ is chosen as the spending function. Standard errors are reported in the parentheses.}
	\label{tab:ate}
\end{table}

%\vspace{-0.3cm}
\subsection{Real data analysis}%\vspace{-0.2cm}
%Online content recommendation services have received extensive attention in the machine learning and statistics literature. These online services strive to make recommendations of advertisements or news articles to individual users by using both the content and user information. 
In this section, we apply the proposed method to a Yahoo! Today Module user click log dataset\footnote{\url{https://webscope.sandbox.yahoo.com/catalog.php?datatype=r&did=49}}, which contains 45,811,883 user visits to the Today Module, during the first ten days in May 2009. 
For the $i$th visit, the dataset contains an ID of the new article recommended to the user, a binary response variable $Y_i$ indicating whether the user clicked the article or not, 
and a five dimensional feature vector summarizing information of the user. Due to privacy concerns, feature definitions and article names were not included in the data. %Here, $Y_i=1$ indicates that the user clicked the article and $Y_i=0$ means the user didn’t click. 
Each feature vector sums up to 1. Therefore, we took the first three and the fifth elements to form the covariates $X_i$. For illustration, we only consider a subset of data that contains visits on May 1st where the recommended article ID is either 109510 or 109520. %There were a total of 50 candidate articles on May 1st. We chose these 
These two articles were being recommended most on that day. This gives us a total of 405888 visits. On the reduced dataset, define $A_i=1$ if the recommended article is 109510 and $A_i=0$ otherwise.

\begin{figure}[!b]
	\centering
	\includegraphics[width=\textwidth,height=0.27\textheight]{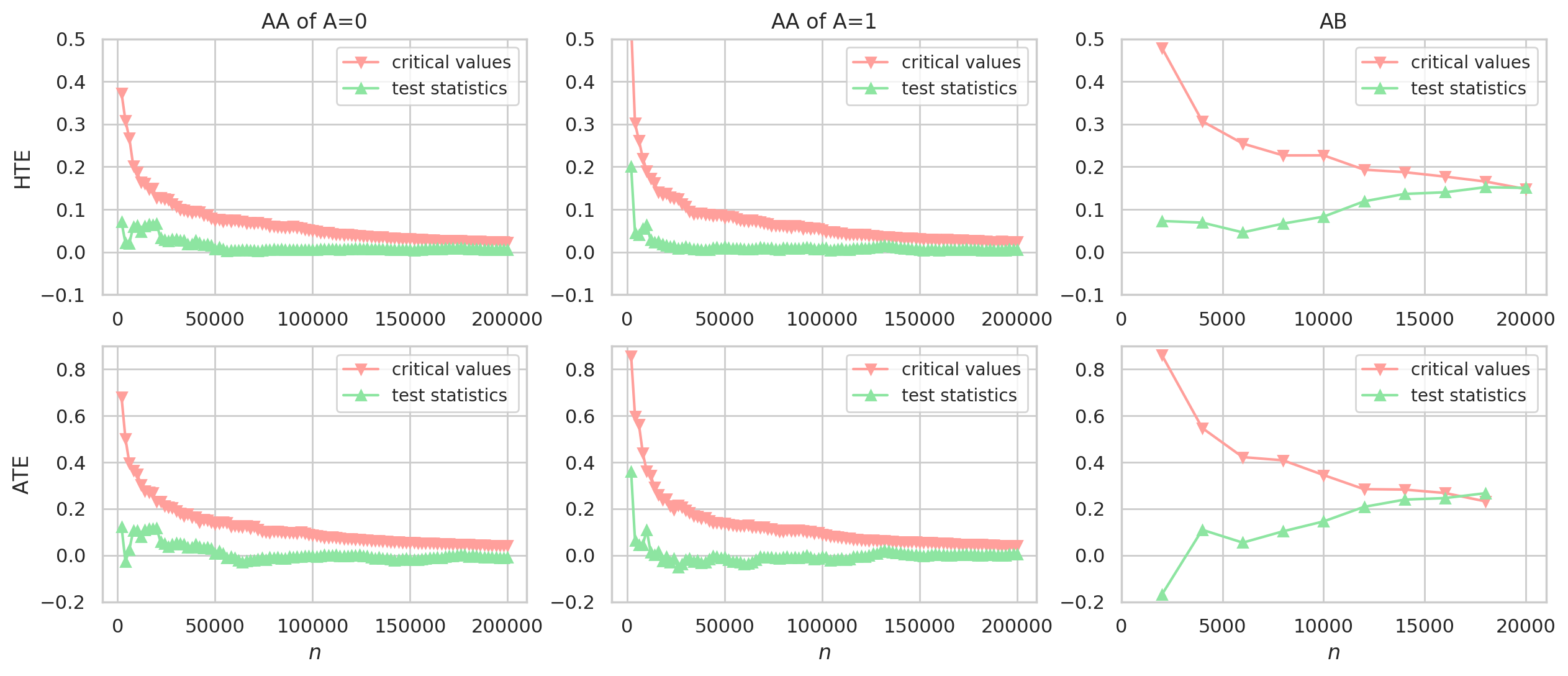}
	\footnotesize
	\caption{Critical values and test statistics.} %across different stages.} 
	\label{fig:qte}
	%	\vspace{-0.5cm}
\end{figure}

We first conduct A/A experiments (which compare these two articles against themselves) to examine the validity of our test.  The A/A experiments are done when every 2000 more users are available, we randomly assign 1000 users to arm A, and the other 1000 users in arm B. We expect our test will not reject $H_0$ in A/A experiments, since the articles being recommended are the same. Then, we conduct A/B experiment to test the QTE of these two articles. The test statistics and their corresponding critical values are plotted in Figure \ref{fig:qte}. On average it takes several seconds to implement our test. It can be seen that our test is able to be reject $H_0$ after obtaining the first one-third of the observations, in the A/B experiment. In the A/A experiments, we fail to reject $H_0$, as expected. 
%so that the new order dispatch strategy can significantly reduce
%239 answer time under certain circumstances.

\section{Proof of Theorem \ref{thm1}}\label{sec:proofthm1}
We present the proof of Theorem \ref{thm1} in this section. Proofs of other theorems are given in the appendix. We begin with some notations. For any matrix $\hbox{Mat}$, we use $\|\hbox{Mat}\|_p$ to denote the matrix norm induced by the corresponding $\ell_p$ norm of vectors, for $1\le p\le +\infty$. For two nonnegative sequences $\{s_{1,n}\}_n$ and $\{s_{2,n}\}_n$, we will use the notation $s_{1,n}\preceq s_{2,n}$ to represent that $s_{1,n}\le \bar{c} s_{2,n}$ for some universal constant $\bar{c}>0$ whose value is allowed to change from place to place. When a matrix $\hbox{Mat}$ is degenerate, $\hbox{Mat}^{-1}$ denotes the Moore-Penrose inverse of $\hbox{Mat}$. For any vector $\psi$, we use $\psi^{(i)}$ to denote its $i$-th element.

Let $n(\cdot)$ be the realization of the counting process $N(\cdot)$. %Similarly, let $K$ be the realization of $K$ and $t_1,t_2,\dots,t_{K}$ be the realizations of $T_1,\dots,T_2,\dots,T_{K}$ that satisfy $0<t_1<t_2<\cdots<t_{K}=T$. 
We will show the assertion in Theorem \ref{thm1} holds for any such realizations that satisfy $n(t_1)<n(t_2)<\cdots<n(t_{K})$. The case where some of the $n(t_k)$'s are the same can be similarly discussed. 

%Under the conditions of Theorem \ref{thm1}, there exists some constant $0<\epsilon<1$ such that
%\begin{eqnarray}\label{epsilon}
%	\pi^*(a,x)\ge \epsilon,\,\,\,\,\forall a\in \{0,1\},x\in \mathbb{X}.
%\end{eqnarray}
For any $j\ge 1$, define $\sigma(\mathcal{F}_j)$ to be the $\sigma$-algebra generated by $\mathcal{F}_j$. For $a\in\{0,1\}$, define
\begin{eqnarray*}
	\widehat{\Sigma}_{a,j}=\frac{1}{j}\sum_{i=1}^{j} \mathbb{I}(A_i=a) \varphi(X_i) \varphi^\top(X_i)\,\,\,\,\hbox{and}\,\,\,\widehat{\beta}_{a,j}=\widehat{\Sigma}_{a,j}^{-1} \left(\frac{1}{j}\sum_{i=1}^j \mathbb{I}(A_i=a)\varphi(X_i) Y_i \right).
\end{eqnarray*}
It is immediate to see that $\widehat{\Sigma}_a(t)=\widehat{\Sigma}_{a,n(t)}$ and $\widehat{\beta}_{a}(t)=\widehat{\beta}_{a,n(t)}$. Define $\delta_n=q n^{-\alpha_0}\log^{\alpha_0} n$. We state the following lemmas before proving Theorem \ref{thm1}. 

%\begin{lemma}\label{somebasiclemma0}
%	For any square matrix $\hbox{Mat}$, we have $\|\hbox{Mat}\|_2\le \|\hbox{Mat}\|_{\infty}$.
%\end{lemma}

\begin{lemma}\label{somebasiclemma1}
	There exists some constant $0<\epsilon_0<1$ such that $\lambda_{\min}[\Mean \varphi(X)\varphi^\top(X)]\ge \epsilon_0$, $\lambda_{\max}[\Mean \varphi(X)\varphi^\top(X)]\le \epsilon_0^{-1}$, $\sup_x \|\varphi(x)\|_2\le \sup_{x} \|\varphi(x)\|_1\le \epsilon_0^{-1}\sqrt{q}$, $\min_{a\in\{0,1\}}\lambda_{\min}[\Sigma_a]\ge \epsilon_0$, $\max_{a\in\{0,1\}} \|\beta_a\|_2\le \epsilon_0^{-1}$, %$\max_{a\in \{0,1\}}|Y^*(a)|\le \epsilon_0^{-1}$ 
	and $\sup_x \max_{a\in \{0,1\}} |Q_0(x,a)|\le \epsilon_0^{-1}$. 
\end{lemma}

\begin{lemma}\label{somebasiclemma2}
	Assume the conditions in Theorem \ref{thm1} hold. Then for any sequence $\{j_n\}_n$ that satisfies $j_n^{\alpha_0} /\log^{\alpha_0} (j_n)\gg q^2$, we have with probability at least $1-O(j_n^{-\alpha_0})$ that for any $a\in\{0,1\}$ and any $k\ge j_n$, 
	\begin{eqnarray}\label{Sigmabasic}
		\|(\widehat{\Sigma}_{a,k}-\Sigma_a)\|_2\preceq q\delta_{k}+\sqrt{q k^{-1}\log k},\\ \label{Sigmabasic1}
		\|(\widehat{\Sigma}_{a,k}^{-1}-\Sigma_a^{-1})\|_2\preceq q\delta_{k}+\sqrt{q k^{-1}\log k}. 
	\end{eqnarray}
\end{lemma}

%\begin{lemma}\label{somebasiclemma3}
%	Assume the conditions in Theorem \ref{thm1} hold. The for any $j\ge 1$, we have with probability at least $1-O(j^{-1/2})$ that for any $a\in\{0,1\}$ and any $k\ge j$,
%	\begin{eqnarray*}
%		\sum_{i=1}^k \{Y_i^*(a)\}^2\preceq k\log k.
%	\end{eqnarray*}
%\end{lemma}

\begin{lemma}\label{somebasiclemma4}
	Assume the conditions in Theorem \ref{thm1} hold. The for any sequence $\{j_n\}_n$ that satisfies $j_n/\log (j_n)\gg q$, we have with probability at least $1-O(j_n^{-1})$ that for any $a\in\{0,1\}$ and any $k\ge j_n$,
	\begin{eqnarray*}
		\left\|\sum_{i=1}^k \varphi(X_i) \mathbb{I}(A_i=a)\{Y_i-Q_0(X_i,a)\}\right\|_2\preceq \sqrt{q k \log k}.
	\end{eqnarray*}
\end{lemma}

\begin{lemma}\label{lemmasomebasic1}
	Assume the conditions in Theorem \ref{thm1} hold. Then for any sequence $\{j_n\}_n$ that satisfies $j_n^{\alpha_0}/\log^{\alpha_0} j_n\gg q^2$, we have with probability at least $1-O(j_n^{-\alpha_0})$ that %for any $a\in \{0,1\}$ and any $k\ge j$,
	\begin{eqnarray*}
		\|\widehat{\beta}_{a,k}-\beta_a\|_2 \preceq q^{1/2} k^{-1/2}\sqrt{\log k},\,\,\,\,\,\,\,\,\forall a\in \{0,1\},\forall k\ge j_n.
	\end{eqnarray*}
\end{lemma}

\begin{lemma}\label{lemmasomebasic2}
	Assume the conditions in Theorem \ref{thm1} hold. Then for any sequence $\{j_n\}_n$ that satisfies $j_n^{\alpha_0}/\log^{\alpha_0} j_n\gg q^2$, we have with probability at least $1-O(j_n^{-\alpha_0})$ that %for any $a\in \{0,1\}$ and any $k\ge j$,
	\begin{eqnarray*}
		\left\|\frac{1}{k}\sum_{i=1}^k \mathbb{I}(A_i=a) \varphi(X_i) \varphi^\top(X_i) \{Y_i-\varphi^\top(X_i) \beta_a\}^2-\Phi_a\right\|_2\preceq q\delta_k+ q^{1/2}k^{-1/2} \sqrt{\log k},\\\forall a\in \{0,1\},k\ge j_n.
	\end{eqnarray*}
\end{lemma}

We now start our proof. We first approximate $\widehat{\beta}_{a,k}-\beta_a$ by a sum of independent mean zero random variables. For $a\in\{0,1\}$, 
\begin{eqnarray*}
	\widehat{\beta}_{a,k}-\beta_a=\widehat{\Sigma}_{a,k}^{-1} \left[\frac{1}{k}\sum_{i=1}^k \mathbb{I}(A_i=a)\varphi(X_i) \{Y_i-Q_0(X_i,a)\}\right]\\+\widehat{\Sigma}_{a,k}^{-1}\left[\frac{1}{k}\sum_{i=1}^k \mathbb{I}(A_i=a)\varphi(X_i)\{Q_0(X_i,a)-\varphi^\top(X_i)\beta_a\}\right].
\end{eqnarray*}
Under the given conditions, using similar arguments in proving (E.40) and (E.41) in \cite{Shi2020}, we can show that the second term is $O(\textrm{err})$. It follows that
\begin{eqnarray}\label{betahat1}
	&&\left\|\widehat{\beta}_{a,k}-\beta_a-\Sigma_a^{-1} \left[\frac{1}{k}\sum_{i=1}^k \mathbb{I}(A_i=a)\varphi(X_i) \{Y_i-Q_0(X_i,a)\}\right]\right\|_2\\ \nonumber
	&\le& \|\widehat{\Sigma}_{a,k}^{-1}-\Sigma_a^{-1}\|_2\left\|\frac{1}{k}\sum_{i=1}^k \mathbb{I}(A_i=a)\varphi(X_i) \{Y_i-Q_0(X_i,a)\}\right\|_2+O(\textrm{err})\\ \nonumber
	&\preceq& (q\delta_{n(t_k)}+\sqrt{q k^{-1}\log k}) q^{1/2} k^{-1/2} \log^{1/2} k+\textrm{err},\,\,\,\,\forall k\ge j_n,
\end{eqnarray}
with probability at least $1-O(j_n^{-\alpha_0})$, by Lemma \ref{somebasiclemma2} and Lemma \ref{somebasiclemma4}. Define
\begin{eqnarray*}
	B^*(t)=\frac{1}{\sqrt{n(t)}} \sum_{i=1}^{n(t)} [\Sigma_1^{-1}\varphi(X_i) A_i \{Y_i-Q_0(X_i,1)\}- \Sigma_0^{-1}\varphi(X_i) (1-A_i) \{Y_i-Q_0(X_i,0)\} ].
\end{eqnarray*}
It follows that
\begin{eqnarray}\label{B*-B}
	\begin{split}
		\|B^*(t_k)-B(t_k)\|_2 \preceq \{q^{3/2} \delta_{n(t_k)}+q\sqrt{n^{-1}(t_k)\log n(t_k)}\} \log^{1/2} n(t_k)\\+\sqrt{n(t_K)}\textrm{err}, \,\,\,\,\forall k\ge 1,
	\end{split}
\end{eqnarray}
with probability at least $1-O(n^{-\alpha_0}(t_1))$. By Lemmas \ref{somebasiclemma1}, \ref{somebasiclemma2} and \ref{lemmasomebasic2}, the denominator in $S(t)$ is of the same order of magnitude as $O(\|\varphi(x)\|_2)$. It follows that
\begin{eqnarray*}
	\left\|\sup_{x\in \mathbb{X}} \frac{\varphi^\top(x) B^*(t_k)}{\sqrt{\sum_{a\in \{0,1\}} \varphi^\top(x)\widehat{\Sigma}_a^{-1}(t_k) \widehat{\Phi}_a(t_k) \widehat{\Sigma}_a^{-1}(t_k) }}-\sup_{x\in \mathbb{X}} \frac{\varphi^\top(x) B(t_k)}{\sqrt{\sum_{a\in \{0,1\}} \varphi^\top(x)\widehat{\Sigma}_a^{-1}(t_k) \widehat{\Phi}_a(t_k) \widehat{\Sigma}_a^{-1}(t_k) }}\right\|_2 \\\le \bar{c} \{q^{3/2} \delta_{n(t_k)}+ q \sqrt{n^{-1}(t_k)\log n(t_k)}\} \sqrt{ \log n(t_k)}+\bar{c} \sqrt{n(t_K)}\textrm{err},\,\,\,\,\forall k\ge 1,
\end{eqnarray*}
with probability at least $1-O(n^{-\alpha_0}(t_1))$, for some constant $\bar{c}>0$, by \eqref{supbasis}. 

%Under the given conditions on $q$ and $n(t_1)$, we have
%\begin{eqnarray*}
%	q\sqrt{n^{-1}(t_k) \log n(t_k)}=o(1),\,\,\,\,\,\,\,\,\forall k\ge 1,
%\end{eqnarray*}
%and hence
%\begin{eqnarray*}
%	\left\|\sup_{x\in \mathbb{X}} \frac{\varphi^\top(x) B^*(t_k)}{\sqrt{\sum_{a\in \{0,1\}} \varphi^\top(x)\widehat{\Sigma}_a^{-1}(t_k) \widehat{\Phi}_a(t_k) \widehat{\Sigma}_a^{-1}(t_k) }}-\sup_{x\in \mathbb{X}} \frac{\varphi^\top(x) B(t_k)}{\sqrt{\sum_{a\in \{0,1\}} \varphi^\top(x)\widehat{\Sigma}_a^{-1}(t_k) \widehat{\Phi}_a(t_k) \widehat{\Sigma}_a^{-1}(t_k) }}\right\|_2\\ \le \bar{c} \{\sqrt{q} \delta_{n(t_k)}+ \sqrt{ n^{-1}(t_k)\log n(t_k)}\}+\bar{c}\sqrt{n(t_K)}\textrm{err}, \,\,\,\,\forall k\ge 1.
%\end{eqnarray*}
Define $S^*(t)$ to be a version of our test with $B(t)$ replaced by $B^*(t)$. For any given $z_1,z_2,\dots,z_{K}$, we obtain
\begin{eqnarray}\nonumber
	&&\hbox{Pr}\left\{ \max_{k\in\{1,\dots,K\}} \left(S^*(t_k)-z_{k,-}^0 \right)\le 0\right\}-O(n^{-\alpha_0}(t_1))\\\label{firstinequality0}&\le& \hbox{Pr}\left\{\max_{k\in\{1,\dots,K\}} \left(S(t_k)-z_k\right)\le 0 \right\}  \\\nonumber
	&\le & 
	\hbox{Pr}\left\{ \max_{k\in\{1,\dots,K\}} \left(S^*(t_k)-z_{k,+}^0 \right)\le 0\right\}+O(n^{-\alpha_0}(t_1)),
\end{eqnarray}
where
\begin{eqnarray*}
	z_{k,-}^0=z_k-\bar{c}\{q^{3/2} \delta_{n(t_k)}\sqrt{\log n(t_k)}+ q\sqrt{n^{-1}(t_k)}\log n(t_k)+\sqrt{n(t_K)}\textrm{err} \}/2,\\
	z_{k,+}^0=z_k+\bar{c}\{q^{3/2}\delta_{n(t_k)}\sqrt{\log n(t_k)}+ q\sqrt{n^{-1}(t_k)}\log n(t_k)+\sqrt{n(t_K)}\textrm{err}\}/2.
\end{eqnarray*}
This completes the first step of the proof.

In the next step, we focus on bounding the difference between $S^*(t)$ and $S^{**}(t)$, the latter being a version of $S^*(t)$ with the denominator replaced with the oracle value 
\begin{eqnarray*}
	\sqrt{\sum_{a\in \{0,1\}} \varphi^\top(x) \Sigma_a^{-1} \Phi_a \Sigma_a^{-1} \varphi(x)}. 
\end{eqnarray*}
Similarly, under Lemmas \ref{somebasiclemma2}, \ref{somebasiclemma4} and \ref{lemmasomebasic2}, we can show that difference $|S^*(t_k)-S^{**}(t_k)|$ can be upper bounded by $O(q^{3/2} \delta_{n(t_k)}\sqrt{\log n(t_k)}+ q\sqrt{n^{-1}(t_k)}\log n(t_k))$, uniformly for any $k$. Combine this together with \eqref{firstinequality0} yields 
\begin{eqnarray}\nonumber
	&&\hbox{Pr}\left\{ \max_{k\in\{1,\dots,K\}} \left(S^{**}(t_k)-z_{k,-} \right)\le 0\right\}-O(n^{-\alpha_0}(t_1))\\\label{firstinequality}&\le& \hbox{Pr}\left\{\max_{k\in\{1,\dots,K\}} \left(S(t_k)-z_k\right)\le 0 \right\}  \\\nonumber
	&\le & 
	\hbox{Pr}\left\{ \max_{k\in\{1,\dots,K\}} \left(S^{**}(t_k)-z_{k,+} \right)\le 0\right\}+O(n^{-\alpha_0}(t_1)),
\end{eqnarray}
where 
\begin{eqnarray*}
	z_{k,-}=z_k-\bar{c}\{q^{3/2} \delta_{n(t_k)}\sqrt{\log n(t_k)}+ q\sqrt{n^{-1}(t_k)}\log n(t_k)+\sqrt{n(t_K)}\textrm{err} \},\\
	z_{k,+}=z_k+\bar{c}\{q^{3/2}\delta_{n(t_k)}\sqrt{\log n(t_k)}+ q\sqrt{n^{-1}(t_k)}\log n(t_k)+\sqrt{n(t_K)}\textrm{err}\}.
\end{eqnarray*}
This completes the second step. 

In the last step, we aim to apply the high-dimensional Gaussian approximation technique developed by \cite{belloni2018} to approximate $S^{**}(t)$ by its Gaussian analogue. 
For any $i\ge 1$, $1\le k\le K$, define a $q$-dimensional vector 
\begin{eqnarray*}
	\xi_{i,k}=\frac{1}{\sqrt{n(t_k)}}[\Sigma_1^{-1}\varphi(X_i) A_i \{Y_i-Q_0(X_i,1)\}- \Sigma_0^{-1}\varphi(X_i) (1-A_i) \{Y_i-Q_0(X_i,0)\}]\mathbb{I}(i\le n(t_k)),
\end{eqnarray*}
or equivalently,
\begin{eqnarray*}
	\xi_{i,k}=\frac{1}{\sqrt{n(t_k)}}[\Sigma_1^{-1}\varphi(X_i) A_i \{Y_i^*(1)-Q_0(X_i,1)\}- \Sigma_0^{-1}\varphi(X_i) (1-A_i) \{Y_i^*(0)-Q_0(X_i,0)\} ]\mathbb{I}(i\le n(t_k)),
\end{eqnarray*}
by Condition (A1). Let $\bm{\xi}_i=(\xi_{i,1}^\top,\xi_{i,2}^\top,\cdots,\xi_{i,K}^\top)^\top$ and $\mathcal{M}_j=\sum_{i=1}^j \bm{\xi}_i$. The sequence $\{\mathcal{M}_i\}_{i\ge 1}$ forms a multivariate martingale with respect to the filtration $\{\sigma(\mathcal{F}_i):i\ge 1\}$, since
\begin{eqnarray*}
	\Mean (\xi_{i,k}|\mathcal{F}_i)=[\{\Mean (\xi_{i,k}|A_i,X_i,\mathcal{F}_i)\}|\mathcal{F}_i]=0,
\end{eqnarray*}
by (A2).  
Let $n(t_0)=0$. For any $i$ such that $n(t_{k-1})<i\le n(t_k)$ for some $1\le k\le K$, we have
\begin{eqnarray*}
	\|\bm{\xi}_i\|_{\infty}\le \frac{1}{\sqrt{n(t_k)}} \{\|\Sigma_1^{-1}\varphi(X_i) \{Y_i^*(1)-Q_0(X_i,1) \beta_1\}\|_2+\|\Sigma_0^{-1}\varphi(X_i) \{Y_i^*(0)-Q_0(X_i,0)\}\|_2\}\\
	\le \sqrt{q}n^{-1/2}(t_k)\epsilon_0^{-2} (2\epsilon_0^{-1}+|Y_i^*(0)|+|Y_i^*(1)|),
	%\\\le \frac{\sqrt{q}}{\epsilon_0^2\sqrt{n(t_k)}} \{|Y_i^*(1)-\varphi^\top(X_i) \beta_1|+|Y_i^*(0)-\varphi^\top(X_i) \beta_0|\}\le \frac{\sqrt{q}}{\epsilon_0^2\sqrt{n(t_k)}} \sum_{a\in \{0,1\}}\{|Y_i^*(a)|+|\varphi^\top(X_i) \beta_a| \},%\le \frac{\sqrt{q}}{\epsilon_0^2\sqrt{n(t_k)}} \{|Y_i^*(1)|+|Y_i^*(0)|+\sqrt{q}\epsilon_0^{-2}\},
\end{eqnarray*}
where the second inequality is due to Lemma \ref{somebasiclemma1}. Under the sub-Gaussianity assumption, $Y^*(0)$ and $Y^*(1)$ have moments of all orders. 
Therefore,
\begin{eqnarray*}
	\Mean \|\bm{\xi}_i\|_{\infty}^3\preceq \frac{q^{3/2}}{n^{3/2}(t_k)}.
\end{eqnarray*}
It follows that
\begin{eqnarray}\label{xithird}
	&&\sum_{i=1}^{n(t_{K})} \Mean\|\bm{\xi}_i\|_{\infty}^3= \sum_{k=1}^{K} \sum_{i=n(t_{k-1})+1}^{n(t_k)} \Mean\|\bm{\xi}_i\|_{\infty}^3\preceq q^{3/2} \sum_{k=1}^{K} \frac{n(t_k)-n(t_{k-1})}{n^{3/2}(t_k)}\\\nonumber&\le& \frac{q^{3/2}}{\sqrt{n(t_1)}}+q^{3/2}\sum_{k=2}^{K} \frac{n(t_k)-n(t_{k-1})}{n^{3/2}(t_k)} 
	\le q^{3/2} n^{-1/2}(t_1)+q^{3/2} \int_{n(t_1)}^{+\infty} x^{-3/2}dx= 3q^{3/2}n^{-1/2}(t_1).
\end{eqnarray}
Define a sequence of independent Gaussian vectors $\{\bm{\eta}_i\}_{i\ge 1}$ that satisfy $\bm{\eta}_i\sim N(0,\Mean (\bm{\xi}_i\bm{\xi}_i^\top|\mathcal{F}_{i-1}) )$ for any $i\ge 1$. Then the distribution of $\bm{\eta}_i$ is the same as 
\begin{eqnarray*}
	\left( \frac{\mathbb{I}(i\le n(t_1))}{\sqrt{n(t_1)}}Z^\top,\frac{\mathbb{I}(i\le n(t_2))}{\sqrt{n(t_2)}}Z^\top,\cdots,\frac{\mathbb{I}(i\le n(t_{K}))}{\sqrt{n(t_{K})}}Z^\top \right),
\end{eqnarray*}
where $Z$ is a $p$-dimensional mean-zero Gaussian vector with covariance matrix
\begin{eqnarray}\label{covariance}
	\begin{split}
		&\Cov [\sum_{a\in\{0,1\}}\Sigma_a^{-1}\varphi(X_i) \mathbb{I}(A_i=a) \{Y_i^*(a)-Q_0(X_i,a) \}|\mathcal{F}_{i-1} ]\\ 
		=&\sum_{a\in\{0,1\}}\Sigma_a^{-1} \Mean [\varphi(X_i) \varphi^\top(X_i) \mathbb{I}(A_i=a) \{Y_i^*(a)-Q_0(X_i,a)\}^2  |\mathcal{F}_{i-1}]\Sigma_a^{-1}\\ 
		=&\sum_{a\in\{0,1\}}\Sigma_a^{-1} \Mean \{\varphi(X_i) \varphi^\top(X_i) \mathbb{I}(A_i=a) \sigma^2(a,X_i)  |\mathcal{F}_{i-1}\} \Sigma_a^{-1}\\ 
		=&\sum_{a\in\{0,1\}}\Sigma_a^{-1} \Mean  \{\varphi(X_i) \varphi^\top(X_i) \pi_{i-1}(a,X_i) \sigma^2(a,X_i)  |\mathcal{F}_{i-1}\}\Sigma_a^{-1}\\
		\equiv &\sum_{a\in\{0,1\}}\Sigma_a^{-1} \Mean^{\mathcal{F}_{i-1}} \pi_{i-1}(a,X) \sigma^2(a,X)  \varphi(X) \varphi^\top(X) \Sigma_a^{-1},
	\end{split}
\end{eqnarray} 
where the second equality follows from (A2) and Lemma \ref{lemmaxyindpast}, the third equality is due to the definition of $\pi_{i-1}$ and the last equality follows from Lemma \ref{lemmaxyindpast}. See the proof of Lemma \ref{lemmakeyequation} for details. 

Similar to \eqref{xithird}, we can show that
\begin{eqnarray}\label{etathird}
	\sum_{i=1}^{n(t_{K})} \Mean\|\bm{\eta}_i\|_{\infty}^3\preceq q^{3/2} n^{-1/2}(t_1).
\end{eqnarray}
Using similar arguments in \eqref{covariance}, we can show that for any $1\le k_1\le k_2\le K$,
\begin{eqnarray*}
	\sum_{i=1}^{n(t_{K})} \Mean \{\xi_{i,k_1}\xi_{i,k_2}^\top|\mathcal{F}_{i-1}\}=\frac{1}{\sqrt{n(t_{k_1})n(t_{k_2})}} \sum_{i=1}^{n(t_{k_1})} \sum_{a\in\{0,1\}}\Sigma_a^{-1} \Mean^{\mathcal{F}_{i-1}} \pi_{i-1}(a,X) \sigma^2(a,X)  \varphi(X) \varphi^\top(X) \Sigma_a^{-1}.
\end{eqnarray*}
Let
\begin{eqnarray*}
	V(k_1,k_2)=\frac{1}{\sqrt{n(t_{k_1})n(t_{k_2})}} \sum_{i=1}^{n(t_{k_1})} \sum_{a\in\{0,1\}}\Sigma_a^{-1}\Mean^{\mathcal{F}_{i-1}} \pi^*(a,X) \sigma^2(a,X)  \varphi(X) \varphi^\top(X) \Sigma_a^{-1}\\
	=\frac{1}{\sqrt{n(t_{k_1})n(t_{k_2})}} \sum_{i=1}^{n(t_{k_1})} \sum_{a\in\{0,1\}}\Sigma_a^{-1} \Phi_a \Sigma_a^{-1}=\frac{\sqrt{n(t_{k_1})}}{\sqrt{n(t_{k_2})}}\sum_{a\in\{0,1\}}\Sigma_a^{-1} \Phi_a \Sigma_a^{-1}.
\end{eqnarray*}
Consider an arbitrary sequence of $\mathbb{R}^{p+1}$ vectors $\{b_k\}_{1\le k\le K}$. Under the given conditions, we have
\begin{eqnarray}\nonumber
	&&\left|b_{k_1}^\top \left(\sum_{i=1}^{n(t_{K})} \Mean (\xi_{i,k_1}\xi_{i,k_2}^\top|\mathcal{F}_{i-1})-V(k_1,k_2)\right) b_{k_2}\right|\\\nonumber&\preceq& \frac{1}{n(t_{k_1})} \sum_{a\in\{0,1\}}\left\|\sum_{i=1}^{n(t_{k_1})} \Mean^{\mathcal{F}_{i-1}}\{\pi_{i-1}(a,X)-\pi^*(a,X)\}\sigma^2(a,X)\varphi(X)\varphi^\top(X) \right\|_2 \|b_{k_1}\|_2\|b_{k_2}\|_2. %\\ \nonumber&\le& \sum_{a\in\{0,1\}}  \left\|\frac{1}{n(t_{k_1})} \sum_{i=1}^{n(t_{k_1})} \{\pi_{i-1}(a,X)-\pi^*(a,X) \} \right\|_2\|b_{k_1}\|_2\|b_{k_2}\|_2.
\end{eqnarray}
Define a matrix $\bm{V}$ as 
\begin{eqnarray}\label{Vdefi}
	\bm{V}=\left(\begin{array}{cccc}
		V(1,1) & V(1,2) & \dots & V(1,K)\\
		V(2,1) & V(2,2) & \dots & V(2,K)\\
		\vdots & \vdots & & \vdots \\
		V(K,1) & V(K,2) & \dots & V(K,K)
	\end{array}\right).
\end{eqnarray}
It follows that
\begin{eqnarray*}%\label{diffcov}
	\left\|\sum_{i=1}^{n(t_{K})} \Mean(\bm{\xi}_{i}\bm{\xi}_{i}^\top|\mathcal{F}_{i-1})-\bm{V} \right\|_2 %\\ \nonumber
	\preceq \sup_{\substack{a\in\{0,1\}\\j\ge n(t_1)}} \left\|\frac{1}{j}\sum_{i=1}^{j} \Mean^{\mathcal{F}_{i-1}}\{\pi_{i-1}(a,X)-\pi^*(a,X)\}\sigma^2(a,X)\varphi(X)\varphi^\top(X) \right\|_2.
\end{eqnarray*}
Using similar arguments in proving \eqref{Sigmabasic}, we can show the RHS of the above equation is upper bounded by 
\begin{eqnarray*}
	\epsilon_0^{-2} q\sup_{\substack{a\in \{0,1\} \\ x\in \mathbb{X}, j\ge n(t_1) }} \left|\frac{1}{j}\sum_{i=1}^{j} \{\pi_{i-1}(a,x)-\pi^*(a,x)\}\right|,
\end{eqnarray*}
and hence by $\epsilon_0^{-2}q\delta_{n(t_1)}$, with probability at least $1-O(n^{-\alpha_0}(t_1))$.
%since
%\begin{eqnarray*}
%	\sum_{j\ge n(t_1)} j^{-(1+\alpha_0)}=n^{-(1+\alpha_0)}(t_1)+\sum_{j>n(t_1)} j^{-(1+\alpha_0)}\le n^{-(1+\alpha_0)}(t_1)+\int_{n(t_1)}^{+\infty} z^{-(1+\alpha_0)}dz\\=n^{-(1+\alpha_0)}(t_1)+\frac{1}{\alpha_0} n^{-\alpha_0}(t_1)\preceq n^{-\alpha_0}(t_1).
%\end{eqnarray*}
Therefore, we have
\begin{eqnarray}\label{mineigen}
	\lambda_{\min}\left[\bm{V}+\delta_{n(t_1)}\bm{I}_{K p\times K p}-\sum_{i=1}^{n(t_{K})} \Mean(\bm{\xi}_{i}\bm{\xi}_{i}^\top|\mathcal{F}_{i-1}) \right]\ge 0,
\end{eqnarray}
with probability at least $1-O(n^{-\alpha_0}(t_1))$, where $\bm{I}_{K p\times K p}$ denotes a $K p\times K p$ identity matrix. 

Moreover, notice that
\begin{eqnarray*}
	\sup_{\substack{a\in\{0,1\}\\ x\in \mathbb{X}, j\ge n(t_1)}} \left|\frac{1}{j} \sum_{i=1}^{j} \{\pi_{i-1}(a,x)-\pi^*(a,x) \} \right|
\end{eqnarray*}
is bounded between $0$ and $1$. For any $a\in\{0,1\}$ and any $z>0$, we have
\begin{eqnarray*}
	&& \Mean \sup_{\substack{a\in\{0,1\}\\x\in \mathbb{X}, j\ge n(t_1)}} \left|\frac{1}{j} \sum_{i=1}^{j} \{\pi_{i-1}(a,x)-\pi^*(a,x) \} \right|\\
	&\le& \Mean \sup_{\substack{a\in\{0,1\}\\x\in \mathbb{X}, j\ge n(t_1)}} \left|\frac{1}{j} \sum_{i=1}^{j} \{\pi_{i-1}(a,x)-\pi^*(a,x) \} \right| \mathbb{I}\left( \sup_{\substack{a\in\{0,1\}\\ x\in \mathbb{X}, j\ge n(t_1)}} \left|\frac{1}{j} \sum_{i=1}^{j} \{\pi_{i-1}(a,x)-\pi^*(a,x) \} \right|\le z \right)\\
	&+& \hbox{Pr}\left( \sup_{\substack{a\in\{0,1\}\\x\in \mathbb{X}, j\ge n(t_1)}} \left|\frac{1}{j} \sum_{i=1}^{j} \{\pi_{i-1}(a,x)-\pi^*(a,x) \} \right|> z \right). %\le \delta_{n(t_{1})}+O(n^{-\alpha_0}(t_{1})).
\end{eqnarray*}
Under the given conditions, we have
\begin{eqnarray*}
	\Mean \sup_{\substack{a\in\{0,1\}\\x\in \mathbb{X}, j\ge n(t_1)}} \left|\frac{1}{j} \sum_{i=1}^{j} \{\pi_{i-1}(a,x)-\pi^*(a,x) \} \right|\preceq \delta_{n(t_{1})}+O(n^{-\alpha_0}(t_{1})).
\end{eqnarray*}
Therefore, we obtain
\begin{eqnarray*}
	\Mean \left\|\sum_{i=1}^{n(t_{K})} \Mean(\bm{\xi}_{i}\bm{\xi}_{i}^\top|\mathcal{F}_{i-1})-\bm{V} \right\|_2\preceq qn^{-\alpha_0}(t_1)+q\delta_{n(t_1)},
\end{eqnarray*}
or 
\begin{eqnarray}\label{expecteddiff}
	\Mean \left\|\sum_{i=1}^{n(t_{K})} \Mean(\bm{\xi}_{i}\bm{\xi}_{i}^\top|\mathcal{F}_{i-1})-\bm{V} \right\|_2\preceq q\delta_{n(t_1)},
\end{eqnarray}
since $n^{-\alpha_0}(t_1)\ll \delta_{n(t_1)}$. 
Combining \eqref{xithird} with \eqref{etathird}, \eqref{mineigen} and \eqref{expecteddiff}, an application of Theorem 2.1 in \cite{belloni2018} yields that
\begin{eqnarray}\label{approximation}
	&&|\Mean \psi(\mathcal{M}_{n(t_{K})})-\Mean \psi(N(0,\bm{V}))|\\ \nonumber
	&\preceq& c_0(\psi) n^{-\alpha_0}(t_1)+c_2(\psi) q \delta_{n(t_1)}+c_3(\psi) q^{3/2} n^{-1/2}(t_1),
\end{eqnarray}
for any thrice differential function $\psi(\cdot)$, and
\begin{eqnarray*}
	c_0(\psi)=\sup_{z,z'\in \mathbb{R}^{pK}} |\psi(z)-\psi(z')|\,\,\hbox{and}\,\,c_i=\sup_{z\in \mathbb{R}^{pK}} \sum_{j_1,\cdots,j_i} |\partial_{j_1}\partial_{j_2}\cdots\partial_{j_i} \psi(z) |,i=2,3,
\end{eqnarray*}
where $\partial_j g(z)$ denotes the partial derivative $\partial g(z)/\partial z^{(j)}$ for any function $\hbox{g}(\cdot)$ and $z^{(j)}$ stands for the $j$-th element of $z$. 

Let $\mathbb{X}_{k,0}$ be an $\varepsilon$-net of $\mathbb{X}$ that satisfies the following: for any $x\in \mathbb{X}$, there exists some $x_0\in \mathbb{X}_0$ such that $\|x-x_0\|_2\le \varepsilon$. Note that we require $\mathbb{X}$ to be a compact set. {\color{black}To simplify the proof, we assume $\mathbb{X}=[0,1]^d$. In cases where $\mathbb{X}\neq [0,1]^d$, we could conduct the min-max normalization to rescale the range of features to $[0,1]$.} Set $\varepsilon=\sqrt{d}/n^4(t_1)$. There exists some $\mathbb{X}_0$ with 
\begin{eqnarray}\label{nocardX0}
	|\mathbb{X}_0|\le n^{4d}(t_1),
\end{eqnarray} 
where $|\mathbb{X}_0|$ denotes the number of elements in $\mathbb{X}_0$. Under Condition (A3), we have
\begin{eqnarray*}
	\sup_{x\in \mathbb{X}}\inf_{x_0\in \mathbb{X}_0}\|\varphi(x)-\varphi(x_0)\|_2\preceq \frac{\sqrt{q}}{n^4(t_1)}.
\end{eqnarray*}
It follows that
\begin{eqnarray}\label{x-x0}
	\sup_{\|\nu\|_2=1} |\sup_{x\in \mathbb{X}}\varphi^\top(x) \nu-\sup_{x\in \mathbb{X}_0}\varphi^\top(x) \nu|\preceq \frac{\sqrt{q}}{n^4(t_1)}.
\end{eqnarray}
Using similar arguments in showing \eqref{B*-B}, we can show the following event occurs with probability at least $1-O(n^{-1}(t_1))$,
\begin{eqnarray*}
	\|B^*(t_k)\|_2\preceq q^{1/2} \log^{1/2} n(t_k),\,\,\,\,\forall k\ge 1.
\end{eqnarray*}
This together with \eqref{x-x0} and the fact that the denominator $\sqrt{\sum_{a\in \{0,1\}} \varphi^\top(x)\Sigma_a^{-1} \Phi_a \Sigma_a^{-1} \varphi(x)}/\|\varphi(x)\|_2$ is uniformly bounded away from zero yields 
\begin{eqnarray*}
	\left|\max_{k\in \{1,\dots,K\}}\sup_{x\in \mathbb{X}}\frac{\varphi^\top(x) B^*(t_k)}{\sqrt{\sum_{a\in \{0,1\}} \varphi^\top(x)\Sigma_a^{-1} \Phi_a \Sigma_a^{-1} \varphi(x)}}-\max_{k\in \{1,\dots,K\}}\sup_{x\in \mathbb{X}_0}\frac{\varphi^\top(x) B^*(t_k)}{\sqrt{\sum_{a\in \{0,1\}} \varphi^\top(x)\Sigma_a^{-1} \Phi_a \Sigma_a^{-1} \varphi(x)}} \right|\\
	\le \max_{k\in \{1,\dots,K\}}\left|\sup_{x\in \mathbb{X}}\frac{\varphi^\top(x) B^*(t_k)}{\sqrt{\sum_{a\in \{0,1\}} \varphi^\top(x)\Sigma_a^{-1} \Phi_a \Sigma_a^{-1} \varphi(x)}}-\sup_{x\in \mathbb{X}_0}\frac{\varphi^\top(x) B^*(t_k)}{\sqrt{\sum_{a\in \{0,1\}} \varphi^\top(x)\Sigma_a^{-1} \Phi_a \Sigma_a^{-1} \varphi(x)}}\right|\\\preceq \frac{\sqrt{q \log n(t_{K})}}{n^4(t_1)},
\end{eqnarray*}
with probability at least $1-O(n^{-1}(t_1))$. Under the given conditions, we have $n(t_1)\gg \max(q, \log n(t_{K}))$. It follows that there exists some constant $\bar{c}^*>0$ such that
\begin{eqnarray}\label{someimportantinequality0}
	\left|\max_{k\in \{1,\dots,K\}}\sup_{x\in \mathbb{X}}\varphi^\top(x) B^*(t_k)-\max_{k\in \{1,\dots,K\}}\sup_{x\in \mathbb{X}_0}\varphi^\top(x) B^*(t_k)\right|\le  \bar{c}^*n^{-2}(t_1),
\end{eqnarray}
with probability at least $1-O(n^{-1}(t_1))$. 

Define
\begin{eqnarray*}
	z_{k,-}^*=z_k-\bar{c}\{q^{3/2} \delta_{n(t_k)}\sqrt{\log n(t_k)}+q \sqrt{n^{-1}(t_k)}\log n(t_k)+\sqrt{n(t_K)}\textrm{err}\} -\bar{c}^* n^{-2}(t_1),\\
	z_{k,+}^*=z_k+\bar{c}\{q^{3/2} \delta_{n(t_k)}\sqrt{\log n(t_k)}+q \sqrt{n^{-1}(t_k)}\log n(t_k)+\sqrt{n(t_K)}\textrm{err}\} +\bar{c}^* n^{-2}(t_1).
\end{eqnarray*}
Combining \eqref{someimportantinequality0} with \eqref{firstinequality} yields
\begin{eqnarray}\nonumber
	\hbox{Pr}\left\{ \max_{k\in\{1,\dots,K\}} \left(\sup_{x\in \mathbb{X}_0} \frac{\varphi^\top(x) B^*(t_k)}{ \sqrt{\sum_{a\in \{0,1\}} \varphi^\top(x) \Sigma_a^{-1} \Phi_a \Sigma_a^{-1} \varphi(x)}}-z_{k,-}^* \right)\le 0\right\}-O(n^{-\alpha_0}(t_1))\\\label{firsthalfinequality}\le \hbox{Pr}\left\{\max_{k\in\{1,\dots,K\}} \left(\sup_{x\in \mathbb{X}} \frac{\varphi^\top(x) B(t_k)}{ \sqrt{\sum_{a\in \{0,1\}} \varphi^\top(x) \widehat{\Sigma}_a^{-1}(t_k) \widehat{\Phi}_a(t_k) \widehat{\Sigma}_a^{-1}(t_k) \varphi(x)}}-z_k\right)\le 0 \right\}  \\\nonumber
	\le 
	\hbox{Pr}\left\{ \max_{k\in\{1,\dots,K\}} \left(\sup_{x\in \mathbb{X}_0} \frac{\varphi^\top(x) B^*(t_k)}{ \sqrt{\sum_{a\in \{0,1\}} \varphi^\top(x) \Sigma_a^{-1} \Phi_a \Sigma_a^{-1} \varphi(x)}}-z_{k,+}^* \right)\le 0\right\}+O(n^{-\alpha_0}(t_1)).
\end{eqnarray}
Notice that $\mathcal{M}_{n(t_{K})}=\{B^*(t_1)^\top,B^*(t_2)^\top,\cdots,B^*(t_{K})^\top\}^\top$. By \eqref{nocardX0} and the fact that the denominator $\sqrt{\sum_{a\in \{0,1\}} \varphi^\top(x) \Sigma_a^{-1} \Phi_a \Sigma_a^{-1} \varphi(x)}\ge \bar{c}\|\varphi(x)\|_2$ for some constant $\bar{c}>0$, there exist a set of vectors $d_1,d_2,\dots,d_{L}\in \mathbb{R}^{qK}$ with $L\le n^{4d}(t_1)  K$, $\max_j \|d_j\|_1\le \epsilon^{-1}$ for some $0<\epsilon<1$ and a function $k(\cdot)$ that maps $\{1,\dots,L\}$ into $\{1,\dots,K\}$ such that 
\begin{eqnarray}\label{firstequality}
	\max_{k\in \{1,\dots,K\}} \left\{\sup_{x\in \mathbb{X}_0} \frac{\varphi^\top(x) B^*(t_k)}{\sqrt{\sum_{a\in \{0,1\}} \varphi^\top(x) \Sigma_a^{-1} \Phi_a \Sigma_a^{-1} \varphi(x)}}-\nu_k\right\}=\max_{ 1\le j\le L} \{d_j^\top \mathcal{M}_{n(t_{K})}-\nu_{k(j)}\},
\end{eqnarray}
for any $\{\nu_k\}_{k=1}^{K}$. 
For any $\eta>0$, $m\in \mathbb{R}^{qK}$, consider the function $\phi_{\eta, \{\nu_k\}_{k}}: \mathbb{R}^{qK}\to \mathbb{R}$, defined as
\begin{eqnarray*}
	\phi_{\eta, \{\nu_k\}_k}(m)=\frac{1}{\eta} \log\left\{\sum_{j=1}^{L}\exp  [\eta \{d_j^\top m-\eta \nu_{k(j)}\}]  \right\}.
\end{eqnarray*}
It has the following property:
\begin{eqnarray*}
	&&\max_{1\le j\le L} \{d_j^\top m-\nu_{k(j)}\}\le \phi_{\eta, \{\nu_k\}_k}(m)\le \max_{1\le j\le L} \{d_j^\top m-\nu_{k(j)}\}+ \eta^{-1} \log L\\
	&\le& \max_{1\le j\le L} \{d_j^\top m-\nu_{k(j)}\}+ \eta^{-1} \{\log K+4d \log n(t_1)\} \\
	&=&\max_{1\le j\le L} [d_j^\top m-\{\nu_{k(j)}-\eta^{-1} \log K- \eta^{-1} 4d \log n(t_1)\}].
\end{eqnarray*}
It follows that
\begin{eqnarray}\label{secondinequality}
	&&\hbox{Pr}\left\{ \max_{k\in \{1,\dots,K \}} \left( \sup_{x\in \mathbb{X}_0} \frac{\varphi^\top(x) B^*(t_k)}{\sqrt{\sum_{a\in \{0,1\}} \varphi^\top(x) \Sigma_a^{-1} \Phi_a \Sigma_a^{-1} \varphi(x)}}-z_{k,+}^* \right)\le 0 \right\}\\ \nonumber
	&\le& \hbox{Pr}\left\{ \phi_{\eta,\{z_{k,+}^{**}\}_k}(\mathcal{M}_{n(t_{K})})\le 0 \right\},
\end{eqnarray}
and that
\begin{eqnarray}\label{thirdinequality}
	&&\hbox{Pr}\left\{ \max_{k\in \{1,\dots,K \}} \left( \sup_{x\in \mathbb{X}_0} \frac{\varphi^\top(x) B^*(t_k)}{\sqrt{\sum_{a\in \{0,1\}} \varphi^\top(x) \Sigma_a^{-1} \Phi_a \Sigma_a^{-1} \varphi(x)}}-z_{k,-}^* \right)\le 0\right\}\\\nonumber&=&\hbox{Pr}\left\{ \max_{k\in \{1,\dots,K \}} \left( \sup_{x\in \mathbb{X}_0} \frac{\varphi^\top(x) B^*(t_k)}{\sqrt{\sum_{a\in \{0,1\}} \varphi^\top(x) \Sigma_a^{-1} \Phi_a \Sigma_a^{-1} \varphi(x)}}-(z_{k,-}^*-3\delta) \right)\le 3\delta \right\}\\ \nonumber&\ge& \hbox{Pr}\left\{ \phi_{\eta,\{z_{k,-}^{**}\}_k}(\mathcal{M}_{n(t_{K})})\le 3\delta \right\},
\end{eqnarray}
where
\begin{eqnarray*}
	z_{k,+}^{**}=z_{k,+}^*+\eta^{-1} \{\log K+4d \log n(t_1) \}\,\,\hbox{and}\,\,z_{k,-}^{**}=z_{k,-}^*-3\delta.
\end{eqnarray*}
The value of $\delta$ will be specified later. In addition, with some calculations, we have
\begin{eqnarray*}
	\partial_j \phi_{\eta,\{\nu_k\}_k}(m)&=&\frac{\sum_{i=1}^{L} d_i^{(j)}\exp \left( \eta[d_i^\top m-\nu_{k(i)}] \right)}{\sum_{i=1}^{L}\exp \left( \eta[d_i^\top m-\nu_{k(i)}] \right)},\\
	\partial_{j_1}\partial_{j_2} \phi_{\eta,\{\nu_k\}_k}(m)&=&\eta\frac{\sum_{i=1}^{L} d_i^{(j_1)} d_i^{(j_2)}\exp \left( \eta[d_i^\top m-\nu_{k(i)}] \right)}{\sum_{i=1}^{L}\exp \left( \eta[d_i^\top m-\nu_{k(i)}] \right)}\\
	&-&\eta\frac{\prod_{l=1,2}\left\{\sum_{i=1}^{L} d_i^{(j_l)}\exp \left( \eta[d_i^\top m-\nu_{k(i)}] \right)\right\}}{\left\{\sum_{i=1}^{L}\exp \left( \eta[d_i^\top m-\nu_{k(i)}] \right)\right\}^2},
\end{eqnarray*}
\begin{eqnarray*}	
	\partial_{j_1}\partial_{j_2}\partial_{j_3} \phi_{\eta,\{\nu_k\}_k}(m)&=&\eta^2\frac{\sum_{i=1}^{L} d_i^{(j_1)} d_i^{(j_2)}d_i^{(j_3)}\exp \left( \eta[d_i^\top m-\nu_{k(i)}] \right)}{\sum_{i=1}^{L}\exp \left( \eta[d_i^\top m-\nu_{k(i)}] \right)}\\
	&-&3\eta^2\frac{\left\{\sum_{i=1}^{L} d_i^{(j_1)}d_i^{(j_2)}\exp \left( \eta[d_i^\top m-\nu_{k(i)}] \right)\right\}}{\left\{\sum_{i=1}^{L}\exp \left( \eta[d_i^\top m-\nu_{k(i)}] \right)\right\}}\\
	&\times& \frac{\left\{\sum_{i=1}^{L} d_i^{(j_3)}\exp \left( \eta[d_i^\top m-\nu_{k(i)}] \right)\right\}}{\left\{\sum_{i=1}^{L}\exp \left( \eta[d_i^\top m-\nu_{k(i)}] \right)\right\}} \\
	&+&2\eta^2\frac{\prod_{l=1,2,3}\left(\sum_{i=1}^{L} d_i^{(j_l)}\exp \left( \eta[d_i^\top m-\nu_{k(i)}] \right)\right)}{\left\{\sum_{i=1}^{L}\exp \left( \eta[d_i^\top m-\nu_{k(i)}] \right)\right\}^3}.
\end{eqnarray*}
Since $\max_i \|d_i\|_1\le \epsilon^{-1}$, we obtain that
\begin{eqnarray}\label{derivatives}
	\sum_j |\partial_j \phi_{\eta,\{\nu_k\}_k}(m)|\le \epsilon^{-1},\,\,\,\,
	\sum_{j_1,j_2} |\partial_{j_1}\partial_{j_2} \phi_{\eta,\{\nu_k\}_k}(m)|\le 2\eta \epsilon^{-2} ,\\\nonumber\sum_{j_1,j_2,j_3} |\partial_{j_1}\partial_{j_2}\partial_{j_3} \phi_{\eta,\{\nu_k\}_k}(m)|\le 6\eta^2 \epsilon^{-3}.
\end{eqnarray}

By Lemma 5.1 of \cite{Cher2016}, for any $\delta>0$, there exists some function $\hbox{g}_{\delta}(\cdot):\mathbb{R}\to \mathbb{R}$ with $\|\hbox{g}_{\delta}'\|_{\infty}\le \delta^{-1}$, $\|\hbox{g}_{\delta}^{''}\|_{\infty}\le K_0\delta^{-2}$, $\|\hbox{g}_{\delta}^{'''}\|_{\infty}\le K_0\delta^{-3}$ for some constant $K_0>0$ such that
\begin{eqnarray*}
	\mathbb{I}(z_0\le 0)\le \hbox{g}_{\delta}(z_0)\le \mathbb{I}(z_0\le 3\delta),\,\,\forall \delta \in \mathbb{R}. 
\end{eqnarray*} 
It follows that
\begin{eqnarray*}
	\mathbb{I}(\phi_{\eta,\{\nu_k\}_k}(m)\le 0)\le \hbox{g} \circ \phi_{\eta,\{\nu_k\}_k}(m)\le \mathbb{I}(\phi_{\eta,\{\nu_k\}_k}(m)\le 3\delta),
\end{eqnarray*}
for any $m\in \mathbb{R}^{qK}$. Combining this together with \eqref{firstequality}, \eqref{secondinequality} and \eqref{thirdinequality}, we obtain that
\begin{eqnarray}\label{fourthinequality}
	\hbox{Pr}\left\{ \max_{k\in \{1,\dots,K \}} \left( \sup_{x\in \mathbb{X}_0} \frac{\varphi^\top(x) B^*(t_k)}{\sqrt{\sum_{a\in \{0,1\}} \varphi^\top(x) \Sigma_a^{-1} \Phi_a \Sigma_a^{-1} \varphi(x)}}-z_{k,+}^* \right)\le 0 \right\}\le \Mean \hbox{g}_{\delta}\circ \phi_{\eta,\{z_{k,+}^{**}\}_k}(\mathcal{M}_{n(t_{K})}),\\\label{fifthinequality}
	\hbox{Pr}\left\{ \max_{k\in \{1,\dots,K \}} \left( \sup_{x\in \mathbb{X}_0} \frac{\varphi^\top(x) B^*(t_k)}{\sqrt{\sum_{a\in \{0,1\}} \varphi^\top(x) \Sigma_a^{-1} \Phi_a \Sigma_a^{-1} \varphi(x)}}-z_{k,-}^* \right)\le 0\right\}\ge \Mean \hbox{g}_{\delta}\circ \phi_{\eta,\{z_{k,-}^{**}\}_k}(\mathcal{M}_{n(t_{K})}).
\end{eqnarray}
Consider the function $\hbox{g}_{\delta}\circ \phi_{\eta,\{\nu_k\}_k}$. Apparently, we have
\begin{eqnarray}\label{c0varphi}
	\sup_{\delta,\eta, \{\nu_k\}_k} c_0(\hbox{g}_{\delta}\circ \phi_{\eta,\{\nu_k\}_k})\le 1.
\end{eqnarray}
By \eqref{derivatives}, we can show that
\begin{eqnarray}\label{c2c3}
	\begin{split}
		&&\sup_{\delta,\eta, \{\nu_k\}_k} c_2(\hbox{g}_{\delta}\circ \phi_{\eta,\{\nu_k\}_k } )\preceq  \delta^{-2}+\delta^{-1}\eta , \\ &&\sup_{\delta,\eta, \{\nu_k\}_k} c_3(\hbox{g}_{\delta}\circ \phi_{\eta,\{\nu_k\}_k } )\preceq  \delta^{-3}+\delta^{-2}\eta+\delta^{-1}\eta^2.
	\end{split}
\end{eqnarray}
Set $\delta=\eta^{-1}\{\log K+4d\log n(t_1)\}$, we obtain
\begin{eqnarray*}
	\sup_{\eta, \{\nu_k\}_k}c_i(\hbox{g}_{\delta}\circ \phi_{\eta,\{\nu_k\}_k } )\preceq \eta^{i} \{\log^i K+\log^i n(t_1)\},\,\,\,\,i=2,3.
\end{eqnarray*}
Combining \eqref{c2c3} together with \eqref{approximation} and \eqref{c0varphi} yields
\begin{eqnarray*}
	&&\sup_{\delta,\eta, \{\nu_k\}_k} |\Mean \hbox{g}_{\delta}\circ \phi_{\eta, \{\nu_k\}_k}(\mathcal{M}_{n(t_{K})})-\Mean \hbox{g}_{\delta}\circ \phi_{\eta, \{\nu_k\}_k}(N(0,\bm{V}))|\\ \nonumber
	&\preceq& n^{-1/2}(t_1) \eta^3 \{\log^3 K+\log^3 n(t_1)\}+ \eta^2 \{\log^2 K+\log^2 n(t_1)\} \delta_{n(t_1)}+n^{-\alpha_0}(t_1). 
\end{eqnarray*}
This together with \eqref{fourthinequality} and \eqref{fifthinequality} yields
\begin{eqnarray}\label{6inequality}
	\hbox{Pr}\left\{ \max_{k\in \{1,\dots,K \}} \left( \sup_{x\in \mathbb{X}_0} \frac{\varphi^\top(x) B^*(t_k)}{\sqrt{\sum_{a\in \{0,1\}} \varphi^\top(x) \Sigma_a^{-1} \Phi_a \Sigma_a^{-1} \varphi(x)}}-z_{k,+}^* \right)\le 0 \right\}-\Mean \hbox{g}_{\delta}\circ \phi_{\eta,\{z_{k,+}^{**}\}_k}(N(0,\bm{V}))\\ \nonumber
	\preceq n^{-1/2}(t_1) \eta^3 \{\log^3 K+\log^3 n(t_1)\}+ \eta^2 \{\log^2 K+\log^2 n(t_1)\} \delta_{n(t_1)}+n^{-\alpha_0}(t_1),\\ \label{7inequality}
	\Mean \hbox{g}_{\delta}\circ \phi_{\eta,\{z_{k,-}^{**}\}_k}(N(0,\bm{V}))-\hbox{Pr}\left\{ \max_{k\in \{1,\dots,K \}} \left( \sup_{x\in \mathbb{X}_0} \frac{\varphi^\top(x) B^*(t_k)}{\sqrt{\sum_{a\in \{0,1\}} \varphi^\top(x) \Sigma_a^{-1} \Phi_a \Sigma_a^{-1} \varphi(x)}}-z_{k,-}^* \right)\le 0\right\}\\ \nonumber
	\preceq n^{-1/2}(t_1) \eta^3 \{\log^3 K+\log^3 n(t_1)\}+ \eta^2 \{\log^2 K+\log^2 n(t_1)\} \delta_{n(t_1)}+n^{-\alpha_0}(t_1).
\end{eqnarray}
Similar to \eqref{secondinequality}-\eqref{fifthinequality}, we can show
\begin{eqnarray*}
	&&\Mean \hbox{g}_{\delta}\circ \phi_{\eta, \{z_{k,+}^{**}\}_k}(N(0,\bm{V}))\le \hbox{Pr}\left( \phi_{\eta, \{z_{k,+}^{**}\}_k}(N(0,\bm{V}))\le 3\delta \right)\\
	&\le& \hbox{Pr}\left( \max_{1\le j\le L} \{d_j^\top N(0,\bm{V})-z_{k(j),+}^{**} \}\le 3\delta \right)=\hbox{Pr}\left( \max_{1\le j\le L} \{d_j^\top N(0,\bm{V})-z_{k(j),+}^{***}\} \le 0 \right),\\
	&&\Mean \hbox{g}_{\delta}\circ \phi_{\eta, \{z_{k,-}^{**}\}_k}(N(0,\bm{V}))\ge \hbox{Pr}\left( \phi_{\eta, \{z_{k,-}^{**}\}_k}(N(0,\bm{V}))\le 0 \right)\\
	&\ge& \hbox{Pr}\left( \max_{1\le j\le L} \{d_j^\top N(0,\bm{V}) -z_{k(j),-}^{***}\} \le 0 \right),
\end{eqnarray*}
where
\begin{eqnarray*}
	z_{k,+}^{***}=z_{k,+}^*+\eta^{-1} \{\log K+4d \log n(t_1) \}+3\delta\,\,\hbox{and}\,\,z_{k,-}^{***}=z_{k,-}^*-\eta^{-1} \{\log K+4d \log n(t_1) \}-3\delta,
\end{eqnarray*}
for each $k$. Let $\sigma(x)=\sum_{a\in \{0,1\}} \varphi^\top(x) \Sigma_a^{-1} \Phi_a \Sigma_a^{-1} \varphi(x)$ and $\widehat{\sigma}(x,t)=\sum_{a\in \{0,1\}} \varphi^\top(x) \widehat{\Sigma}_a^{-1}(t) \widehat{\Phi}_a(t) \widehat{\Sigma}_a^{-1}(t) \varphi(x)$. 
Notice that for any $\{\nu_k\}_k$, we have
\begin{eqnarray*}
	\hbox{Pr}\left\{\max_{k\in\{1,\dots,K\}} \left(\sup_{x\in \mathbb{X}_0} \sigma^{-1}(x)\varphi^\top(x) G(t_k) -\nu_{k}\right)\le 0 \right\}=\hbox{Pr}\left( \max_{1\le j\le L} \{d_j^\top N(0,\bm{V})-\nu_{k(j)}\}\le 0\right).
\end{eqnarray*}
This together with \eqref{6inequality} and \eqref{7inequality} yields
\begin{eqnarray*}
	\hbox{Pr}\left\{ \max_{k\in \{1,\dots,K \}} \left( \sup_{x\in \mathbb{X}_0} \frac{\varphi^\top(x) B^*(t_k)}{\sigma(x)}-z_{k,+}^* \right)\le 0 \right\}-\hbox{Pr}\left\{\max_{k\in\{1,\dots,K\}} \left(\sup_{x\in \mathbb{X}_0}  \frac{\varphi^\top(x) G(t_k)}{\sigma(x)} -z_{k,+}^{***}\right)\le 0 \right\}\\ \nonumber
	\preceq n^{-1/2}(t_1) \eta^3 \{\log^3 K+\log^3 n(t_1)\}+ \eta^2 \{\log^2 K+\log^2 n(t_1)\} \delta_{n(t_1)}+n^{-\alpha_0}(t_1),\\
	\hbox{Pr}\left\{\max_{k\in\{1,\dots,K\}} \left(\sup_{x\in \mathbb{X}_0} \frac{\varphi^\top(x) G(t_k)}{\sigma(x)} -z_{k,-}^{***}\right)\le 0 \right\}-\hbox{Pr}\left\{ \max_{k\in \{1,\dots,K \}} \left( \sup_{x\in \mathbb{X}_0} \frac{\varphi^\top(x) B^*(t_k)}{\sigma(x)}-z_{k,-}^* \right)\le 0\right\}\\ \nonumber
	\preceq n^{-1/2}(t_1) \eta^3 \{\log^3 K+\log^3 n(t_1)\}+ \eta^2 \{\log^2 K+\log^2 n(t_1)\} \delta_{n(t_1)}+n^{-\alpha_0}(t_1).
\end{eqnarray*}
In view of \eqref{firsthalfinequality}, we have shown that
\begin{eqnarray*}
	\hbox{Pr}\left\{ \max_{k\in \{1,\dots,K \}} \left( \sup_{x\in \mathbb{X}} \frac{\varphi^\top(x) B(t_k)}{\sigma(x)}-z_{k} \right)\le 0 \right\}-\hbox{Pr}\left\{\max_{k\in\{1,\dots,K\}} \left(\sup_{x\in \mathbb{X}_0} \frac{\varphi^\top(x) G(t_k)}{\sigma(x)} -z_{k,+}^{***}\right)\le 0 \right\}\\ \nonumber
	\preceq n^{-1/2}(t_1) \eta^3 \{\log^3 K+\log^3 n(t_1)\}+ \eta^2 \{\log^2 K+\log^2 n(t_1)\} \delta_{n(t_1)}+n^{-\alpha_0}(t_1),\\
	\hbox{Pr}\left\{\max_{k\in\{1,\dots,K\}} \left(\sup_{x\in \mathbb{X}_0} \frac{\varphi^\top(x) G(t_k)}{\sigma(x)} -z_{k,-}^{***}\right)\le 0 \right\}-\hbox{Pr}\left\{ \max_{k\in \{1,\dots,K \}} \left( \sup_{x\in \mathbb{X}} \frac{\varphi^\top(x) B(t_k)}{\sigma(x)}-z_{k} \right)\le 0\right\}\\ \nonumber
	\preceq n^{-1/2}(t_1) \eta^3 \{\log^3 K+\log^3 n(t_1)\}+ \eta^2 \{\log^2 K+\log^2 n(t_1)\} \delta_{n(t_1)}+n^{-\alpha_0}(t_1).
\end{eqnarray*}
%where $\mathbb{X}_0$ is defined in \eqref{equality0}. 
%The covariance matrix $\Cov(G(t_k))$ is given by $\sum_{a\in\{0,1\}} \Sigma_a^{-1} \Phi_a \Sigma_a^{-1}$ and is nonsingular by Lemma \ref{somebasiclemma1}. In addition, we have $\|\varphi(x)\|_2\ge \bar{c},\forall x\in \mathbb{X}_0$, by Condition A3. Thus, there exists some constant $c_*>0$ such that
%\begin{eqnarray*}
%	c_*\le \varphi^\top(x)\left( \sum_{a\in\{0,1\}} \Sigma_a^{-1} \Phi_a \Sigma_a^{-1} \right)^{1/2} \varphi(x),\,\,\,\,\forall x\in \mathbb{X}_0.
%\end{eqnarray*}
By Theorem 1 of \cite{cherno2017}, we obtain that
\begin{eqnarray*}
	&&\hbox{Pr}\left\{\max_{k\in\{1,\dots,K\}} \left(\sup_{x\in \mathbb{X}_0} \frac{\varphi^\top(x) G(t_k)}{\sigma(x)} -z_{k,+}^{***}\right)\le 0 \right\}-\hbox{Pr}\left\{\max_{k\in\{1,\dots,K\}} \left(\sup_{x\in \mathbb{X}_0} \frac{\varphi^\top(x) G(t_k)}{\sigma(x)} -z_{k,-}^{***}\right)\le 0 \right\}\\
	&\preceq & \eta^{-1} \{\log K+\log n(t_1)\}^{3/2}+q^{3/2} \delta_{n(t_1)} \{\log K+\log n(t_1)\}+ q\sqrt{ n^{-1}(t_1)} \{\log K+\log n(t_1)\}^{3/2}\\
	&+& \sqrt{n(t_K)}\textrm{err}\{\log K+\log n(t_1)\}^{1/2}.
\end{eqnarray*}
%Notice that $\{\log K+\log n(t_1)\}^{1/2}\le \log^{1/2} K+\log^{1/2} \log n(t_1)$. 
%By H{\"older}'s inequality, we have
%\begin{eqnarray*}
%	\{n^{-b}(t_1) \log^b n(t_1)\} \sqrt{\log K}\preceq n^{-b}(t_1) \{\log^{b+1/2} n(t_1)+ \log^{b+1/2} K\},
%\end{eqnarray*}
%for any $b>0$. It follows that
%\begin{eqnarray*}
%	\hbox{Pr}\left\{\max_{k\in\{1,\dots,K\}} \left(\sup_{x\in \mathbb{X}_0} \varphi^\top(x) G(t_k) -z_{k,+}^{***}\right)\le 0 \right\}-\hbox{Pr}\left\{\max_{k\in\{1,\dots,K\}} \left(\sup_{x\in \mathbb{X}_0} \varphi^\top(x) G(t_k) -z_{k,-}^{***}\right)\le 0 \right\}\\
%	\preceq \eta^{-1} \{\log^{3/2} K+\log^{3/2} n(t_1)\}+q^2\delta_{n(t_1)} n^{-1/2}(t_1) \{\log K+\log n(t_1)\}
%	+ q^{3/2} n^{-1}(t_1) \{\log^{3/2} K+\log^{3/2} n(t_1)\}.
%\end{eqnarray*}
Thus, we obtain
\begin{eqnarray*}
	&&\left|\hbox{Pr}\left\{ \max_{k\in \{1,\dots,K \}} \left( \sup_{x\in \mathbb{X}} \varphi^\top(x) B(t_k)-z_{k} \right)\le 0 \right\}-\hbox{Pr}\left\{ \max_{k\in \{1,\dots,K \}} \left( \sup_{x\in \mathbb{X}} \varphi^\top(x) G(t_k)-z_{k} \right)\le 0 \right\}\right|\\
	&\preceq& n^{-1/2}(t_1) \eta^3 \{\log^3 K+\log^3 n(t_1)\}+ \eta^2 \{\log^2 K+\log^2 n(t_1)\} \delta_{n(t_1)}+n^{-\alpha_0}(t_1)\\
	&+&\eta^{-1} \{\log K+\log n(t_1)\}^{3/2}+q^{3/2} \delta_{n(t_1)} \{\log K+\log n(t_1)\}
	+ q\sqrt{ n^{-1}(t_1)} \log n(t_1) \{\log K+\log n(t_1)\}^{3/2}\\
	&+& \sqrt{n(t_K)}\textrm{err}\{\log K+\log n(t_1)\}^{1/2}.
	%\\\preceq n^{-1/2}(t_1) q^{3} \eta^3 \{\log^3 K+\log^3 n(t_1)\}+ q^2\eta^2 \{\log^2 K+\log^2 n(t_1)\} \{ n^{-\alpha_0}(t_1)+\delta_{n(t_1)}\}\\+\eta^{-1} \{\log K+\log n(t_1)\}.
\end{eqnarray*}
Setting $\eta=\min(n^{1/8}(t_1)\log^{-3/8} \{K n(t_1)\}, n^{-\alpha_0/3}(t_1) \log^{-\alpha_0/3-1/6}\{ K n(t_1) \} )$ yields the desired results. The proof is hence completed.  

\section{Discussion}%\vspace{-0.2cm}
{\color{black}
	\subsection{Growing number of basis functions}\label{sec:grow}
	In the current proposal, we use the same set of basis functions at each interim stage. It would be more desirable to allow the number of basis functions $q$ to increase with $k$ to deal with the model approximation error. 
	
	In this section, we extend our proposal to allow $q$ to vary with $k$. We focus on the case where the maximum number of interim stages $K$ is known and the basis functions to be used at each interim stage are predetermined. Let $\{\varphi_k\}_k$ denote the sets of basis functions and $q_k$ the dimension of $\varphi_k$ for any $k$. Let $\widehat{\Sigma}_a^{(k)}(t)=N^{-1}(t)\sum_{i=1}^{N(t)}\mathbb{I}(A_i=a)\varphi_k(X_i)\varphi_k^\top(X_i)$. 
	To implement the resulting test, we need to compute the regression coefficients 
	\begin{eqnarray*}
		\widehat{\beta}_a^{(k)}(t_k)=\{\widehat{\Sigma}_a^{(k)}(t_k)\}^{-1} \left\{ \frac{1}{N(t_k)}\sum_{i=1}^{N(t_k)} \mathbb{I}(A_i=a) \varphi_k(X_i) Y_i \right\},
	\end{eqnarray*} 
	at the $k$th interim stage. To allow for online updating, at the $k$th interim stage, we not only compute $\widehat{\Sigma}_a^{(k)}(t_k)$ and $N^{-1}(t_k)\sum_{i=1}^{N(t_k)} \mathbb{I}(A_i=a) \varphi_k(X_i) Y_i$, but $\{\widehat{\Sigma}_a^{(\kappa)}(t_k)\}_{k< \kappa \le K}$ and $\{\gamma_{a,\kappa}(t_k)\}_{k<\kappa \le K}$ as well where $\gamma_{a,\kappa}(t_k)={N(t_k)}^{-1}\sum_{i=1}^{N(t_k)} \mathbb{I}(A_i=a) \varphi_{\kappa}(X_i) Y_i$. These quantities allow us to compute $\widehat{\beta}_a^{(\kappa)}(t_{\kappa})$ for $\kappa>k$, without storing historical data. As such, the proposed test statistic at the $k$th interim stage is given by
	\begin{eqnarray*}
		\sup_x \frac{\varphi_k^\top(x) \{\widehat{\beta}_1^{(k)}(t_k)-\widehat{\beta}_0^{(k)}(t_k)\}}{\widehat{s.e}[\varphi_k^\top(x) \{\widehat{\beta}_1^{(k)}(t_k)-\widehat{\beta}_0^{(k)}(t_k)\}]},
	\end{eqnarray*}
	where $\widehat{s.e}[\cdot]$ denotes some consistent standard error estimator. 
	
	In addition, we extend the proposed bootstrap algorithm to determine the stopping boundary. We aim to construct bootstrap statistics $\{\widehat{\beta}_a^{(k),\tiny{\textrm{MB}^*}}(t_k)\}_k$ to approximate the distribution of $\{\widehat{\beta}_a^{(k)}(t_k)\}_k$. To allow for online updating, at the $k$th interim stage, we compute 
	$\{\widehat{\beta}_a^{(k),\tiny{\textrm{MB}^*}}(t_k), \widehat{\beta}_a^{(k+1),\tiny{\textrm{MB}^*}}(t_k), \cdots, \widehat{\beta}_a^{(K),\tiny{\textrm{MB}^*}}(t_k)\}^\top$ as
	\begin{eqnarray}\label{eqn:boot}
		\frac{1}{N(t_k)}\sum_{j=1}^{k} \left(\sum_{i=N(t_{j-1})+1}^{N(t_j)}  \mathbb{I}(A_i=a) \phi_{[k:K]}(X_i,Y_i) \phi_{[k:K]}^\top(X_i,Y_i)   \right)^{1/2} e_{j,a},
	\end{eqnarray}
	where $\phi_{[k:K]}(X_i,Y_i)$ denotes the vector
	\begin{eqnarray*}
		\left\{ [\{\widehat{\Sigma}_a^{(k)}(t_j)\}^{-1} \varphi_k(X_i)\{Y_i-\varphi_k(X_i)\}]^{\top}, \cdots,[\{\widehat{\Sigma}_a^{(K)}(t_j)\}^{-1} \varphi_K(X_i)\{Y_i-\varphi_K(X_i)\}]^{\top} \right\}^\top,
	\end{eqnarray*}
	and $e_{j,a}$ denotes a multivariate normal random vector with zero mean and identity covariance matrix. This yields the bootstrap test statistic
	\begin{eqnarray*}
		\sup_x \frac{\varphi_k^\top(x) \{\widehat{\beta}_1^{(k), \tiny{\textrm{MB}^*}}(t_k)-\widehat{\beta}_0^{(k),\tiny{\textrm{MB}^*}}(t_k)\}}{\widehat{s.e}[\varphi_k^\top(x) \{\widehat{\beta}_1^{(k)}(t_k)-\widehat{\beta}_0^{(k)}(t_k)\}]},
	\end{eqnarray*} 
	based on which the $\alpha$-spending approach is applicable (see Equation \eqref{estimatingequation} for details). The resulting test is valid, under certain regularity conditions on $\{\varphi_k\}_k$. See Appendix \ref{secAE} for details. 
	
	Finally, we discuss some drawbacks of the resulting test. Compared to our proposed test in the main text, it would be much more computationally intensive to implement such a test. For instance, in order to compute the test statistic, at the $k$th interim stage, we need to compute not only $\gamma_{a,k}(t_k)$ and $\widehat{\Sigma}_a^{(k)}(t_k)$, but $\{\gamma_{a,\kappa}(t_k)\}_{\kappa>k}$ and $\{\widehat{\Sigma}_a^{(\kappa)}(t_k)\}_{\kappa>k}$ as well. More important, in order to compute the bootstrap statistic, we need to do a Cholesky decomposition on a $(\sum_{j=k}^K q_k)\times (\sum_{j=k}^K q_k)$ matrix at the $k$th interim stage (see Equation \eqref{eqn:boot}). In cases where $K-k$ is large, this is much more computationally intensive than the proposed procedure that only requires to do a Cholesky decomposition on a $q\times q$ matrix.

	\subsection{Extensions to the two-sided test}
	In this paper, we focus on the null hypothesis $H_0^a$ that the heterogeneous treatment effect (HTE) is nonpositive for any realization of the baseline covariates. It is also interesting to consider the two-sided null hypothesis that HTE is either always nonpositive ($H_0^a$), or always nonnegative (denoted by $H_0^b$). Notice that the latter corresponds to a union of $H_0^a$ and $H_0^b$. To test such a null, one can separately test $H_0^a$ and $H_0^b$ using the proposed test, obtain the corresponding p-values $p_a$ and $p_b$, and derive the p-value using the union-intersection principle, i.e., $p=\max(p_a,p_b)$. }

\section*{Acknowledgement}
The authors thank the AE, and two anonymous reviewers for their constructive comments, which have led to a significant improvement of the proposed methodology. 

\clearpage

\appendix
\section{More on the basis function}\label{secAE}
%\subsection{Condition (A3)}
\noindent (A3)(i) Assume $\lambda_{\min}[\Mean \varphi(X)\varphi^\top(X)]\asymp 1$,  $\lambda_{\max}[\Mean \varphi(X)\varphi^\top(X)]\asymp 1$, $\sup_{x} \|\varphi(x)\|_1=O(q^{1/2})$, $\liminf_q \inf_{x\in \mathbb{X}} \|\varphi(x)\|_2>0$. In addition, assume
\begin{eqnarray}\label{eqn:Lip}
	\sup_{\substack{x,y\in \mathbb{X} \\ x\neq y }} \frac{\|\varphi(x)-\varphi(y)\|_2}{\|x-y\|_2} \preceq q^{1/2}.
\end{eqnarray}
\noindent (ii) Suppose
\begin{eqnarray}\label{minimaxAE}
	\textrm{err}\equiv \inf_{\beta_0,\beta_1\in \mathbb{R}^q}\sup_{x\in \mathbb{X}, a\in \{0,1\}} |Q_0(x,a)-Q(x,a;\beta_0,\beta_1)|=o(\{N(T)\}^{-1/2}).
\end{eqnarray}
%with probability tending to $1$. 
%\noindent (A4). Assume $\mathbb{X}=[0,1]^d$ where $d$ denotes to the dimension of the covariates. 

When a tensor-product B-spline is used \citep[see Section 6 of][for a brief overview of tenor-product B-splines]{Chen2015}, (A3)(i) is automatically satisfied. 
Specifically, $\lambda_{\min}[\Mean \varphi(X)\varphi^\top(X)]\asymp 1$,  $\lambda_{\max}[\Mean \varphi(X)\varphi^\top(X)]\asymp 1$ follow from Theorem 3.3 of \citep{Burman1989}. $\sup_{x} \|\varphi(x)\|_1=O(q^{1/2})$ follows by noting that the absolute value of each element in $\varphi(x)$ is bounded by $O(q^{1/2})$ and that the number of nonzero elements in $\varphi(x)$ is finite. $\liminf_q \inf_{x\in \mathbb{X}} \|\varphi(x)\|_2>0$ follows from the arguments used in the proof of Lemma E.4 of \cite{Shi2020}. The last condition in \eqref{eqn:Lip} holds by noting that each function in the vector $\varphi(\cdot)$ is Lipschitz continuous when a tensor-product B-spline is used. 
%\citep[see for example,][]{Chen2015}. 

%\subsection{On the approximation error}
%The proposed test remains valid as long as the approximation error satisfies

%with probability tending to $1$. 
%In the following, we introduce some sufficient conditions for \eqref{minimaxAE}.

Suppose the Q-function $Q_0(\cdot,a)$ is $p$-smooth \citep[see the definition of $p$-smoothness in][]{stone1982}, for $a\in [0,1]$. When a tensor-product B-spline or Wavelet basis is used for $\varphi(\cdot)$, then there exist some $\beta_0^*$ and $\beta_1^*$ that satisfy
\begin{eqnarray*}
	\inf_{\beta_0,\beta_1\in \mathbb{R}^q}\sup_{x\in \mathbb{X}, a\in \{0,1\}} |Q_0(x,a)-Q(x,a;\beta_0,\beta_1)|=O(q^{-p/d}).
\end{eqnarray*}
See Section 2.2 of \cite{Huang1998} for detailed discussions on the approximation power of these basis functions. Condition \eqref{minimaxAE} is thus automatically satisfied when
\begin{eqnarray*}
	q\gg \{N(T)\}^{d/(2p)}.
\end{eqnarray*}

{\color{black}As we have commented, the normalized test requires a weaker condition than the unnormalized test without standardization. Specifically, if we use the unnormalized test, we would require the approximation error to decay at a rate at $o\{q^{-1/2}N^{-1/2}(t)\}$, strictly faster than the RHS of \eqref{minimaxAE} when $q$ grows to infinity. To elaborate, the denominator in the normalized test is of the same order of magnitude as $N^{-1/2}(t) \|\phi(x)\|_2$, uniformly in $x$. This ensures the bias of the ratio $\varphi(x)^T (\widehat{\beta}_1-\widehat{\beta}_0)/\widehat{s.e.}[\varphi(x)^T (\widehat{\beta}_1-\widehat{\beta}_0)]$ is of the same order of magnitude as the bias of $\widehat{\beta}_1-\widehat{\beta}_0$, uniformly in $x$, eliminate the effect of the approximation error. Without standardization, the bias of the test would be of the same order of magnitude as the bias of $q^{1/2} (\widehat{\beta}_1-\widehat{\beta}_0)$. 
}
%Alternatively, one could impose the following additional assumption
%\begin{eqnarray}\label{eqn:pa}
%	\|P_a\|_{\infty}=\sup_{h:h\neq 0} \frac{\|P_a h\|_{\infty}}{\|h\|_{\infty}}=O(1),
%\end{eqnarray}
%where $P_a h$ denotes the projection of $h$ onto the sieve space, i.e., 
%\begin{eqnarray*}
%	P_a h (x)=\varphi^\top(x) \left\{\Mean \mathbb{I}(A=a) \varphi(X)\varphi^\top(X) \right\}^{-1} \left\{\Mean \mathbb{I}(A=a)\varphi(X) h(X) \right\},
%\end{eqnarray*}
%for any function $h$. We remark that \eqref{eqn:pa} holds when B-spline or wavelet basis functions are used \citep{Chen2015}. In cases where \eqref{eqn:pa} is satisfied, the assumption on the approximation error in (A3)(ii) could be relaxed to the following:
%\begin{eqnarray*}
%	\textrm{err}=o(\{N(T)\}^{-1/2}). 
%\end{eqnarray*}

Finally, we present the regularity conditions on $\{\varphi_k\}_k$ to ensure the validity of the proposed test in Section \ref{sec:grow}. They are very similar to those imposed in (A3).

\smallskip

\noindent (A3*)(i) For any $k$, assume $\lambda_{\min}[\Mean \varphi_k(X)\varphi_k^\top(X)]\asymp 1$,  $\lambda_{\max}[\Mean \varphi_k(X)\varphi_k^\top(X)]\asymp 1$, $\sup_{x} \|\varphi_k(x)\|_1=O(q_k^{1/2})$, $\liminf_{k} \inf_{x\in \mathbb{X}} \|\varphi_k(x)\|_2>0$. In addition, assume
\begin{eqnarray*}
	\sup_{\substack{x,y\in \mathbb{X} \\ x\neq y }} \frac{\|\varphi_k(x)-\varphi_k(y)\|_2}{\|x-y\|_2} \preceq q_k^{1/2}.
\end{eqnarray*}
\noindent (ii) Suppose
\begin{eqnarray*}
	\inf_{\beta \in \mathbb{R}^{q_k}}\sup_{x\in \mathbb{X}, a\in \{0,1\}} |Q_0(x,a)-\varphi_k^\top(x) \beta |=o(\{N(t_k)\}^{-1/2}).
\end{eqnarray*}
%with probability tending to $1$. 
%\section{Evaluation of ATE in the real data}\label{secATE}
%Furthermore, we also check whether on average 1) the key metric, answer time, is better, and 2) the other three metrics are not worse. In order to answer this question, we also make an one-time analysis on average treatment effects of the four metrics at the end of the first week. That is, we consider the following test for each metric (multiple testing issue is ignored here).
%\begin{equation*}%\label{ATE}
%H_0: \tau_0 = 0\,\,\,\,\hbox{vs}\,\,\,\, H_1: \tau_0 > 0,
%\end{equation*}
%where $\tau_0 = \Mean \{Y^{*}(1) - Y^{*}(0)\}$. We apply the Horvitz-Thompson estimator \cite{horvitz1952generalization} to estimating $\tau_0$,
%\begin{equation*}
%\widehat{\tau}_0=n^{-1}\sum_{i=1}^n \frac{Y_i A_i}{\widehat\pi(X_i)}-
%n^{-1}\sum_{i=1}^n \frac{Y_i (1-A_i)}{1-\widehat\pi(X_i)},
%\end{equation*}
%where the propensity score $\pi(X_i) = \prob(A_i=1|X_i)$ is estimated via a logistic regression model, i.e. 
%$\pi_{\theta}(X_i) = \exp(\theta_0 + X_i^\top\theta_1)/\{1+\exp(\theta_0 + X_i^\top\theta_1)\}$, and $\widehat{\pi}(X_i)$ 
%can then be obtained by plugging in the maximum likelihood estimator $\widehat{\theta}$. The standard error of $\widehat{\tau}_0$ is obtained via 400 bootstrap estimates. The estimated ATEs (standard errors in brackets) for the four metrics are 0.10(0.03), 0.00(0.02), 0.04(0.03) and 0.03(0.02), with corresponding p values 0.00, 0.89, 0.15, 0.14.

%\section{Testing the average treatment effects}
%\subsection{The algorithm}

%\section{Additional Theoretical Results}

\section{Proofs}

%\subsection{Lemma \ref{lemmakeyequation} and its proof}

\subsection{Proof of Lemma \ref{lemmakeyequation}}
Set $\mathcal{F}_{0}=\emptyset$. We state the following lemma before proving Lemma \ref{lemmakeyequation}.
\begin{lemma}\label{lemmaxyindpast}
	For any $j\ge 1$, $(X_j,Y_j^*(0),Y_j^*(1)) \independent \mathcal{F}_{j-1}$. 
\end{lemma}
For any $a\in \{0,1\}$, $i\ge 1$, notice that
\begin{eqnarray*}
	\Mean \mathbb{I}(A_i=a) \{Y_i-Q_0(X_i,a)\}=\Mean \mathbb{I}(A_i=a) \{Y_i^*(a)-Q_0(X_i,a) \}\\=\Mean \Mean^{X_i,\mathcal{F}_{i-1}}[\mathbb{I}(A_i=a) \{Y_i^*(a)-Q_0(X_i,a) \}],
\end{eqnarray*}
where the first equation is due to Assumption (A1) and $\Mean^{X_i,\mathcal{F}_{i-1}}$ denotes the conditional expectation given $\mathcal{F}_{i-1}$ and $X_i$.  By Assumption (A2), we have
\begin{eqnarray*}
	\Mean^{X_i,\mathcal{F}_{i-1}}[\mathbb{I}(A_i=a) \{Y_i^*(a)-Q_0(X_i,a) \}]=\{\Mean^{X_i,\mathcal{F}_{i-1}} \mathbb{I}(A_i=a)\} [\Mean^{X_i,\mathcal{F}_{i-1}} \{Y_i^*(a)-Q_0(X_i,a) \}].
\end{eqnarray*}
The second term on the RHS equals zero due to Lemma \ref{lemmaxyindpast} and our model assumption $\Mean \{Y_i^*(a)|X_i\}=Q_0(X_i,a) $. The proof is hence completed. 

%\subsection{Proof of Theorem \ref{thm1}}

\subsection{Proof of Lemma \ref{lemmaxyindpast}}
The assertion trivially holds for $j=1$. We prove it holds for any $j\ge 2$, by induction. By (A2), we have $(X_j,Y_j^*(0),Y_j^*(1)) \independent A_1|X_1$. Since $(X_j,Y_j^*(0),Y_j^*(1))\independent (X_1,Y_1^*(0),Y_1^*(1))$, this further implies $(X_j,Y_j^*(0),Y_j^*(1)) \independent A_1$ and hence $(X_j,Y_j^*(0),Y_j^*(1))\independent (X_1,A_1,Y_1^*(0),Y_1^*(1))$. By (A1), $Y_1$ is completely determined by $A_1$, $Y_1^*(0)$ and $Y_1^*(1)$. Therefore, we obtain $(X_j,Y_j^*(0),Y_j^*(1))\independent \mathcal{F}_1$. 

Suppose we have shown that $(X_j,Y_j^*(0),Y_j^*(1))\independent \mathcal{F}_k$ for some $k<j-1$. To prove $(X_j,Y_j^*(0),Y_j^*(1))\independent \mathcal{F}_{k+1}$, it suffices to show $(X_j,Y_j^*(0),Y_j^*(1))\independent (X_{k+1},A_{k+1},Y_{k+1})$. By (A1), $Y_{k+1}$ is determined by $A_{k+1}$, $Y_{k+1}^*(0)$ and $Y_{k+1}^*(1)$. Since $(X_j,Y_j^*(0),Y_j^*(1))\independent (X_{k+1},Y_{k+1}^*(0),Y_{k+1}^*(1))$, it suffices to show $(X_j,Y_j^*(0),Y_j^*(1))\independent A_{k+1}$. This is implied by $(X_j,Y_j^*(0),Y_j^*(1))\independent A_{k+1}| X_{k+1},\mathcal{F}_{k}$ and that $(X_j,Y_j^*(0),Y_j^*(1))\independent X_{k+1},\mathcal{F}_k$. The proof is hence completed.  

%\subsection{Proof of Lemma \ref{somebasiclemma0}}
%Notice that $\|\hbox{Mat}\|_2\le \sqrt{\|\hbox{Mat}\|_1\|\hbox{Mat}\|_{\infty}}$. Since $\hbox{Mat}$ is a square matrix, we have $\|\hbox{Mat}\|_1=\|\hbox{Mat}\|_{\infty}$. The assertion thus follows. 

\subsection{Proof of Lemma \ref{somebasiclemma1}}
The assertions 
\begin{eqnarray}\label{mineigensigma}
	\epsilon_0\le \lambda_{\min}[\Mean \varphi(X) \varphi^\top(X)]\le \lambda_{\max}[\Mean \varphi(X) \varphi^\top(X)] \le \epsilon_0^{-1},
\end{eqnarray}
and
\begin{eqnarray}\label{supbasis}
	\sup_{x} \|\varphi(x)\|_1\le \epsilon_0^{-1}\sqrt{q},
\end{eqnarray}
for some $0<\epsilon_0<1$ are directly implied by the conditions that $\lambda_{\min}[\Mean \varphi(X) \varphi^\top(X)]\asymp 1$, $\lambda_{\max}[\Mean \varphi(X) \varphi^\top(X)]\asymp 1$, $\sup_{x} \|\varphi(x)\|_1\le \epsilon_0^{-1}\sqrt{q}$. Since $\|\varphi(x)\|_2\le \|\varphi(x)\|_1$, we obtain 
$\sup_{x} \|\varphi(x)\|_2\le\sup_{x} \|\varphi(x)\|_1\le \epsilon_0^{-1}\sqrt{q}$.

Under the condition $\inf_{a,x} \pi^*(a,x)>0$, we can similarly show that 
$\lambda_{\min}[\Sigma_a]\ge \epsilon_0$ for some $\epsilon_0>0$. 
%Suppose $\lambda_{\min}[\Mean \varphi(X) \varphi^\top(X))=0$. Then there exists some vector $\nu \in \mathbb{R}^{p+1}$ such that $\Mean (\nu^\top \varphi(X))^2=0$. Therefore, we have $\Var(\nu^\top \varphi(X))=0$ and hence $\Var(X^\top \nu^{(-1)})=0$, where $\nu^{(-1)}$ denotes the sub-vector of $\nu$ formed by its last $p$ elements. However, this violates the assumption that the covariance matrix of $X$ is non-degenerate. 

%Since $|Y^*(0)|$ and $|Y^*(1)|$ are bounded, there exists some constant $0<\epsilon_0<1$ that satisfies $\max_{a\in \{0,1\}}|Y^*(a)|\le \epsilon_0^{-1}$. 
Notice that $Q_0(x,a)=\Mean \{Y^*(a)|X=x\}$. Under the condition that $\Mean [\{Y^*(a)\}^2|X]$ is bounded, we obtain $\sup_{x\in \mathbb{X}}\max_{a\in \{0,1\}}|Q_0(x,a)|\le \epsilon_0^{-1}$ for some $0<\epsilon<1$.  

Notice that $\beta_a=\Sigma_a^{-1}\Mean \varphi^\top(X) Y^*(a)$. Since $\lambda_{\min}[\Sigma_a]$ is bounded away from $0$, it suffices to show $\|\Mean \varphi^\top(X) Y^*(a)\|_2=O(1)$, or equivalently,
\begin{eqnarray*}
	\sup_{\nu \in \mathbb{R}^p, \|\nu \|_2=1} |\Mean \nu^\top \varphi(X) Y^*(a)|=O(1).
\end{eqnarray*}
By Cauchy-Schwarz inequality, it suffices to show
\begin{eqnarray*}
	\sup_{\nu \in \mathbb{R}^p, \|\nu \|_2=1} \Mean |Y^*(a)|^2 \Mean |\nu^\top \varphi(X)|^2=O(1).
\end{eqnarray*}
%Since $|Y^*(a)|=O(1)$ almost surely, 
We have by the condition $\lambda_{\max}[\Mean \varphi(X) \varphi^\top(X)]=O(1)$ that
\begin{eqnarray*}
	\sup_{\nu \in \mathbb{R}^p, \|\nu \|_2=1}\Mean |\nu^\top \varphi(X)|^2=\sup_{\nu \in \mathbb{R}^p, \|\nu \|_2=1} \nu^\top \Mean \varphi(X) \varphi^\top(X) \nu\le \lambda_{\max}[ \Mean \varphi(X) \varphi^\top(X)]=O(1). 
\end{eqnarray*} 
The sub-Gaussianity of $Y^*(a)$ implies that it has bounded second moment. 
The proof is hence completed.

\subsection{Proof of Lemma \ref{somebasiclemma2}}
%We will prove the results regarding $\widehat{\Sigma}_{1,j}$ and $\widehat{\beta}_{1,j}$ only. Results regarding $\widehat{\Sigma}_{0,j}$ and $\widehat{\beta}_{0,j}$ can be similarly discussed. 
\subsubsection{Proof of \eqref{Sigmabasic}}
Notice that
\begin{eqnarray}\label{somebasiceq0}
	\|j(\widehat{\Sigma}_{1,j}-\Sigma_1)\|_2=\left\|\sum_{i=1}^j \{A_i \varphi(X_i) \varphi^\top(X_i) -\Mean^{\mathcal{F}_{i-1}} \pi_{i-1}(1,X) \varphi(X) \varphi^\top(X)\}\right\|_2\\ \nonumber
	+ j\left\|\Mean^{\mathcal{F}_{i-1}} \varphi(X) \varphi^\top(X) \left(\frac{1}{j}\sum_{i=1}^j\pi_{i-1}(1,X)-\pi^*(1,X)\right)\right\|_2. 
\end{eqnarray}
By Lemma \ref{somebasiclemma1}, we have
\begin{eqnarray*}
	\left\|\Mean^{\mathcal{F}_{i-1}} \varphi(X) \varphi^\top(X) \left(\frac{1}{j}\sum_{i=1}^j\pi_{i-1}(1,X)-\pi^*(1,X)\right)\right\|_2\le \varepsilon_0^{-2}q \Mean^{\mathcal{F}_{i-1}} \left|\frac{1}{j}\sum_{i=1}^j\pi_{i-1}(1,X)-\pi^*(1,X)\right|\\
	\le \varepsilon_0^{-2} q^2 j^{-\alpha_0} \log^{\alpha_0} j,\,\,\,\,\forall j\ge j_n,
\end{eqnarray*}
with probability at least $1-O(j_n^{-\alpha_0})$. 

Consider the first term on the RHS of \eqref{somebasiceq0}. 
For any $i\ge 1$, define $M_i=\varphi(X_i) \varphi^\top(X_i) \{A_i-\pi_{i-1}(1,X_i)\}$. Notice that $\{M_i\}_{i\ge 1}$ forms a martingale difference sequence with respect to the filtration $\{\sigma(\mathcal{F}_{i-1}):i\ge 2\}$, since
\begin{eqnarray}\label{proofkeyeq1}
	&&\Mean [\varphi(X_i) \varphi^\top(X_i) \{A_i-\pi_{i-1}(X_i)\}|\mathcal{F}_{i-1}]\\\nonumber
	&=&\Mean^{\mathcal{F}_{i-1}} [\Mean (\varphi(X_i) \varphi(X_i)^\top \{A_i-\pi_{i-1}(X_i)\}|\mathcal{F}_{i-1},X_i)]=0, %\Mean^{X_i} \varphi(X_i)^{(q_1)} \varphi(X_i)^{(q_2)} \pi_{i-1}(1,X_i)\\
	%=\Mean \{\varphi(X)^{(q_1)} \varphi(X)^{(q_2)} \pi_{i-1}(1,X)|\},
\end{eqnarray}
where $\Mean^{\mathcal{F}_i,X_i}$ denotes the conditional expectation given $X_i$ and $\mathcal{F}_i$. Here, the first equality is due to that $X_i\independent \mathcal{F}_{i-1}$, implied by Lemma \ref{lemmaxyindpast}. Under the given conditions on the basis function $\varphi(\cdot)$, using similar arguments in proving Equation (C.15) of \cite{Shi2020}, we can show that the following event occurs with probability at least $1-O(j^{-2})$,
\begin{eqnarray*}
	\left\|\sum_{i=1}^j M_i\right\|_2\preceq \sqrt{q j\log (j)}.
\end{eqnarray*}
Notice that $\sum_{k\ge j} k^{-2}\le j^{-2}+ \sum_{k> j} \{k(k-1)\}^{-1}=j^{-2}+j^{-1}$. Thus, the following occurs with probability at least $1-O(j_n^{-1})$,
\begin{eqnarray}\label{inequality2}
	\left\|\sum_{i=1}^j \{A_i \varphi(X_i) \varphi^\top(X_i) -\Mean^{\mathcal{F}_{i-1}} \pi_{i-1}(1,X) \varphi(X) \varphi^\top(X)\}\right\|_2\preceq \sqrt{qj\log j},\,\,\,\,\forall j\ge j_n.
\end{eqnarray}
It follows that
\begin{eqnarray*}
	\|(\widehat{\Sigma}_{1,k}-\Sigma_1)\|_2\preceq q\delta_k+\sqrt{q k^{-1}\log k},\,\,\,\,\forall k\ge j_n,
\end{eqnarray*}
with probability at least $1-O(j^{-\alpha_0})$. Similarly, we can show
\begin{eqnarray*}
	\|(\widehat{\Sigma}_{0,k}-\Sigma_0)\|_2\preceq q\delta_k+\sqrt{q k^{-1}\log k},\,\,\,\,\forall k\ge j_n,
\end{eqnarray*}
with probability at least $1-O(j_n^{-\alpha_0})$. The proof is hence completed. 

\subsubsection{Proof of \eqref{Sigmabasic1}}
When $j_n$ satisfies $j_n^{\alpha_0}/\log^{\alpha_0} (j_n)\gg q^2$, it follows from \eqref{Sigmabasic} and \eqref{mineigensigma} that
\begin{eqnarray}\label{mineigensigma1}
	\lambda_{\min}[\widehat{\Sigma}_{a,k}]\ge \lambda_{\min}[\Sigma_a]-\|\widehat{\Sigma}_{a,k}-\Sigma_a\|_2\ge 2^{-1} \varepsilon_0,\,\,\,\,\forall k\ge j_n,
\end{eqnarray}
with probability at least $1-O(j_n^{-\alpha_0})$. Combining \eqref{mineigensigma} with \eqref{mineigensigma1} and \eqref{Sigmabasic}, we obtain
\begin{eqnarray*}
	\|\widehat{\Sigma}_{a,k}^{-1}-\Sigma_{a}^{-1}\|_2=\|\widehat{\Sigma}_{a,k}^{-1}(\widehat{\Sigma}_{a,k}-\Sigma_{a})\Sigma_{a}^{-1}\|_2\le \lambda_{\min}[\Sigma_a]\lambda_{\min}[\widehat{\Sigma}_{a,k}) \|\widehat{\Sigma}_{a,k}-\Sigma_{a}\|_2\\\preceq q\delta_k+\sqrt{q k^{-1}\log k},\,\,\,\,\forall k\ge j_n,
\end{eqnarray*}
with probability at least $1-O(j_n^{-\alpha_0})$. %, for any $j\ge j_0$. When $j<j_0$, the event defined in \eqref{Sigmabasic1} also holds with probability at least $1-j_0 j^{-1}\le 0$. 
The proof is hence completed. 

\subsection{Proof of Lemma \ref{somebasiclemma4}}
For any $l\in\{1,\dots,q\}$ and $i\ge 1$, define $M_i(l)=\varphi^{(l)}(X_i) A_i \{Y_i-Q_0(X_i,a)\}$. Here, $\varphi^{(l)}(X_i)$ corresponds to the $l$-th element of $\varphi(X_i)$. 
%For any $i\ge 1$, define $W_i=\varphi(X_i) A_i \{Y_i-\varphi^\top(X_i) \beta_1\}$. 
Similar to \eqref{proofkeyeq1}, we can show $\{M_i(l)\}_{i\ge 1}$ forms a martingale difference sequence with respect to the filtration $\{\sigma(\mathcal{F}_{i-1}):i\ge 1\}$. 
By \eqref{proofkeyeq1}, we have for any $l$,
\begin{eqnarray}\label{proofkeyq1.5}
	\Mean \{\varphi^{(l)}(X_i)\}^2\le \lambda_{\max}[\varphi(X_i) \varphi^\top(X_i)]\le \epsilon_0^{-1}.
\end{eqnarray}
Similar to Lemma \ref{somebasiclemma1}, we can show that $\sigma^2(a,x)$ is uniformly bounded by $4\epsilon^{-1}$ for some $\epsilon>0$ as well. It follows that
\begin{eqnarray*}
	&&\Mean \{M_i^2(l)|\mathcal{F}_{i-1}\}=\Mean [\{\varphi^{(l)}(X_i)\}^2 A_i \{Y_i^*(1)-Q_0(X_i,1)\}^2|\mathcal{F}_{i-1}]\\&\le& \Mean [\{\varphi^{(l)}(X_i)\}^2\{Y_i^*(1)-Q_0(X_i,1)\}^2|\mathcal{F}_{i-1}]
	%\le \Mean [\{Y_i^*(1)-\varphi^\top(X_i)\beta_1\}^2|\mathcal{F}_{i-1}]\\
	=\Mean \sigma^2(1,X_i) \{\varphi^{(l)}(X_i)\}^2\\
	&\le& 4\epsilon_0^{-2}\Mean \{\varphi^{(l)}(X_i)\}^2\le 4\epsilon_0^{-3},
\end{eqnarray*}
where the first equality is due to (A1), the first inequality is due to that $A$ is bounded between $0$ and $1$, the second equality follows from Lemma \ref{lemmaxyindpast}, the second inequality follows from Lemma \ref{somebasiclemma1}, and the last inequality is due to \eqref{proofkeyq1.5}. It follows that
\begin{eqnarray}\label{expectedquadratic}
	\sum_{i=1}^k \Mean \{M_i^2(l)|\mathcal{F}_{i-1}\}\le 4k\epsilon_0^{-3}. 
\end{eqnarray}
Similarly, by (A1) and Lemma \ref{somebasiclemma1}, we have
\begin{eqnarray}\label{quadratic0}
	\sum_{i=1}^k M_i^2(l)\le 2 \sum_{i=1}^k \{\psi^{(l)}(X_i) \}^2 (\epsilon_0^{-2}+Y_i^2).
\end{eqnarray}
Under sub-Gaussianity, $Y_i^2$ has bounded sub-exponential tail. Similar to \eqref{inequality2}, it follows from Bernstein's inequality \citep[Lemma 2.2.11][]{van1996} and Bonferroni's inequality that, with probability at least $1-O(j^{-1})$ that
\begin{eqnarray*}
	\sum_{i=1}^k [M_i^2(l)- \Mean \{M_i^2(l)|\mathcal{F}_{i-1}\}]\preceq \sqrt{qk\log k},\,\,\,\,\forall k\ge j.
\end{eqnarray*}
Thus, for any sequence $j_n$ that satisfies $j_n/\log (j_n)\gg q$, we have by \eqref{expectedquadratic} that
\begin{eqnarray*}
	\sum_{i=1}^k M_i^2(l)+\sum_{i=1}^k \Mean \{M_i^2(l)|\mathcal{F}_{i-1}\}\le \bar{c} k,\,\,\,\,\forall k\ge j_n,
\end{eqnarray*}
for some constant $\bar{c}>0$, with probability at least $1-O(j_n^{-1})$. It follows that
\begin{eqnarray*}
	&&\hbox{Pr}\left(  \bigcap_{k\ge j_n}\{|\sum_{i=1}^k M_i(l)|\le 2\sqrt{\bar{c}k\log k} \}  \right)\\
	&\ge& \hbox{Pr}\left(\left\{ \bigcap_{k\ge j_n} \{|\sum_{i=1}^k M_i(l)|\le 2\sqrt{\bar{c}k\log k}\} \right\}\bigcap  \left\{ \bigcap_{k\ge j_n} \{\sum_{i=1}^k [M_i^2(l)+ \{M_i^2(l)|\mathcal{F}_{i-1}\}]\le \bar{c}k \} \right\} \right)\\
	&-&O(j_n^{-1})\ge \hbox{Pr}\left( \left\{ \bigcap_{k\ge j_n} \{\sum_{i=1}^k [M_i^2(l)+ \{M_i^2(l)|\mathcal{F}_{i-1}\}]\le \bar{c}k \} \right\} \right)-O(j_n^{-1})\\
	&-&\hbox{Pr}\left(\left\{ \bigcup_{k\ge j_n} \{|\sum_{i=1}^k M_i(l)|> 2\sqrt{\bar{c}k\log k}\} \right\}\bigcap  \left\{ \bigcap_{k\ge j_n} \{\sum_{i=1}^k [M_i^2(l)+ \{M_i^2(l)|\mathcal{F}_{i-1}\}]\le \bar{c}k \} \right\} \right)\\
	&\ge&1-\hbox{Pr}\left(\left\{ \bigcup_{k\ge j_n} \{|\sum_{i=1}^k M_i(l)|> 2\sqrt{\bar{c}k\log k}\} \right\}\bigcap  \left\{ \bigcap_{k\ge j_n} \{\sum_{i=1}^k [M_i^2(l)+\{M_i^2(l)|\mathcal{F}_{i-1}\}]\le \bar{c}k\} \right\} \right)\\
	&-&O(j_n^{-1}).
\end{eqnarray*}
By Bonferroni's inequality and Theorem 2.1 of \cite{Bercu2008}, we have
\begin{eqnarray}\nonumber
	&&\hbox{Pr}\left(  \bigcap_{k\ge j_n}\{|\sum_{i=1}^k M_i(l)|\le 2\sqrt{\bar{c}k\log k}\}  \right)\ge 1-O(j_n^{-1})\\ \nonumber
	&-&\sum_{k\ge j_n} \hbox{Pr}\left( \{|\sum_{i=1}^k M_i(l)|> 2\sqrt{\bar{c}k\log k}\}\bigcap \left\{ \bigcap_{k'\ge j_n} \{\sum_{i=1}^{k'} [M_i^2(l)+\{M_i^2(l)|\mathcal{F}_{i-1}\}]\le \bar{c}k' \} \right\} \right)\\ \nonumber
	&\ge&1-O(j_n^{-1})-\sum_{k\ge j_n} \hbox{Pr}\left( \{|\sum_{i=1}^k M_i(l)|> 2\sqrt{\bar{c}k\log k}\}\bigcap \left\{ \sum_{i=1}^k [M_i^2(l)+\{M_i^2(l)|\mathcal{F}_{i-1}\}]\le \bar{c}k \right\} \right)\\\label{lastterm}
	&\ge&1-O(j_n^{-1})-2\sum_{k\ge j_n}\exp\left( -\frac{4\bar{c}k\log k}{2\bar{c}k} \right)=1-O(j_n^{-1})-\sum_{k\ge j_n} 2k^{-2}.
\end{eqnarray}
The last term on the RHS of \eqref{lastterm} is $1-O(j_n^{-1})$. To summarize, we have shown that the following event occurs with probability at least $1-O(j_n^{-1})$,
\begin{eqnarray*}
	\bigcap_{k\ge j_n}\left\{|\sum_{i=1}^k M_i(l)|\le 2\sqrt{\bar{c}k\log k}\right\}.
\end{eqnarray*}
By Bonferroni's inequality, we have
\begin{eqnarray*}
	\bigcap_{k\ge j_n}\left\{ \left\|\sum_{i=1}^k \varphi(X_i) A_i\{Y_i-Q_0(X_i,1)\} \right\|_2\le 2\sqrt{\bar{c}q k\log k} \right\},
\end{eqnarray*}
with probability at least $1-O(j_n^{-1/2})$. Similarly, we can show
\begin{eqnarray*}
	\bigcap_{k\ge j_n}\left\{ \left\|\sum_{i=1}^k \varphi(X_i) (1-A_i)\{Y_i-Q_0(X_i,0)\} \right\|_2\le c \sqrt{q k \log k}\right\},
\end{eqnarray*}
for some constant $c>0$, with probability at least $1-O(j_n^{-1})$. The proof is hence completed.

\subsection{Proof of Lemma \ref{lemmasomebasic1}}
Combining Lemma \ref{somebasiclemma4} with Lemma \ref{somebasiclemma1} yields that
\begin{eqnarray*}
	\left\|\Sigma_a^{-1}\left(\frac{1}{k}\sum_{i=1}^k \mathbb{I}(A_i=a)\varphi(X_i) \{Y_i-Q_0(X_i,a) \}\right) \right\|_2\preceq q^{1/2} k^{-1/2} \sqrt{\log k},\,\,\,\,\forall k\ge j_n,a\in\{0,1\},
\end{eqnarray*}
with probability at least $1-O(j_n^{-1})$. Combining this together with \eqref{betahat1} yields that
\begin{eqnarray*}
	\|\widehat{\beta}_{a,k}-\beta_a\|_2\preceq q^{1/2} k^{-1/2}\sqrt{\log k},\,\,\,\,\forall k\ge j_n,a\in\{0,1\},
\end{eqnarray*}
with probability at least $1-O(j_n^{-1})$. The proof is hence completed. 

\subsection{Proof of Lemma \ref{lemmasomebasic2}}
Notice that 
\begin{eqnarray}\label{secondterm}
	\begin{split}
		&\left\|\frac{1}{k}\sum_{i=1}^k \mathbb{I}(A_i=a)\varphi(X_i) \varphi^\top(X_i) \{Y_i-Q_0(X_i,a)\}^2-\Phi_a \right\|_2\\ 
		\le& \left\|\frac{1}{k}\sum_{i=1}^k \mathbb{I}(A_i=a)\varphi(X_i) \varphi^\top(X_i) [\{Y_i-Q_0(X_i,a)\}^2-\sigma^2(a,X_i)] \right\|_2\\ 
		+&\left\|\frac{1}{k}\sum_{i=1}^k \mathbb{I}(A_i=a)\varphi(X_i) \varphi^\top(X_i) \sigma^2(a,X_i)-\Phi_a  \right\|_2.
	\end{split}
\end{eqnarray}
Similar to the proof of Lemma \ref{somebasiclemma2}, we can show that the second term on the RHS of \eqref{secondterm} is of the order $O(q \delta_k+\sqrt{q k^{-1}\log k})$, for any $a\in\{0,1\}$ and any $k\ge j_n$, with probability at least $1-O(j_n^{-\alpha_0})$. As for the first term, notice that each element in the matrix
\begin{eqnarray}\label{somematrix}
	\frac{1}{k}\sum_{i=1}^k \mathbb{I}(A_i=a)\varphi(X_i) \varphi^\top(X_i) [\{Y_i-Q_0(X_i,a)\}^2-\sigma^2(a,X_i) ]
\end{eqnarray}
corresponds to a martingale with respect to the filtration $\{\sigma(\mathcal{F}_{i-1}):i\ge 1\}$, under (A1) and (A2). 
Using similar arguments in proving Lemma E.2 of \cite{Shi2020}, we can show that
\begin{eqnarray*}
	\left\|\frac{1}{k}\sum_{i=1}^k \mathbb{I}(A_i=a)\varphi(X_i) \varphi^\top(X_i) [\{Y_i-Q_0(X_i,a) \}^2-\sigma^2(a,X_i)]\right\|_{2}\preceq q^{1/2}k^{-1/2} \sqrt{\log k},\,\,\,\,\\\forall a\in \{0,1\},k\ge j_n,
\end{eqnarray*}
with probability at least $1-O(j_n^{-1})$. 

Finally, we focus on providing an upper bound for the bias term
\begin{eqnarray*}
	\left\|\frac{1}{k}\sum_{i=1}^k \mathbb{I}(A_i=a)\varphi(X_i) \varphi^\top(X_i) [\{Y_i-Q_0(X_i,a)\}^2-\{Y_i-\varphi^\top(X_i)\beta_a\}^2]\right\|\\
	\le 2\left\|\frac{1}{k}\sum_{i=1}^k \mathbb{I}(A_i=a)\varphi(X_i) \varphi^\top(X_i) [\{Y_i-Q_0(X_i,a)\}\{Q_0(X_i,a)-\varphi^\top(X_i)\beta_a\}]\right\|\\
	+\left\|\frac{1}{k}\sum_{i=1}^k \mathbb{I}(A_i=a)\varphi(X_i) \varphi^\top(X_i)\right\| o\{(NT)^{-1/2}\}.
\end{eqnarray*}
Using similar arguments, the first term on the RHS can be upper bounded by $Cq^{1/2}k^{-1/2} \sqrt{\log k}$ for some constant $C>0$. The second term is $o\{(NT)^{-1/2}\}$. The proof is hence completed.

%, or when $\inf_{x\in \mathbb{X}} |\varphi^\top(x) (\beta_1-\beta_0)|>0$, for $a\in \{0,1\}$. 
\subsection{Proof of Lemma \ref{thm5}}
We begin by providing an upper bound for $\max_{a\in \{0,1\}} \|\widehat{\beta}_{a,k}-\beta_a\|_2$. With some calculations, we have
\begin{eqnarray*}
	\max_{a\in \{0,1\}} \|\widehat{\beta}_{a,k}-\beta_a\|_2=\max_{a\in \{0,1\}} \frac{1}{k}\left\|\widehat{\Sigma}_{a,k}^{-1} \left(\sum_{i=1}^k \varphi(X_i) \mathbb{I}(A_i=a)\{Y_i- \varphi^\top(X_i) \beta_a\} \right)\right\|_2\\
	\le \max_{a\in \{0,1\}} \left\|\widehat{\Sigma}_{a,k}^{-1}\right\|_2 \max_{a\in \{0,1\}} \frac{1}{k}\left\|\sum_{i=1}^k \varphi(X_i) \mathbb{I}(A_i=a)\{Y_i-\varphi^\top(X_i) \beta_a\}\right\|_2.
\end{eqnarray*}
Using similar arguments in proving Theorem \ref{thm1}, we obtain with probability at least $1-O(j_n^{-1})$ that
\begin{eqnarray}\label{inequality1}
	\max_{a\in \{0,1\}} \frac{1}{k}\left\|\sum_{i=1}^k \varphi(X_i) \mathbb{I}(A_i=a) \{Y_i- \varphi^\top(X_i) \beta_a\} \right\|_2\preceq q^{1/2} k^{-1/2}\sqrt{\log k},\,\,\,\,\forall k\ge j_n.
\end{eqnarray}
Similarly, we can show with probability at least $1-O(j_n^{-1})$ that
\begin{eqnarray}\label{inequality1.5}
	\max_{a\in \{0,1\}} \frac{1}{k}\left\|\sum_{i=1}^k \varphi(X_i) \mathbb{I}(A_i=a) \{Y_i- \varphi^\top(X_i) \beta_a\} \right\|_2\preceq q^{1/2} k^{-1/2}\sqrt{\log j_n},\\\nonumber\forall 1 \le k<j_n.
\end{eqnarray}

Similar to \eqref{somebasiceq0}, we have
\begin{eqnarray*}
	\max_{a\in \{0,1\}} \lambda_{\min}[\widehat{\Sigma}_{a,k}]\ge 	\min_{a\in \{0,1\}} \lambda_{\min}\left( \Mean^{\mathcal{F}_{i-1}} \varphi(X) \varphi^\top(X) \frac{1}{k}\sum_{i=1}^k\pi_{i-1}(a,X) \right)\\
	-\max_{a\in \{0,1\}}\frac{1}{k}\left\|\sum_{i=1}^k \{\mathbb{I}(A_i=a) \varphi(X_i) \varphi^\top(X_i) -\Mean^{\mathcal{F}_{i-1}} \pi_{i-1}(a,X) \varphi(X) \varphi^\top(X)\}\right\|_2.
\end{eqnarray*}
Using similar arguments in proving \eqref{inequality2}, we can show that
\begin{eqnarray}\label{inequality3}
	\max_{a\in \{0,1\}} \left\|\sum_{i=1}^k \{\mathbb{I}(A_i=a) \varphi(X_i) \varphi^\top(X_i) -\Mean^{\mathcal{F}_{i-1}} \pi_{i-1}(a,X) \varphi(X) \varphi^\top(X)\}\right\|_2\preceq \sqrt{qk\log k},\\ \nonumber 
	\forall k\ge j_n,
\end{eqnarray}
with probability at least $1-O(j_n^{-1})$. Similarly, we can show
\begin{eqnarray}\label{inequality3.5}
	\max_{a\in \{0,1\}} \left\|\sum_{i=1}^k \{\mathbb{I}(A_i=a) \varphi(X_i) \varphi^\top(X_i) -\Mean^{\mathcal{F}_{i-1}} \pi_{i-1}(a,X) \varphi(X) \varphi^\top(X)\}\right\|_2\preceq \sqrt{qk\log j_n},\\ \nonumber 
	\forall 1\le k< j_n,
\end{eqnarray}
with probability at least $1-O(j_n^{-1})$. 

Without loss of generality, assume $\varepsilon_0\le 1/2$. Notice that we have $\pi_{i-1}(a,x)\ge \varepsilon_0$, for any $a\in \{0,1\}$, $x\in \mathbb{X}$ and $i\ge N_0$. This together with Lemma \eqref{somebasiclemma1} implies that
\begin{eqnarray*}
	\inf_{a\in \{0,1\},n\ge j_n} \lambda_{\min}\left( \Mean^{\mathcal{F}_{i-1}} \varphi(X) \varphi^\top(X) \frac{1}{n}\sum_{i=1}^n\pi_{i-1}(a,X) \right)\ge \frac{n-N_0}{n} \varepsilon_0\ge \frac{j-N_0}{j} \varepsilon_0.
\end{eqnarray*}
Combining this together with \eqref{inequality3} and \eqref{inequality3.5} yields
\begin{eqnarray*}
	\max_{a\in \{0,1\}} \lambda_{\min}[\widehat{\Sigma}_{a,k}]\ge \frac{\varepsilon_0}{2},\,\,\,\,\forall k\ge L^*\sqrt{q \log j_n},
\end{eqnarray*}
for some constant $L^*\ge 1$, with probability at least $1-O(j_n^{-1})$. This together with \eqref{inequality1} and \eqref{inequality1.5} yields that
\begin{eqnarray*}
	\max_{a\in \{0,1\}} \|\widehat{\beta}_{a,k}-\beta_a\|_2 \preceq q^{1/2} k^{-1/2}\sqrt{\log \max(k,j_n)},\,\,\,\,\forall k\ge L^*\sqrt{q\log j_n},
\end{eqnarray*}
%for some constant $\bar{L}>0$, 
with probability at least $1-O(j_n^{-1})$.

By Condition (A3), we have
\begin{eqnarray}\label{inequality4}
	|\varphi^\top(X) (\widehat{\beta}_{1,k}-\widehat{\beta}_{0,k}-\beta_1+\beta_0)|\le \bar{L}q k^{-1/2} \log^{1/2} \max(k,j_n), \,\,\,\,\forall k\ge L^*\sqrt{q \log j_n},
\end{eqnarray}
for some constant $\bar{L}>0$, with probability at least $1-O(j_n^{-1})$. 

For any $z_1,z_2\in \mathbb{R}$, we have $\mathbb{I}(z_1>0)\neq \mathbb{I}(z_2>0)$ only when $|z_1-z_2|\ge |z_2|$. Hence, under the event defined in \eqref{inequality4}, the event $\mathbb{I}\{\varphi^\top(X) (\widehat{\beta}_{1,k}-\widehat{\beta}_{0,k})>0 \}\neq \mathbb{I}\{\varphi^\top(X) (\beta_1-\beta_0)>0 \}$ occurs only when 
\begin{eqnarray*}
	|\varphi^\top(X) (\beta_1-\beta_0)|\le |\varphi^\top(X) (\widehat{\beta}_{1,k}-\widehat{\beta}_{0,k}-\beta_1+\beta_0)|\le \bar{L} q k^{-1/2} \sqrt{\log \max(k,j_n)},
\end{eqnarray*}
for any $k\ge j_n$. Since $Q_0(X,1)-Q_0(X,0)$ can be well-approximated by $\varphi^\top(X) (\beta_1-\beta_0)$, the above event occurs only when
\begin{eqnarray*}
	|Q_0(X,1)-Q_0(X,0)|\le  2\bar{L} q k^{-1/2} \sqrt{\log \max(k,j_n)}.
\end{eqnarray*}
Under the given conditions, we have
\begin{eqnarray}\label{inequality5}
	\prob\left(	|Q_0(X,1)-Q_0(X,0)|\le 2\bar{L}q k^{-1/2} \log^{1/2} \max(k,j_n)\right)\le 2\bar{L}L_0q k^{-1/2} \log^{1/2} \max(k,j_n). 
\end{eqnarray}
Notice that when $\mathbb{I}\{\varphi^\top(X) (\widehat{\beta}_{1,k}-\widehat{\beta}_{0,k})>0 \}= \mathbb{I}\{\varphi^\top(X) (\beta_1-\beta_0)>0 \}$, we have $\pi_{k}(a,X)=\pi^*(a,X)$. Thus, we obtain $\pi_{k}(a,X)=\pi^*(a,X)$ if $|\varphi^\top(X) (\beta_1-\beta_0)|>2 \bar{L}q k^{-1/2} \sqrt{\log \max(k,j_n)}$, for any $k\ge L^*\sqrt{q \log j_n}$. Set $k_0=L^* \sqrt{q \log j_n}$. By \eqref{inequality4} and \eqref{inequality5}, we have with probability at least $1-O(j_n^{-1})$ that
\begin{eqnarray*}
	&&\sum_{a\in\{0,1\}} \Mean^{\mathcal{F}_{i-1}} \left|\sum_{i=1}^k \{\pi_{i-1}(a,X)-\pi^*(a,X)\}\right|
	\le \sum_{a\in\{0,1\}} \sum_{i=1}^{k_0} \Mean^{\mathcal{F}_{i-1}} |\pi_{i-1}(a,X)-\pi^*(a,X)|\\
	&+&  \sum_{a\in\{0,1\}} \sum_{i=k_0+1}^{k} \Mean^{\mathcal{F}_{i-1}} |\pi_{i-1}(a,X)-\pi^*(a,X)|\le 2L^* \sqrt{q\log j_n} \\
	&+&\sum_{a\in\{0,1\}} \sum_{i=k_0+1}^{k} \Mean^{\mathcal{F}_{i-1}} |\pi_{i-1}(a,X)-\pi^*(a,X)| \mathbb{I}\{|\varphi^\top(X) (\beta_1-\beta_0)|> 2\bar{L}q i^{-1/2} \log^{1/2} i\}\\
	&+&\sum_{a\in\{0,1\}} \sum_{i=k_0+1}^k \Mean^{\mathcal{F}_{i-1}} |\pi_{i-1}(a,X)-\pi^*(a,X)| \mathbb{I}\{|\varphi^\top(X) (\beta_1-\beta_0)|\le 2\bar{L}q i^{-1/2} \log^{1/2} i\}\\
	&\le &  2L^* \sqrt{q\log j_n}+ \sum_{a\in\{0,1\}} \sum_{i=k_0+1}^n \prob\left(	|\varphi^\top(X) (\beta_1-\beta_0)|\le \bar{L}q i^{-1/2} \sqrt{\log i}\right)\preceq q k^{1/2}\log^{1/2} k,\,\,\,\,\forall k\ge j_n.
\end{eqnarray*}
The proof is hence completed. 

\subsection{Proof of Theorem \ref{thm3}}
%We state the following lemmas before presenting the proof. 
%\begin{lemma}\label{lemmasomebasic0}
%	Assume the conditions in Theorem \ref{thm3} hold. Then for any $j\ge 1$, we have with probability at least $1-O(j^{-1/5})$ that for any $a\in\{0,1\}$ and any $k\ge j$,
%	\begin{eqnarray*}
%		\sum_{i=1}^k \{Y_i^*(a)\}^2\preceq k,\,\,\,\,\,\,\,\,\sum_{i=1}^k \{Y_i^*(a)\}^4\preceq k. 
%	\end{eqnarray*}
%\end{lemma}

Similar to the proof of Theorem \ref{thm1}, we will show that the assertion in Theorem \ref{thm3} holds for any $n(\cdot)$ that correspond to the realizations of $N(\cdot)$ that satisfy $n(t_1)<n(t_2)<\cdots<n(t_{K})$. For any $1\le k_1\le k_2\le K$, define
\begin{eqnarray*}
	\widehat{V}(k_1,k_2)=\sqrt{n(t_{k_1})n(t_{k_2})}\Cov\left(\widehat{\beta}_1^{\tiny{\hbox{MB}}*}(t_{k_1})-\widehat{\beta}_0^{\tiny{\hbox{MB}}*}(t_{k_1}), \widehat{\beta}_1^{\tiny{\hbox{MB}}*}(t_{k_2})-\widehat{\beta}_0^{\tiny{\hbox{MB}}*}(t_{k_2})|\{(X_i,A_i,Y_i)\}_{i=1}^{+\infty}\right)\\
	=\frac{1}{\sqrt{n(t_{k_1})n(t_{k_2})}}\sum_{a=0}^1 \sum_{j=1}^{k_1} \sum_{i=n(t_{j-1})+1}^{n(t_j)}\widehat{\Sigma}_a^{-1}(t_j) \mathbb{I}(A_i=a) \varphi(X_i) \varphi^\top(X_i) \{Y_i-\varphi^\top(X_i) \widehat{\beta}_a(t_j)\}^2 \widehat{\Sigma}_a^{-1}(t_j),
\end{eqnarray*} 
and 
\begin{eqnarray*}
	\widehat{\bm{V}}=\left(\begin{array}{cccc}
		\widehat{V}(1,1) & \widehat{V}(1,2) & \dots & \widehat{V}(1,K)\\
		\widehat{V}(2,1) & \widehat{V}(2,2) & \dots & \widehat{V}(2,K)\\
		\vdots & \vdots & & \vdots \\
		\widehat{V}(K,1) & \widehat{V}(K,2) & \dots & \widehat{V}(K,K)
	\end{array}\right).
\end{eqnarray*}
We aim to bound the entrywise $\ell_{\infty}$ norm of $\widehat{\bm{V}}-\bm{V}$ where $\bm{V}$ is defined in \eqref{Vdefi}. %For any $1\le q_1,q_2\le q$, $1\le k_1,k_2\le k$, let $\widehat{V}^{(q_1,q_2)}(k_1,k_2)$ and $V^{(q_1,q_2)}(k_1,k_2)$ be the $(q_1,q_2)$-th element of $\widehat{V}(k_1,k_2)$ and $V(k_1,k_2)$, respectively. 
It suffices to bound $\max_{1\le k_1\le k_2\le K}\sup_{b_1,b_2\in \mathbb{R}^{p+1},\|b_1\|_2=\|b_2\|_2=1} |b_1^T \{\widehat{V}(k_1,k_2)-V(k_1,k_2)\} b_2|=\max_{1\le k_1\le k_2\le K}\|\widehat{V}(k_1,k_2)-V(k_1,k_2)\|_2$. For any $k_1,k_2$, we decompose $\widehat{V}(k_1,k_2)-V(k_1,k_2)$ as
\begin{eqnarray*}
	\widehat{V}(k_1,k_2)-V(k_1,k_2)=\widehat{V}(k_1,k_2)-\widehat{V}^*(k_1,k_2)+\widehat{V}^{*}(k_1,k_2)-\widehat{V}^{**}(k_1,k_2)+\widehat{V}^{**}(k_1,k_2)-V(k_1,k_2),
\end{eqnarray*}
where
\begin{eqnarray*}
	\widehat{V}^*(k_1,k_2)=\frac{1}{\sqrt{n(t_{k_1})n(t_{k_2})}} \sum_{a=0}^1 \sum_{j=1}^{k_1}\sum_{i=n(t_{j-1})+1}^{n(t_j)}\Sigma_a^{-1}\mathbb{I}(A_i=a)\varphi(X_i) \varphi^\top(X_i) \{Y_i-\varphi^\top(X_i) \widehat{\beta}_a(t_j)\}^2 \Sigma_a^{-1},\\
	\widehat{V}^{**}(k_1,k_2)=\frac{1}{\sqrt{n(t_{k_1})n(t_{k_2})}}\sum_{a=0}^1 \sum_{j=1}^{n(t_{k_1})}\Sigma_a^{-1}\mathbb{I}(A_i=a)\varphi(X_i)\varphi^\top(X_i) \{Y_i-\varphi^\top(X_i) \beta_a^*\}^2 \Sigma_a^{-1}.
\end{eqnarray*}
By Lemma \ref{somebasiclemma1} and Lemma \ref{lemmasomebasic2}, we obtain that
\begin{eqnarray}\nonumber
	&&\max_{1\le k_1\le k_2\le K} \|\widehat{V}^{**}(k_1,k_2)-V(k_1,k_2) \|_{2}\\\nonumber
	&\le& \max_{1\le k_1\le K}\sum_{a=0}^1 \left\|\frac{1}{n(t_{k_1})}\sum_{j=1}^{n(t_{k_1})}\Sigma_a^{-1}\mathbb{I}(A_i=a)\varphi(X_i)\varphi^\top(X_i) \{Y_i-\varphi^\top(X_i) \beta_a^*\}^2 \Sigma_a^{-1}- \Sigma_a^{-1}\Phi_a\Sigma_a^{-1} \right\|_2\\ \nonumber
	&\le& \max_{1\le k_1\le K}\frac{1}{ \epsilon_0^2} \sum_{a=0}^1 \left\|\frac{1}{n(t_{k_1})}\sum_{j=1}^{n(t_{k_1})}\mathbb{I}(A_i=a)\varphi(X_i)\varphi^\top(X_i) \{Y_i-\varphi^\top(X_i) \beta_a^*\}^2 -\Phi_a \right\|_2 \\\label{somematrix1}
	&\preceq& q\delta_{n(t_1)}+q^{1/2} n^{-1/2}(t_1)\sqrt{\log n(t_1)},
\end{eqnarray}
with probability at least $1-O(n^{-\alpha_0}(t_1))$. 

Notice that
\begin{eqnarray*}
	\sqrt{n(t_{k_1})n(t_{k_2})}\widehat{V}^*(k_1,k_2)
	= \sum_{a=0}^1 \sum_{j=1}^{k_1}\sum_{i=n(t_{j-1})+1}^{n(t_j)}\Sigma_a^{-1}\mathbb{I}(A_i=a)\varphi(X_i) \varphi^\top(X_i) \{Y_i-\varphi^\top(X_i) \widehat{\beta}_a(t_j)\}^2 \Sigma_a^{-1}\\
	= \sum_{a=0}^1 \sum_{j=1}^{k_1}\sum_{i=n(t_{j-1})+1}^{n(t_j)}\Sigma_a^{-1}\mathbb{I}(A_i=a)\varphi(X_i) \varphi^\top(X_i) \{Y_i-\varphi^\top(X_i) \beta_a^*+\varphi^\top(X_i) \beta_a^*-\varphi^\top(X_i) \widehat{\beta}_a(t_j)\}^2 \Sigma_a^{-1}\\
	=\sum_{a=0}^1 \sum_{j=1}^{k_1}\sum_{i=n(t_{j-1})+1}^{n(t_j)}\Sigma_a^{-1}\mathbb{I}(A_i=a)\varphi(X_i) \varphi^\top(X_i) \{\varphi^\top(X_i) \beta_a^*-\varphi^\top(X_i) \widehat{\beta}_a(t_j)\}^2\Sigma_a^{-1}\\
	+2 \sum_{a=0}^1 \sum_{j=1}^{k_1} \sum_{i=n(t_{j-1})+1}^{n(t_j)}\Sigma_a^{-1}\mathbb{I}(A_i=a)\varphi(X_i) \varphi^\top(X_i) \{Y_i-\varphi^\top(X_i) \beta_a^*\}\varphi^\top(X_i) \{\beta_a^*-\widehat{\beta}_a(t_j)\}\Sigma_a^{-1}\\
	+\sqrt{n(t_{k_1})n(t_{k_2})}\widehat{V}^{**}(k_1,k_2).
\end{eqnarray*}
It follows that
\begin{eqnarray*}
	&&\max_{1\le k_1\le k_2\le K} \left\|\widehat{V}^*(k_1,k_2)-\widehat{V}^{**}(k_1,k_2)\right\|_2\\
	&\le& \max_{1\le k_1\le K} \frac{1}{n(t_{k_1})}\left\|\sum_{a=0}^1 \sum_{j=1}^{k_1}\sum_{i=n(t_{j-1})+1}^{n(t_j)}\Sigma_a^{-1}\mathbb{I}(A_i=a)\varphi(X_i) \varphi^\top(X_i) \{\varphi^\top(X_i) \beta_a^*-\varphi^\top(X_i) \widehat{\beta}_a(t_j)\}^2\Sigma_a^{-1}\right\|_2\\
	&+& \max_{1\le k_1\le K}  \frac{2}{n(t_{k_1})} \left\|\sum_{a=0}^1 \sum_{j=1}^{k_1} \sum_{i=n(t_{j-1})+1}^{n(t_j)}\Sigma_a^{-1}\mathbb{I}(A_i=a)\varphi(X_i) \varphi^\top(X_i) \{Y_i-\varphi^\top(X_i) \beta_a\}\varphi^\top(X_i) \{\beta_a^*-\widehat{\beta}_a(t_j)\}\Sigma_a^{-1}\right\|_2.
\end{eqnarray*}
By Lemma \ref{somebasiclemma1}, we obtain that
\begin{eqnarray}\label{somematrix2}
	&&\max_{1\le k_1\le k_2\le K} \left\|\widehat{V}^*(k_1,k_2)-\widehat{V}^{**}(k_1,k_2)\right\|_2\\ \nonumber
	&\preceq & \max_{\substack{1\le k_1\le K\\ a\in \{0,1\} }} \frac{1}{n(t_{k_1})} \Big\| \underbrace{\sum_{j=1}^{k_1}\sum_{i=n(t_{j-1})+1}^{n(t_j)}\mathbb{I}(A_i=a)\varphi(X_i) \varphi^\top(X_i) \{\varphi^\top(X_i) \beta_a^*-\varphi^\top(X_i) \widehat{\beta}_a(t_j)\}^2}_{\Psi_{1,a,k_1}}\Big\|_2\\ \nonumber
	&+& \max_{\substack{1\le k_1\le K\\ a\in \{0,1\} }} \frac{2}{n(t_{k_1})} \Big\|\underbrace{\sum_{j=1}^{k_1} \sum_{i=n(t_{j-1})+1}^{n(t_j)}\mathbb{I}(A_i=a)\varphi(X_i) \varphi^\top(X_i) \{Y_i-\varphi^\top(X_i) \beta_a^*\}\varphi^\top(X_i) \{\beta_a^*-\widehat{\beta}_a(t_j)\}}_{\Psi_{2,a,k_1}}\Big\|_2.
\end{eqnarray}
By Lemmas \ref{somebasiclemma1} and \ref{lemmasomebasic1}, we have with probability at least $1-O(n^{-1}(t_1))$ that
\begin{eqnarray}\label{secondandhalfinequality}
	\frac{1}{n(t_{k_1})}\|\Psi_{1,a,k_1}\|_2\preceq q^2 n^{-1}(t_1) \log\{n(t_1)\} \left\|\frac{1}{n(t_{k_1})}\sum_{i=1}^{n(t_{k_1})}\mathbb{I}(A_i=a)\varphi(X_i) \varphi^\top(X_i)\right\|_2,\\\nonumber
	\forall 1\le k_1\le K,a\in \{0,1\}.
\end{eqnarray}
Similar to Lemma \ref{somebasiclemma2}, we can show there exists some constant $c_*>0$ that
\begin{eqnarray}\label{secondandthreequarterinequality}
	\frac{1}{n(t_{k_1})}\left\|\sum_{i=1}^{n(t_{k_1})}[\mathbb{I}(A_i=a)\varphi(X_i) \varphi^\top(X_i)-\Mean^{\mathcal{F}_{i-1}}\{\mathbb{I}(A_i=a)\varphi(X_i) \varphi^\top(X_i)\}]\right\|_2\\\nonumber\le c_* \{q \delta_{n(t_{k_1})}+q^{1/2}n^{-1/2}(t_{k_1})\sqrt{\log n(t_{k_1})} \},\,\,\,\,\forall 1\le k_1\le K, a\in \{0,1\},
	%	\left\|\sum_{i=1}^{n(t_{k_1})}\{\mathbb{I}(A_i=a)-\pi_{i-1}(a,X_i)\}\varphi(X_i) \varphi^\top(X_i)\right\|_2\le \bar{c} \{\delta_{n(t_{k_1})}+q^{1/2}n^{-1/2}(t_{k_1})\sqrt{\log n(t_{k_1})} \},\,\,\,\,\forall 1\le k_1\le K.
\end{eqnarray}
with probability at least $1-O(n^{-1}(t_1))$. By Lemma \ref{somebasiclemma1}, we can show with probability at least $1-O(n^{-1}(t_1))$ that
\begin{eqnarray*}
	\max_{1\le k_1\le K}\frac{1}{n(t_{k_1})}\left\|\sum_{i=1}^{n(t_{k_1})}\Mean^{\mathcal{F}_{i-1}}\{\mathbb{I}(A_i=a)\varphi(X_i) \varphi^\top(X_i)\}\right\|_2=O(1).
\end{eqnarray*}
This together with \eqref{secondandhalfinequality} and \eqref{secondandthreequarterinequality} yields
\begin{eqnarray}\label{Psi1ak1}
	n^{-1}(t_{k_1})\|\Psi_{1,a,k_1}\|_2\preceq q^2 n^{-1}(t_1)\log \{n(t_1)\},\,\,\,\,\forall 1\le k_1\le K,a \in \{0,1\},
\end{eqnarray}
with probability at least $1-O(n^{-1}(t_1))$. 

Moreover, using similar arguments in proving Lemma \ref{lemmasomebasic2}, we can show that for any $1\le k_1\le K$, the following event occurs with probability at least $1-O(n^{-2}(t_{k_1}) )$, 
\begin{eqnarray*}
	\frac{1}{n(t_{k_1})} \left\|\sum_{i=1}^{n(t_{k_1})}\mathbb{I}(A_i=a)\varphi(X_i) \varphi^\top(X_i) \{Y_i-\varphi^\top(X_i) \beta_a^*\}\varphi^{(l)}(X_i)\right\|_2\preceq q^{1/2} n^{-1/2}(t_{k_1})\sqrt{\log n(t_{k_1})},\\
	\forall 1\le l\le q.
\end{eqnarray*}
Since $\sum_{k_1=1}^{K} n^{-2}(t_{k_1})\le n^{-1}(t_1)$, we obtain with probability at least $1-O(n^{-1}(t_1))$ that
\begin{eqnarray*}
	\frac{1}{n(t_{k_1})} \left\|\sum_{i=1}^{n(t_{k_1})}\mathbb{I}(A_i=a)\varphi(X_i) \varphi^\top(X_i) \{Y_i-\varphi^\top(X_i) \beta_a\}\varphi^{(l)}(X_i)\right\|_2\preceq q^{1/2} n^{-1/2}(t_{k_1})\sqrt{\log n(t_{k_1})},\\\forall 1\le l\le q, 1\le k_1\le K.
\end{eqnarray*}
In addition, it follows from Lemma \ref{lemmasomebasic1} that
\begin{eqnarray*}
	n^{-1}(t_{k_1})\|\Psi_{2,a,k_1}\|_2\preceq q^{3/2} n^{-1}(t_1)\log \{n(t_1)\},\,\,\,\,\forall 1\le k_1\le K,a \in \{0,1\}.
\end{eqnarray*}

This together with \eqref{Psi1ak1} yields that
\begin{eqnarray*}%\label{somematrix4}
	\max_{1\le k_1\le k_2\le K} \left\|\widehat{V}^*(k_1,k_2)-\widehat{V}^{**}(k_1,k_2)\right\|_2\preceq q^2 n^{-1}(t_1)\log n(t_1),
\end{eqnarray*}
with probability at least $1-O(n^{-1}(t_1))$. Under the given conditions, we have
\begin{eqnarray}\label{somematrix4}
	\max_{1\le k_1\le k_2\le K} \left\|\widehat{V}^*(k_1,k_2)-\widehat{V}^{**}(k_1,k_2)\right\|_2\preceq q^{1/2} n^{-1/2}(t_1)\log^{1/2} n(t_1),
\end{eqnarray}
with probability at least $1-O(n^{-1}(t_1))$.

Moreover, with some calculations, we can show that
\begin{eqnarray*}
	&&\max_{1\le k_1\le k_2\le K} \left\|\widehat{V}(k_1,k_2)-\widehat{V}^*(k_1,k_2)\right\|_2\le 	\sum_{a=0}^1 \max_{j\ge 1} \|\Sigma_a^{-1}-\widehat{\Sigma}_a^{-1}(t_j)\|_2\\
	&\times& 
	\max_{1\le k_1\le K} \frac{2}{n(t_{k_1})} \left\| \sum_{j=1}^{k_1}\sum_{i=n(t_{j-1})+1}^{n(t_j)}\mathbb{I}(A_i=a)\varphi(X_i) \varphi^\top(X_i) \{Y_i-\varphi^\top(X_i) \widehat{\beta}_a(t_j)\}^2 \Sigma_a^{-1} \right\|_2\\
	&+&\sum_{a=0}^1 \max_{1\le k_1\le K} \frac{1}{n(t_{k_1})} \left\| \sum_{j=1}^{k_1}\sum_{i=n(t_{j-1})+1}^{n(t_j)}\Sigma_a^{-1}\mathbb{I}(A_i=a)\varphi(X_i) \varphi^\top(X_i) \{Y_i-\varphi^\top(X_i) \widehat{\beta}_a(t_j)\}^2 \Sigma_a^{-1} \right\|_2\\
	&\times & \max_{j\ge 1} \|\Sigma_a^{-1}-\widehat{\Sigma}_a^{-1}(t_j)\|_2^2 .
\end{eqnarray*}
In view of Lemma \ref{somebasiclemma1} and Lemma \ref{somebasiclemma2}, we have with probability at least $1-O(n^{-\alpha_0}(t_1))$ that
\begin{eqnarray*}
	&&\max_{1\le k_1\le k_2\le K} \left\|\widehat{V}(k_1,k_2)-\widehat{V}^*(k_1,k_2)\right\|_2\le O(1) (q\delta_{n(t_1)}+\sqrt{q n^{-1}(t_1)\log n(t_1)})\\
	&\times & \max_{1\le k_1\le K} \frac{1}{n(t_{k_1})}\Big\|\underbrace{\sum_{j=1}^{k_1}\sum_{i=n(t_{j-1})+1}^{n(t_j)}\mathbb{I}(A_i=a)\varphi(X_i) \varphi^\top(X_i) \{Y_i-\varphi^\top(X_i) \widehat{\beta}_a(t_j)\}^2}_{\Psi_{3,a,k_1}}\Big\|_2,
\end{eqnarray*}
where $O(1)$ denotes some positive constant. Similar to \eqref{somematrix1} and \eqref{somematrix4}, we can show with probability at least $1-O(n^{-\alpha_0}(t_1))$ that
\begin{eqnarray*}
	\max_{a\in \{0,1\}} \max_{1\le k_1\le K} \left\|\frac{1}{n(t_{k_1})} \Psi_{3,a,k_1}-\Psi_a\right\|_2=o(1).
\end{eqnarray*}
Similar to Lemma \ref{somebasiclemma1}, we can show $\max_{a\in \{0,1\}}\|\Psi_a\|_2=O(1)$. It follows that
\begin{eqnarray*}
	\max_{1\le k_1\le k_2\le K} \left\|\widehat{V}(k_1,k_2)-\widehat{V}^*(k_1,k_2)\right\|_2\preceq q\delta_{n(t_1)}+\sqrt{q n^{-1}(t_1)\log n(t_1)},
\end{eqnarray*}
with probability at least $1-O(n^{-\alpha_0}(t_1))$. Combining this together with \eqref{somematrix1} and \eqref{somematrix4}, we obtain with probability at least $1-O(n^{-\alpha_0}(t_1))$ that
\begin{eqnarray*}
	\max_{1\le k_1\le k_2\le K} \left\|\widehat{V}(k_1,k_2)-V(k_1,k_2)\right\|_2\preceq q\delta_{n(t_1)}+\sqrt{q n^{-1}(t_1)\log n(t_1)}.
\end{eqnarray*}
Consider the function $\hbox{g}_{\delta}\circ \phi_{\eta, \{\nu_k\}_k}$ defined in the proof of Theorem \ref{thm1}. We fix $\delta=\eta^{-1} \{\log K + 4d\log n(t_1) \}$. Based on Lemma A2 in \cite{belloni2018}, we have with probability at least $1-O(n^{-\alpha_0}(t_1))$ that
\begin{eqnarray*}
	\sup_{\{\nu_k \}_k}\left|\Mean^* \hbox{g}_{\delta}\circ \phi_{\eta, \{\nu_k\}_k}(N(0,\bm{\widehat{V}}))-\Mean \hbox{g}_{\delta}\circ \phi_{\eta, \{\nu_k\}_k}(N(0,\bm{V}))\right|\\ \nonumber
	\preceq \eta^{2}\{\log^2 K+\log^2 n(t_1)\} \left(q\delta_{n(t_1)}+\sqrt{q n^{-1}(t_1)\log n(t_1)}\right),
\end{eqnarray*}
where $\Mean^*$ denotes the expectation conditional on the observed data. For a given set of thresholds $\{\nu_k\}_k$, using similar arguments in proving \eqref{secondinequality}, \eqref{thirdinequality}, \eqref{fourthinequality} and \eqref{fifthinequality}, we can show with probability at least $1-O(n^{-\alpha_0}(t_1))$ that
\begin{eqnarray*}
	&&\prob^*\left\{ \max_{k\in\{1,\dots,K\}} \left(\sqrt{n(t_k)}\widehat{S}^{\tiny{\hbox{MB}}*}-\nu_k\right)\le 0 \right\}\le \Mean^* \hbox{g}_{\delta}\circ \phi_{\eta, \{\nu_{k,+}\}_k}(N(0,\bm{\widehat{V}}))\\
	&\le& \Mean \hbox{g}_{\delta}\circ \phi_{\eta, \{\nu_{k,+}\}_k}(N(0,\bm{V}))+O(1) \eta^{2}\{\log^2 K+\log^2 n(t_1)\} \left(q\delta_{n(t_1)}+\sqrt{q n^{-1}(t_1)\log n(t_1)}\right)\\
	&\le&\prob\left\{ \max_{k\in\{1,\dots,K\}} \left(\sup_{x\in \mathbb{X}_0} \varphi^\top(x) G(t_k)-\nu_{k,+}^* \right)\le 0\right\}\\
	&+&O(1) \eta^{2}\{\log^2 K+\log^2 n(t_1)\} \left(q\delta_{n(t_1)}+\sqrt{q n^{-1}(t_1)\log n(t_1)}\right),
\end{eqnarray*}
and
\begin{eqnarray*}
	&&\prob^*\left\{ \max_{k\in\{1,\dots,K\}} \left(\sqrt{n(t_k)}\widehat{S}^{\tiny{\hbox{MB}}*}-\nu_k\right)\le 0 \right\}\ge \Mean^* \hbox{g}_{\delta}\circ \phi_{\eta, \{\nu_{k,-}\}_k}(N(0,\bm{\widehat{V}}))\\
	&\ge& \Mean \hbox{g}_{\delta}\circ \phi_{\eta, \{\nu_{k,-}\}_k}(N(0,\bm{V}))-O(1) \eta^{2}\{\log^2 K+\log^2 n(t_1)\} \left(q\delta_{n(t_1)}+\sqrt{q n^{-1}(t_1)\log n(t_1)}\right)\\
	&\ge& \prob\left\{ \max_{k\in\{1,\dots,K\}} \left(\sup_{x\in \mathbb{X}_0} \varphi^\top(x) G(t_k)-\nu_{k,-}^* \right)\le 0\right\}\\
	&-&O(1) \eta^{2}\{\log^2 K+\log^2 n(t_1)\} \left(q\delta_{n(t_1)}+\sqrt{q n^{-1}(t_1)\log n(t_1)}\right),
\end{eqnarray*}
where $O(1)$ denotes some positive constant, and
\begin{eqnarray*}
	&&\nu_{k,+}=\nu_k + \eta^{-1}\{4d\log n(t_1)+\log K\}+\bar{c}^* n^{-2}(t_1),\,\,\,\,\nu_{k,+}^*=\nu_{k,+}+3\eta^{-1}\{4d\log n(t_1)+\log K\},\\
	&&\nu_{k,-}=\nu_k - 3\eta^{-1}\{4d\log n(t_1)+\log K\}-\bar{c}^* n^{-2}(t_1),\,\,\,\,\nu_{k,-}^*=\nu_{k,-}-\eta^{-1}\{4d\log n(t_1)+\log K\}.
\end{eqnarray*}
By Theorem 2 of \cite{cherno2017}, we obtain that
\begin{eqnarray*}
	\hbox{Pr}\left\{\max_{k\in\{1,\dots,K\}} \left(\sup_{x\in \mathbb{X}_0} \frac{\varphi^\top(x) G(t_k)}{\sigma(x,t_k)} -\nu_{k,+}^{*}\right)\le 0 \right\}-\hbox{Pr}\left\{\max_{k\in\{1,\dots,K\}} \left(\sup_{x\in \mathbb{X}_0} \frac{\varphi^\top(x) G(t_k)}{\sigma(x,t_k)} -\nu_{k,-}^{*}\right)\le 0 \right\}\\
	\preceq \eta^{-1}\{\log^{3/2} n(t_1)+\log^{3/2} K\}+\bar{c}^* n^{-2}(t_1) \{\log^{1/2} n(t_1)+\log^{1/2} K\}. 
\end{eqnarray*}
It follows that
\begin{eqnarray*}
	\sup_{\{\nu_k\}_k} \left| \prob^*\left\{ \max_{k\in\{1,\dots,K\}} \left(\sqrt{n(t_k)}\widehat{S}^{\tiny{\hbox{MB}}*}-\nu_k\right)\le 0 \right\}-\prob\left\{ \max_{k\in\{1,\dots,K\}} \left(\sup_{x\in \mathbb{X}} \frac{\varphi^\top(x) G(t_k)}{\sigma(x,t_k)}-\nu_{k} \right)\le 0\right\}\right|\\
	\preceq \eta^{2}\{\log^2 K+\log^2 n(t_1)\} \left(q\delta_{n(t_1)}+\sqrt{q n^{-1}(t_1)\log n(t_1)}\right)\\
	+\eta^{-1}\{\log^{3/2} n(t_1)+\log^{3/2} K\}+\bar{c}^* n^{-2}(t_1) \{\log^{1/2} n(t_1)+\log^{1/2} K\},
\end{eqnarray*}
with probability at least $1-O(n^{-\alpha_0}(t_1))$. Set $$\eta=\min[q^{-1/3}n^{\alpha_0/3}(t_1)\log^{-(1+2\alpha_0)/6} \{K n(t_1)\} , q^{-1/6} n^{1/6}(t_1) \log^{-1/3} \{K n(t_1)\}],$$ 
we obtain the desired result.

\subsection{Proof of Theorem \ref{thm6}}
We will show that under the current conditions, the stopping boundaries $\{\widehat{z}_k\}_k$ are upper bounded by $c\sqrt{\log N(t_1)}$ for some constant $c>0$, with probability tending to $1$. Using similar arguments in proving Lemma \ref{somebasiclemma4} and Equation \eqref{someimportantinequality0}, we can show that
\begin{eqnarray*}
	\left| \sup_{x\in \mathbb{X}} \frac{\sqrt{N(t_K)}\varphi^\top(x) \{\widehat{\beta}_1(t_K)-\beta_1^*-\widehat{\beta}_0(t_K)+\beta_0^*\}}{\widehat{s.e.}[\varphi^\top(x) \{\widehat{\beta}_1(t_K)-\widehat{\beta}_0(t_K)\}]} \right|
\end{eqnarray*}
is upper bounded by $O\{\sqrt{\log N(t_1)}\}$ as well, with probability tending to $1$. 

Under the alternative hypothesis and the assumption on the approximation error, 
\begin{eqnarray*}
	\sqrt{N(t_K)}\sup_{x\in \mathbb{X}} \varphi^\top(x)(\beta_1^*-\beta_0^*)\gg \sqrt{q\log \{N(t_1)\}}. 
\end{eqnarray*} 
Using similar arguments in the proof of Theorem \ref{thm1}, the standard error estimator is of the order of magnitude $O(\sqrt{q})$, uniformly for any $x$. 
It follows that 
\begin{eqnarray*}
	\sqrt{N(t_K)}\sup_{x\in \mathbb{X}} \frac{\varphi^\top(x)(\beta_1^*-\beta_0^*)}{\widehat{s.e.}[\varphi^\top(x) \{\widehat{\beta}_1(t_K)-\widehat{\beta}_0(t_K)\}]}>\widehat{z}_K,
\end{eqnarray*}
with probability tending to 1. The proof is hence completed.

It remains to show that the stopping boundaries are upper bounded by $c\sqrt{q\log N(t_1)}$. We will prove this assertion by induction. We first notice that $\widehat{z}_1$ satisfies
\begin{eqnarray*}
	\prob\left(\left.\sup_{x\in \mathbb{X}} \frac{\varphi^\top(x) \widehat{S}^{\textrm{MB}*}(t_1)}{\widehat{s.e.}[\varphi^\top(x) \{\widehat{\beta}_1(t_K)-\widehat{\beta}_0(t_K)\}]} -\widehat{z}_1> 0\right|\textrm{Data} \right)=\alpha(t_1).
\end{eqnarray*}
%by Theorem \ref{thm4}. 
The conditional variance of $\varphi^\top(x)\widehat{S}^{\textrm{MB}*}(t_1)/\|\varphi(x)\|_2$ is upper bounded by $O(1)$. The denominator is lower bounded by $c \|\varphi(x)\|_2$ for some constant $c>0$. 
As such, there exists some constant $c_1>0$ such that
\begin{eqnarray*}
	\prob\left(\left.\sup_{x\in \mathbb{X}} \frac{\varphi^\top(x) \widehat{S}^{\textrm{MB}*}(t_1)}{\widehat{s.e.}[\varphi^\top(x) \{\widehat{\beta}_1(t_1)-\widehat{\beta}_0(t_1)\}]}>c_1 \sqrt{\log N(t_1)}\right|\textrm{Data} \right)\le \frac{1}{N^C(t_1)}.
\end{eqnarray*}
As such, $\widehat{z}_1\le c_1 \sqrt{\log N(t_1)}$.  

Suppose we have shown that $\{\widehat{z}_j\}_{j=1}^k$ are upper bounded by $c_k\sqrt{\log N(t_1)}$. We aim to show $\widehat{z}_{k+1}$ is upper bounded by $c_{k+1}\sqrt{\log N(t_1)}$ for some constant $c_{k+1}>0$. A key observation is that
\begin{eqnarray*}
	\prob\left(\left.\sup_{x\in \mathbb{X}} \frac{\varphi^\top(x) \widehat{S}^{\textrm{MB}*}(t_{k+1})}{\widehat{s.e.}[\varphi^\top(x) \{\widehat{\beta}_1(t_1)-\widehat{\beta}_0(t_1)\}]} -\widehat{z}_{k+1}> 0\right|\textrm{Data} \right)\\\ge 	\prob\left(\left.\max_{1\le j\le k+1} \sup_{x\in \mathbb{X}} \frac{\varphi^\top(x) \widehat{S}^{\textrm{MB}*}(t_j)}{\widehat{s.e.}[\varphi^\top(x) \{\widehat{\beta}_1(t_1)-\widehat{\beta}_0(t_1)\}]} -\widehat{z}_j> 0\right|\textrm{Data} \right)\\-\prob\left(\left.\max_{1\le j\le k} \sup_{x\in \mathbb{X}} \frac{\varphi^\top(x) \widehat{S}^{\textrm{MB}*}(t_j)}{\widehat{s.e.}[\varphi^\top(x) \{\widehat{\beta}_1(t_1)-\widehat{\beta}_0(t_1)\}]} -\widehat{z}_j> 0\right|\textrm{Data} \right)=\alpha(t_{k+1})-\alpha(t_k).
\end{eqnarray*}
Since $\alpha(t_{k+1})-\alpha(t_k)\ge N^C(t_1)$, using similar arguments in proving $\widehat{z}_1\le c_1 \sqrt{\log N(t_1)}$, we can show that $\widehat{z}_{k+1}\le c_{k+1}\sqrt{\log N(t_1)}$. This shows that the power of the proposed test approaches to one. Similarly, one can show that the stopping time is upper bounded by $t_k$, as long as  
\begin{eqnarray*}
	\sqrt{N(t_k)}\sup_{x\in \mathbb{X}} \varphi^\top(x)(\beta_1^*-\beta_0^*)\gg \sqrt{q\log \{N(t_1)\}}. 
\end{eqnarray*} 

\section{Comparison of the baseline method}\label{secbaseline}
We first introduce the test based on LIL. 
Consider our test statistic $S(t)$. Under $H_0$, it can be bounded from above by
\begin{eqnarray}\label{eqn:bound}
	\sup_{x\in \mathbb{X}}\varphi^\top (x)\{\widehat{\beta}_1(t)-\beta_1^*-\widehat{\beta}_0(t)+\beta_0^* \}.
\end{eqnarray}
It suffices to provide an upper bound for the above expression. By Cauchy-Schwarz inequality, \eqref{eqn:bound} can be upper bounded by
\begin{eqnarray*}
	\sup_{x\in \mathbb{X}}\|\varphi (x)\|_2 \| \widehat{\beta}_1(t)-\beta_1^*-\widehat{\beta}_0(t)+\beta_0^*\|_2.
\end{eqnarray*}
It suffices to provide anytime upper bound for $\| \widehat{\beta}_1(t)-\beta_1^*-\widehat{\beta}_0(t)-\beta_0^*\|_2$.

Recall that
\begin{eqnarray*}
	&&\widehat{\beta}_1(t)-\beta_1^*-\widehat{\beta}_0(t)+\beta_0^*\\
	&=&\frac{1}{N(t)}\sum_{i=1}^{N(t)} [\mathbb{I}(A_i=1)\widehat{\Sigma}_1^{-1}(t)\varphi(X_i)\{Y_i-\varphi^\top(X_i) \beta_1^*\} - \mathbb{I}(A_i=0)\widehat{\Sigma}_0^{-1}(t)\varphi(X_i)\{Y_i-\varphi^\top(X_i) \beta_0^*\} ].
\end{eqnarray*}
The above expression is asymptotically equivalent to
\begin{eqnarray*}
	\frac{1}{N(t)}\sum_{i=1}^{N(t)} [\mathbb{I}(A_i=1)\Sigma_1^{-1}\varphi(X_i)\{Y_i-\varphi^\top(X_i) \beta_1^*\} - \mathbb{I}(A_i=0)\Sigma_0^{-1}\varphi(X_i)\{Y_i-\varphi^\top(X_i) \beta_0^*\} ].
\end{eqnarray*}
By the law of iterated logarithm, the $\ell$-th dimension of the above expression can be upper bounded by
\begin{eqnarray*}
	N^{-1/2}(t)\sqrt{2\sigma_{\ell}^2 \log \log \{N(t)\} } ,
\end{eqnarray*}
where $\sum_{\ell} \widehat{\sigma}_{\ell}^2$ can be consistently estimated by
\begin{eqnarray*}
	\frac{1}{N(t)}\sum_{i=1}^{N(t)} \|\mathbb{I}(A_i=1)\widehat{\Sigma}_1^{-1}(t)\varphi(X_i)\{Y_i-\varphi^\top(X_i) \widehat{\beta}_1(t)\} - \mathbb{I}(A_i=0)\widehat{\Sigma}_0^{-1}(t)\varphi(X_i)\{Y_i-\varphi^\top(X_i) \widehat{\beta}_0(t)\} \|_2^2.
\end{eqnarray*}
As such, the finite error bound is given by
\begin{eqnarray*}
	\sup_{x\in \mathbb{X}}\|\varphi (x)\|_2 \frac{\sqrt{2\log \log \{N(t)\}}}{\sqrt{N(t)}}\times \\
	\sqrt{\frac{1}{N(t)}\sum_{i=1}^{N(t)} \|\mathbb{I}(A_i=1)\widehat{\Sigma}_1^{-1}(t)\varphi(X_i)\{Y_i-\varphi^\top(X_i) \widehat{\beta}_1(t)\} - \mathbb{I}(A_i=0)\widehat{\Sigma}_0^{-1}(t)\varphi(X_i)\{Y_i-\varphi^\top(X_i) \widehat{\beta}_0(t)\} \|_2^2}.
\end{eqnarray*}

{\color{black}We next discuss the test based on AVT. At time $t$, we compute the following test statistic
	\begin{eqnarray*}
		\sqrt{\frac{\widehat{\sigma}^2/(N_0^{-1}(t)+N_1^{-1}(t))}{\widehat{\sigma}^2/(N_0^{-1}(t)+N_1^{-1}(t))+\tau^2}}\exp\left[\frac{\tau^2 \{N_0^{-1}(t)\sum_{i=1}^{N(t)} (1-A_i)Y_i-N_1^{-1}(t)\sum_{i=1}^{N(t)} A_i Y_i\}^2 }{2\{\widehat{\sigma}^2/(N_0^{-1}(t)+N_1^{-1}(t))\}\{\widehat{\sigma}^2/(N_0^{-1}(t)+N_1^{-1}(t))+\tau^2\} }\right],
	\end{eqnarray*}
	where $N_a(t)=\sum_{i=1}^{N(t)} \mathbb{I}(A_i=a)$ and $\widehat{\sigma}^2$ is the pooled variance estimator $\{N(t)-2\}^{-1} [\{N_0(t)-1\} \widehat{\sigma}_0^2+\{N_1(t)-1\} \widehat{\sigma}_1^2]$ where $\widehat{\sigma}_a^2$ denotes the sampling variance estimator based on $\{Y_i\}_{\mathbb{I}(A_i=a)}$. The constant $\tau$ corresponds to a hyperparameter and we fix $\tau=1$ in our implementation. 
}
\clearpage
\bibliography{CHTE}

\end{document}